\documentclass[fleqn,nonblindrev]{informs4}

\OneAndAHalfSpacedXI 

\usepackage{endnotes}
\let\footnote=\endnote
\let\enotesize=\small   
\def\notesname{Endnotes}%
\def\makeenmark{$^{\theenmark}$}
\def\enoteformat{\rightskip0pt\leftskip0pt\parindent=1.75em
	\leavevmode\llap{\theenmark.\enskip}}
\usepackage{booktabs}
\usepackage[most]{tcolorbox}
\tcbset{
  colback=gray!5,
  colframe=gray!60,
  boxrule=0.5pt,
  arc=3pt,
  left=6pt,
  right=6pt,
  top=6pt,
  bottom=6pt
}
\usepackage{soul} 
\usepackage{xcolor}
\usepackage{booktabs}
\usepackage{enumitem}

\usepackage{graphicx}
\usepackage{eqndefns-left}
\usepackage{multirow}
\usepackage{hhline}
\usepackage{adjustbox}
\usepackage[small, margin=1cm]{caption}
\usepackage{appendix}
\usepackage{color}
\definecolor{strcolor}{rgb}{0.6, 0.2, 0.6}
\definecolor{commentcolor}{rgb}{0.3125, 0.5, 0.3125}
\definecolor{keycol}{rgb}{0, 0, 1}

\usepackage{bbm}

\usepackage{listings}
\usepackage{url}

\newcommand {\bea}{\begin{eqnarray}}
	\newcommand {\eea}{\end{eqnarray}}

\newcommand {\pos}[1]{\paren{#1}^+}

\newtheorem{algorithm}{Algorithm}

\def\blot{\quad \mbox{$\vcenter{ \vbox{ \hrule height.4pt
				\hbox{\vrule width.4pt height.9ex \kern.9ex \vrule width.4pt}
				\hrule height.4pt}}$}}

\usepackage{natbib}
\bibpunct[, ]{(}{)}{,}{a}{}{,}%
\def\bibfont{\fontsize{8}{9.5}\selectfont}%
\def\todo#1{\textcolor{red}{TODO: #1}}

\usepackage{amsmath,amsfonts,bm,mathtools}

\def\eqref#1{equation~\ref{#1}}

\def\1{\bm{1}}

\def\rvh{{\mathbf{h}}}

\def\rvz{{\mathbf{z}}}

\def\ve{{\bm{e}}}

\def\vk{{\bm{k}}}

\def\vq{{\bm{q}}}

\def\vs{{\bm{s}}}

\def\vv{{\bm{v}}}

\def\vx{{\bm{x}}}

\def\vz{{\bm{z}}}

\def\vbeta{{\bm{\beta}}}

\def\mZ{{\bm{Z}}}

\DeclareMathAlphabet{\mathsfit}{\encodingdefault}{\sfdefault}{m}{sl}
\SetMathAlphabet{\mathsfit}{bold}{\encodingdefault}{\sfdefault}{bx}{n}

\newtheorem{lem}{Lemma}

\newif\ifshownotes
\shownotestrue

\def\todo#1{\ifshownotes\textcolor{red}{TODO: #1}\fi}

\usepackage{amsmath}

\makeatletter
\def\Dskips#1#2#3#4{\abovedisplayskip 0.5em \@plus#2em \@minus#3em
   \abovedisplayshortskip \z@ \@plus#2em
   \belowdisplayshortskip 0.3em \@plus#2em \@minus#2em
   \belowdisplayskip \abovedisplayskip}%
\makeatother

\usepackage{algorithm,algorithmic}
 \usepackage{subcaption}
\renewcommand{\algorithmiccomment}[1]{\bgroup\hfill//~#1\egroup}

\usepackage{multirow, booktabs}

\usepackage{listings}
\TheoremsNumberedThrough     
\ECRepeatTheorems

\EquationsNumberedThrough    

\usepackage{pifont}
\newcommand{\cmark}{\ding{51}}
\newcommand{\xmark}{\ding{55}}

\gdef\AQ#1{}
\gdef\CQ#1{}
\begin{document}
	
\def\COPYRIGHTHOLDER{INFORMS}%
\def\COPYRIGHTYEAR{2017}%
\def\DOI{\fontsize{7.5}{9.5}\selectfont\sf\bfseries\noindent https://doi.org/10.1287/opre.2017.1714\CQ{Word count = 9740}}

\RUNAUTHOR{Wang et~al.} %

\RUNTITLE{Can Explanations Improve Recommendations? Evidence from Prediction-Informed Explanations}
\TITLE{Can Explanations Improve Recommendations? Evidence from Prediction-Informed Explanations}

\ARTICLEAUTHORS{

\AUTHOR{Yuyan Wang\textsuperscript{1}\thanks{We thank participants at the AI in Management Conference, the ICML 2025 Interpretability Workshop, NeurIPS 2025 Efficient Reasoning Workshop, BizAI Conference 2026, Marketing Science 2026, Netflix Research Seminar, and Stanford Quant Marketing WIP Seminar for helpful comments and suggestions. All errors are our own.}, Pan Li\textsuperscript{2}, Minmin Chen\textsuperscript{3}}
\AFF{\textsuperscript{1}Stanford Graduate School of Business, \textsuperscript{2}Scheller College of Business, Georgia Institute of Technology, \textsuperscript{3}Google Inc.}






}
	

\ABSTRACT{
Recommender systems are central to how modern digital platforms match consumers with content, yet state-of-the-art models are often black-box, limiting their interpretability and adoption in many business settings. Existing explainable AI approaches typically treat explanations as separate or post-hoc tasks, often at the cost of predictive accuracy. We challenge this practice and propose that explanations, when incorporated as an \emph{integral} part of the learning process, can improve \emph{both} explainability and predictive performance. We introduce RecPIE (Recommendation with Prediction-Informed Explanations), a framework that jointly learns personalized natural-language explanations and recommendation predictions. Central to RecPIE's design is our proposed \emph{usefulness} reward, which uses downstream recommendation accuracy to guide explanation generation, in contrast to existing literature that focuses on explanation quality objectives alone, whose generated explanations do not necessarily improve predictive accuracy. Specifically, guided by this reward, RecPIE fine-tunes an LLM using LoRA through reinforcement learning (PPO) to generate \emph{prediction-informed explanations}, which are then embedded into a discriminative recommender to improve predictions (\emph{explanation-informed predictions}), creating a mutually reinforcing learning loop. We evaluate RecPIE on large-scale point-of-interest recommendation from Google Maps, a challenging recommendation setting where consumer preferences span diverse place categories and are long-tailed, requiring the recommender system to generalize across categories and to data-sparse scenarios. RecPIE improves predictive accuracy by 1.7--3.6\% over state-of-the-art baselines and outperforms the strongest full-data baselines using only 25\% of the training data. The gains are largest exactly where black-box AI systems struggle most: consumers with niche tastes and sparse histories, where firms face a particularly difficult targeting problem with limited information. In a human study with 566 participants, RecPIE's explanations are preferred 61.5\% of the time (compared with 16.6\% for the strongest baseline), and are rated closer to consumer-elicited reasons. Together, these results suggest that insights about consumer behavior and AI decisions need not be imposed externally or generated post hoc; they can be learned as an integral component of the AI systems, allowing the systems themselves to discover which insights are useful for improving the objective the platform ultimately cares about.
}








	
	%
	
\maketitle

\section{Introduction}
\label{sec:intro}



Recommender systems are central to modern digital platforms, shaping how consumers discover products, content, and services \citep{adomavicius2005toward, ghose2012designing, ricci2010introduction, schafer1999recommender}. Despite their strong performance, these systems are typically black boxes, offering limited insight into why a recommendation is made or why a consumer chooses a particular product. In recent years, there has been a growing demand for explainability driven by the need for transparency, regulatory concerns, and implications for user trust and adoption \citep{bauer2023expl, song2025customer}. Recommender systems are therefore expected to achieve high accuracy and meaningful explainability simultaneously.
 
Existing approaches in the explainable AI (XAI) literature typically treat prediction and explanation as separate, disjoint tasks \citep{nauta2023anecdotal}. As a result, they often introduce a trade-off: explanations improve interpretability at the cost of predictive performance \citep{mohammadi2025regulating, zhang2023comprehensible, linardatos2020explainable}. However, as firms rely on accurate targeting to generate business value, they are often unwilling to sacrifice accuracy for explainability. As a result, most platforms continue to rely on black-box models as their recommender system \citep{zhai2024actions, covington2016deep}, and rely on post-hoc explanation tools that are often unreliable \citep{ragodos2024model}. 

In this paper, we argue that this trade-off is not inherent. Instead, we propose that explanations can be designed as an \emph{integral} component of the learning process and, when properly aligned with prediction outcomes, can improve \emph{both} interpretability and performance. We introduce \textbf{RecPIE} (Recommendation with Prediction-Informed Explanations), a framework that \emph{jointly} optimizes recommendation prediction and explanation generation. The key idea is to embed explanation generation directly into the learning loop: predictions guide the generation of explanations (\textbf{prediction-informed explanations}), and these explanations are fed back to improve subsequent predictions (\textbf{explanation-informed predictions}). This bidirectional structure allows both objectives to mutually reinforce each other.

Our approach is motivated by recent advances in large language models (LLMs), which show that explicit reasoning and natural-language explanations can improve performance in many domains \citep{wei2022chain, wang2025works}. However, using such reasoning in recommender systems presents two key challenges. First, unlike domains such as mathematics or coding, where ground-truth reasoning is well defined, consumer decisions are inherently personalized: different users may choose the same product for different reasons. As a result, ground-truth explanations are not observed, and naively generated explanations may be noisy or misleading due to LLM hallucination \citep{xu2024hallucination, kalai2025language}, potentially degrading performance \citep{lebovitz2021ai, susarla2023janus, aagerfalk2022artificial, brucks2025prompt}. Second, recommender systems are fundamentally \emph{discriminative}, while LLMs are \emph{generative} models, which typically perform poorly when used directly for prediction \citep{ye2025lola, tsai2024leveraging}, raising the question of how generative reasoning can be effectively integrated into discriminative decision-making scenarios.

RecPIE addresses these challenges by integrating a generative explanation component with a discriminative recommendation component and learning which explanations are \emph{useful} for prediction. To address the first challenge, the absence of ground-truth explanations and the risk of LLM hallucination, RecPIE aligns the two components through a \emph{shared downstream objective} and a shared personalized soft prompt. Specifically, it optimizes LLM-generated explanations for \emph{usefulness}, defined as their contribution to downstream predictive performance: explanations that improve predictions are reinforced, while those that do not are penalized. This enables the system to learn personalized explanations grounded in observable outcomes without requiring ground-truth explanations. To address the second challenge, the mismatch between generative and discriminative models, RecPIE incorporates LLM-generated explanations as inputs to a discriminative recommender rather than using the LLM for direct prediction. This integration enables explanation-informed predictions to outperform standard prediction-only methods. The two components, prediction-informed explanations and explanation-informed predictions, are trained alternately, with the LLM fine-tuned on a customized usefulness reward through reinforcement learning with proximal policy optimization (PPO) \citep{schulman2017proximal}. This process iteratively refines both personalized explanations and predictive performance.

We evaluate RecPIE in a large-scale industrial setting using Google Maps point-of-interest recommendation data \citep{li2022uctopic, yan2023personalized}, where the goal is to predict a consumer's preference for a future candidate place given her past visitation history. This is one of the most challenging recommendation scenarios as it often requires cross-category reasoning (e.g. inferring a preference for an art gallery from past visits to restaurants). We compare RecPIE with 21 state-of-the-art recommendation baselines on both the recommendation accuracy and explanation quality. On recommendation accuracy, RecPIE improves over the best baseline in each metric: \textbf{3.28\%} in RMSE, \textbf{3.56\%} in MAE, and \textbf{1.67\%} in AUC. Such gains are economically significant, as it has been shown that even improvements of less than 0.5\% in recommendation accuracy can translate to revenue increases of millions of dollars at platform scale \citep{wang2025recommending, gomez2015netflix, li2024variety, chen2024background}. On explanation quality, human evaluations with 566 participants show that RecPIE's explanations are preferred 61.5\% of the time (versus 16.6\% for the most preferred baseline). In addition, we provide evidence that RecPIE's explanations are faithful both to the user and to the model, and they exhibit significantly higher coverage of key aspects identified in participants' ground-truth explanations, suggesting that RecPIE is able to capture the underlying factors driving consumer decision-making.

RecPIE also substantially improves learning efficiency and reduces data requirements. It outperforms the strongest full-data baselines using only 25\% of the training data, and converges to semantically meaningful item representations up to four times faster. The gains are largest where black-box recommender systems struggle most: consumers with niche tastes and sparse histories, where firms face a particularly difficult targeting problem with limited information. Finally, we shed light on the potential mechanism behind RecPIE's gains and provide evidence that the performance gains mainly come from LLMs' reasoning capabilities, not from external knowledge or summarization skills.

Our work contributes to the design and deployment of AI decision systems in digital platforms. First, we introduce a new perspective on explainability in recommender systems. Instead of treating explanations as separate or post-hoc tasks, we incorporate them as an \emph{integral} component of the learning process, enabling AI systems to \emph{learn} which explanations of consumer behavior are useful for predicting future preferences. By aligning explainability with predictive accuracy, RecPIE helps mitigate a long-standing trade-off that has limited firms' adoption of explainable AI \citep{bauer2023expl, song2025customer, mohammadi2025regulating}. RecPIE also supports efficient deployment by allowing explanations to be precomputed and cached, reducing real-time LLM inference costs while enabling platforms to retain control over sensitive consumer data, including location histories, to protect consumer privacy.

Second, our work highlights the value of incorporating behavioral and structural insights into the design of AI systems to improve data efficiency and generalization. Modern AI increasingly faces data limitations, as acquiring additional high-quality data becomes costly and difficult \citep{sutskever2024fossilfuel}. As a result, these systems must extrapolate to unseen consumers, products, and environments, where black-box models often struggle to capture the underlying data-generating process and generalize effectively \citep{wang2024going}. RecPIE addresses this challenge by learning useful explanations for predicting consumer behavior, allowing it to extract informative signals from limited data. RecPIE outperforms the strongest baseline using only a fraction of the training data and is particularly effective in data-sparse settings and for consumers with niche tastes, while demonstrating strong generalizability to previously unseen consumers and places. 
These advantages make RecPIE particularly promising for data-scarce platforms and those seeking to expand to new consumers, products, and regions \citep{ibragimov2025transfer}.
 
The rest of the paper is organized as follows. Section~\ref{sec:related_work} reviews related literature. Section~\ref{sec:formulation} discusses the motivation. Section~\ref{sec:framework} presents the RecPIE framework in detail. Sections~\ref{sec:context}--\ref{sec:understanding} report the empirical setting and results. Section~\ref{sec:exp_additional_privacy_latency} discusses practical deployment considerations. Section~\ref{sec:discussion} concludes with a discussion.

\section{Related Work}
\label{sec:related_work}



\subsection{Explainable AI in Digital Platforms}
\label{sec:related_xai}

Explanations play a central role in intelligent systems by improving user trust, understanding, and adoption \citep{meske2022explainable, gregor1999explanations, bauer2023expl, martens2014explaining}. Building on this, explainable AI (XAI) aims to make machine learning models interpretable \citep{brasse2023explainable, adadi2018peeking, rudin2019stop, nauta2023anecdotal, mohammadi2025regulating, sisodia2025generative} and is seeing growing adoption in digital platforms \citep{song2025customer, chen2025beyond}.

Existing XAI methods fall into three broad categories. \emph{Post-hoc explanations} generate explanations for trained models \citep{chen2018learning, oramas2017visual, wagner2019interpretable, lundberg2017unified}, but are often unreliable and do not improve predictive performance \citep{ragodos2024model}. \emph{Intrinsically interpretable} or white-box models build interpretability into the architecture \citep{mohammadi2025regulating, pan2021explainable, subramanian2020obtaining}, but typically sacrifice accuracy \citep{zhang2023comprehensible}. \emph{Supervised explanation training} uses ground-truth explanations during learning \citep{camburu2018snli, liu2018towards, sun2020dual}, but requires costly and subjective human annotations. Appendix~\ref{appen:related_work_expl_recsys} summarizes the literature on these techniques applied to recommender systems. In summary, prior XAI literature has documented an accuracy--explainability trade-off, whereby improving interpretability can come at the expense of predictive or economic performance \citep{bauer2021expl, muller2016utilizing, rai2020explainable, zhang2023comprehensible, hartmann2023more, burrell2016machine}. For example, \citet{zhang2023comprehensible} demonstrates in a field experiment that restricting a firm to interpretable policies reduces profits by 7.5\%, or 23 cents per customer. In contrast, we demonstrate that jointly optimizing explanation generation and prediction under a shared objective can improve \emph{both} explanation quality and predictive accuracy, suggesting that explainability need not come at the cost of performance and offering a promising approach to mitigating this long-standing trade-off.

In evaluating explanations, existing XAI research focuses on qualities of explanations themselves, such as \emph{plausibility}, whether an explanation is convincing to a human, or \emph{faithfulness}, whether it reflects the reasoning that underlies a decision \citep{herman2017promise, jacovi2020towards, lyu2024faithful}. However, we propose a \emph{usefulness} objective, which measures how much the explanations contribute to the accuracy of the downstream prediction task, and show that such usefulness objectives can improve both the explainability and prediction accuracy while preserving plausibility and faithfulness of the generated explanations.

\subsection{LLMs and Explanations in Recommender Systems}
\label{sec:related_llm_rec}

Building on the success of LLMs, a growing body of work explores their use in recommender systems \citep{wu2024survey, xing2025llms}. Existing methods fall into three main streams. The first uses LLMs as feature extractors or augmenters, generating embeddings from user and item information for downstream recommenders \citep{peng2023gpt, ren2024representation, wang2024llm}. The second repurposes or fine-tunes LLMs as recommender systems, drawing on user profiles, behavioral prompts, and task instructions \citep{he2023large, huang2023recommender, wang2024rethinking, bao2023tallrec, zhai2024actions, deng2025onerec, he2026plum}. These approaches focus on using LLMs to improve user or item understanding, and do not leverage LLMs for explanations. The third stream is to use LLMs for explanation generation, such as grounding the explanations in the recommender system \citep{leng2026interpretable, wang2018reinforcement}, or prompting or fine-tuning LLMs based on explanation objectives \citep{wang2023llm4vis, tsai2024leveraging} or ground-truth explanations \citep{yang2024llm2er, li2023personalized, li2023prompt} that are difficult to obtain at scale. These approaches focus on the quality of the generated explanations themselves, and do \emph{not} use explanations to \emph{improve} recommendation performance. Closest to our work is ExpCTR \citep{yu2025expctr}, which fine-tunes LLMs using alignment rewards to generate explanations for CTR prediction. Unlike RecPIE, its reward does not directly optimize explanations for downstream predictive \emph{usefulness}, and its explanation generator lacks the collaborative filtering signals enabled by our shared learnable soft prompts.

\subsection{Point-of-Interest (POI) Recommender Systems}

Our empirical setting is point-of-interest (POI) recommender systems, which predict consumers' preferences for candidate places based on their visitation histories. This is arguably one of the hardest recommendation problems as it requires cross-category generalization and learning from sparse, long-tailed interactions \citep{ye2011exploiting, yang2014modeling}. Existing research extends collaborative filtering with geographical and social signals \citep{ye2011exploiting, cheng2012fused}, while subsequent neural approaches model spatiotemporal dependencies through recurrent and attention-based architectures \citep{liu2016predicting, feng2018deepmove, sun2020where, luo2021stan}. More recently, \citet{li2024large} leverage pretrained LLMs to encode heterogeneous location-based social network data. Despite these advances, POI recommendation decisions remain largely uninterpretable. RecPIE addresses this limitation by learning explicit, \emph{personalized} explanations that improve both interpretability and predictive accuracy.

\section{Research Context and Motivation}
\label{sec:formulation}


\subsection{Point-of-Interest Recommender Systems}  
\label{sec:formulation_poi}

\emph{Point-of-interest (POI) recommendation} is an important application of modern recommender systems. As consumers increasingly rely on digital platforms to decide where to go, personalized location recommendations have become a central feature of services such as TripAdvisor and Google Maps. We show the representative interfaces of both platforms in Appendix~\ref{appen:poi_interfaces}. 

\label{para:poi_defn}Mathematically, POI recommendation follows the standard formulation of personalized recommender systems \citep{covington2016deep}: identifying a small set of relevant items from a large catalog and presenting them in a ranked list tailored to each consumer's preferences and context. Let $i$ index consumers, $j$ index places, and $\vz$ denote contextual information (e.g., time and location). For each candidate place $j$, the system predicts a preference score $\hat{y}(i,j,\vz)$ corresponding to a target outcome $y$, which is operationalized as a rating \citep{yang2019exploring, yan2018personalized}:
\begin{equation}
  \label{eqn:ml_recsys}
  \hat{y}(i,j,\vz) = f(\vx_i, \vx_j, \vx_{ij}, \{j_{i_1},\ldots,j_{i_n}\}, \vz),
\end{equation}
where $\vx_i$ and $\vx_j$ denote consumer and place features, $\vx_{ij}$ captures consumer–place interaction features, and $\{j_{i_1},\ldots,j_{i_n}\}$ represents consumer $i$'s visitation history. 

\begin{figure}[hbtp!]
    \begin{subfigure}[b]{0.43\textwidth}
        \centering
        \includegraphics[width=\textwidth]{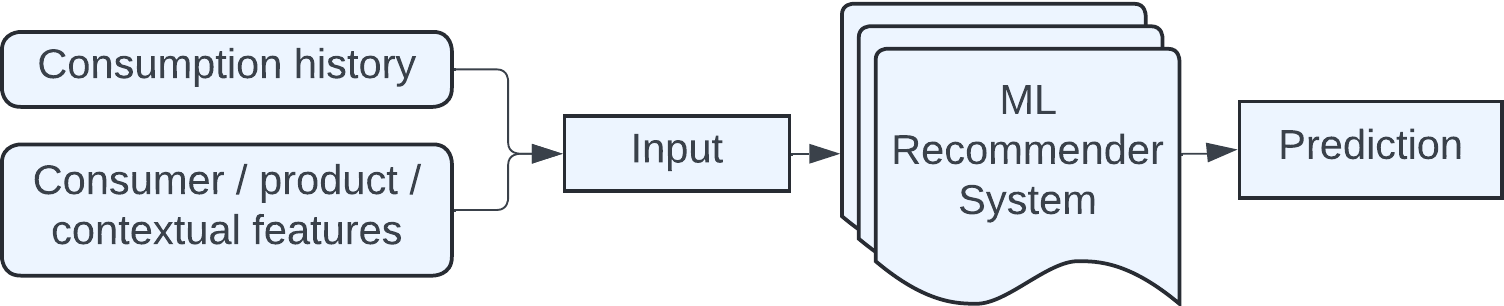}
        \caption{A typical recommender system.}
        \label{fig:illustration_baseline}
    \end{subfigure}
    \hfill
     \centering
    \begin{subfigure}[b]{0.53\textwidth}
        \centering
        \includegraphics[width=\textwidth]{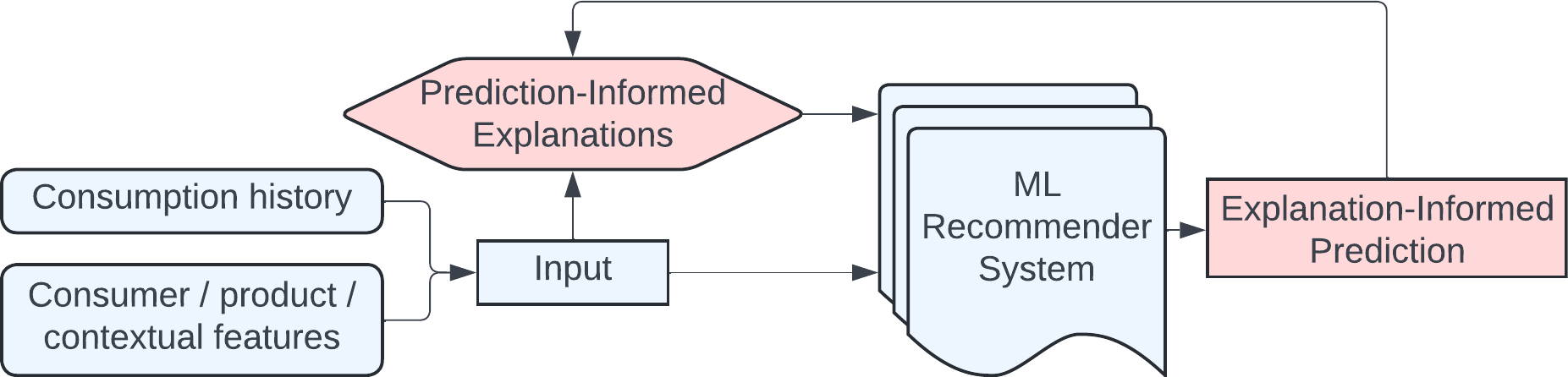}
        \caption{RecPIE (detailed architecture in Figure~\ref{fig:llm_recsys_diagram}).}
        \label{fig:illustration_framework}
    \end{subfigure}
    \caption{Comparison between our proposed framework, RecPIE, and a typical recommender system.}
  \label{fig:comparison}
 \vspace{-0.2cm}
\end{figure}

Figure~\ref{fig:illustration_baseline} illustrates a typical recommender system. While $f(\cdot)$ may be any machine learning model, in this paper we adopt a state-of-the-art Transformer-based architecture \citep{vaswani2017attention} to encode the consumer's consumption sequence $\{j_{i_1},\ldots,j_{i_n}\}$.\endnote{We also conduct robustness checks using alternative model architectures. Details of our implementation of $f(\cdot)$ are provided in Section~\ref{sec:framework_dnn}.} Candidate places are then ranked in descending order of $\hat{y}(i,j,\vz)$ to generate personalized POI recommendations \citep{liu2009learning}.

POI recommendation is arguably one of the most challenging recommendation problems for three reasons. First, it requires \emph{cross-category generalization}. Unlike many established recommender systems which operate within a coherent item domain such as videos \citep{covington2016deep} or restaurants \citep{wang2025recommending}, in POI recommendation, a single consumer's history may span museums, restaurants and retail stores, requiring cross-category understanding instead of relying on within-category similarity.

Second, visits are driven by \emph{heterogeneous} and \emph{unobserved} intents. Two consumers may visit the same place for fundamentally different reasons, and even the same consumer may choose a restaurant for its food, atmosphere, or location. These latent intents are difficult and costly to elicit at scale, making personalized reasoning about \emph{why} a place may appeal to a consumer critical to POI recommender systems.

Third, POI interactions are highly \emph{long-tailed}. For example, in the Foursquare NYC check-in dataset \citep{yang2014modeling}, 78.0\% of venues receive five or fewer check-ins, while the top 1\% of venues account for 19.1\% of all check-ins. Because conventional recommender systems rely heavily on repeated interactions, the fact that most places only receive a few visits makes POI recommendation particularly challenging.



\subsection{The Promise and Challenges with LLM-Generated Explanations}
\label{sec:llm_promise_challenges}

Recent advances in LLMs suggest that explicit reasoning can significantly improve learning efficiency and performance \citep{wang2025works, wei2022chain, guo2025deepseek}. This is particularly appealing for data-sparse recommendation problems such as POI recommendation. We illustrate two promising examples in Appendix~\ref{appen:llm_examples}. When asked why a consumer might or might not purchase orange juice given her prior purchase history, GPT-4o identified the semantic connection between oranges and orange juice in the positive case (which would require a conventional recommender system many co-purchase observations to learn) and the distinction between whole-fruit and processed juice in the negative one, among many possibly irrelevant product details such as packaging and logo. When asked why a consumer whose history spans a garage, fine-dining restaurants, and a symphony venue might visit an art museum, Claude Opus~4.8 connected these places through possible preferences for fine art, cultural experiences, and architectural design.

However, directly applying LLM reasoning to POI recommendation presents two challenges. First, a plausible explanation is not necessarily accurate or useful for prediction \citep{herman2017promise, jacovi2020towards}. The same LLM can generate multiple distinct explanations that all sound plausible (see Appendix~\ref{appen:llm_hallucination} for an example), yet they do not all reflect the actual reason behind the consumer's choice. Relying on hallucinated or incorrect explanations \citep{xu2024hallucination, kalai2025language} can hurt the recommendation performance. The challenge is therefore to determine \emph{which} explanations should be trusted and incorporated into the recommender system. Second, there is a mismatch between LLMs and recommender systems. LLMs are inherently \emph{generative}, whereas recommendation is a \emph{discriminative} task that requires selecting which item to recommend. Prior research suggests that LLMs often underperform discriminative models when used directly for prediction and recommendation \citep{ye2025lola, tsai2024leveraging}. The challenge is therefore how to leverage the reasoning capabilities of generative LLMs while preserving the predictive advantages of discriminative recommenders.

Therefore, our research questions are: In the absence of ground-truth (personalized) explanations, can we still learn \emph{useful} explanations that improve recommendations? And how can generative LLMs be combined with discriminative recommenders? Here, we use the term \emph{explanations} to refer to personalized natural-language rationales that describe potential reasons for a consumer’s predicted preferences.

\subsection{Conceptual Motivation: Recommendation and Explanation as Reinforcing Tasks}
\label{sec:conceptual_motivation}

To address the challenges above, we reconsider the role of explanations in recommender systems. Existing approaches treat explanation generation as a \emph{separate} or post-hoc task. When ground-truth explanations are unavailable, however, existing approaches are not able to distinguish explanations that are \emph{useful} for recommendation from those that are merely plausible, and therefore there is no guarantee that the explanations will improve the performance of the recommender system. Motivated by this, we consider an alternative design where explanations are an \emph{integral} part of recommendation learning, and define the \emph{usefulness} of an explanation by whether incorporating it improves downstream recommendation accuracy. This focus aligns with research evaluating explanations by their contribution to learning and decision performance \citep{pruthi2022evaluating, gregor1999explanations, tintarev2012evaluating}.  

\phantomsection
\label{para:why_usefulness}
We focus on usefulness because predictive accuracy is the central objective for industrial recommender systems as it directly ties to business outcomes such as engagement, conversion, and revenue. Existing XAI methods either optimize explanation quality separately from recommendation accuracy, or trade off accuracy for explainability \citep{zhang2023comprehensible}. Such a trade-off can limit practical adoption when firms are reluctant to sacrifice recommendation performance for explainability. By making explainability complementary to predictive performance, we aim to reduce a key barrier to the adoption of explainable recommender systems.

\emph{What does a useful explanation capture and why can it help prediction?} Among all explanations an LLM can generate, some may capture decision-relevant structure more effectively than others, even when the consumer's true motivation is unobserved. Such explanations can help the recommender identify informative preference dimensions and generalize to future predictions, particularly in data-sparse or out-of-distribution settings, such as niche places and consumer preferences, where black-box approaches often struggle. This motivates a mutually reinforcing learning process: prediction accuracy provides the learning signal for identifying useful explanations, which we call \emph{prediction-informed explanations}; these explanations then feed back into the recommender to improve predictions, which we call \emph{explanation-informed predictions}. Therefore, explanations and predictions can benefit each other instead of trading off with each other. 
 
\emph{What are good candidates for useful explanations?} We argue that LLMs, having been trained across diverse environments, are more likely than approaches without LLMs to encode implicit knowledge of the key variables that drive consumer decision-making, and therefore to generate \emph{useful} explanations. Incorporating this knowledge into the learning process can improve data efficiency and predictive performance, particularly in data-sparse settings such as niche products and consumer tastes. We formalize these insights in Appendix~\ref{appen:theory} by building on high-dimensional statistical learning theory. Because the analysis relies on stylized models, it does not provide formal guarantees for the full LLM-based framework. Rather, it serves as a motivation for our proposed approach of jointly learning prediction-informed explanations and explanation-informed predictions. We describe the resulting RecPIE framework in detail in the next section.
 
\phantomsection
\label{sec:usefulness_faithfulness}
\noindent \paragraph{\textbf{Remark 1.}} It is worth noting that our notion of \emph{usefulness} is different from qualities commonly emphasized in the XAI literature: \emph{plausibility}, whether an explanation is convincing to humans, and \emph{faithfulness}, whether it accurately reflects the factors underlying a model's prediction \citep{herman2017promise, jacovi2020towards, lyu2024faithful}. While plausibility and faithfulness capture important qualities of explanations, they target qualities of the explanations themselves, independent of the prediction, and therefore neither directly measures an explanation's contribution to downstream predictive accuracy, which is what our usefulness objective captures. That being said, these objectives are not necessarily in conflict. Explanations that capture decision-relevant preference structures may not only improve predictions of future consumer behavior, but also provide meaningful interpretations of consumer preferences. Although optimizing usefulness does not theoretically guarantee plausibility or faithfulness, our empirical results suggest  that optimizing for usefulness also leads to improvements in empirical measures of plausibility and faithfulness.

\section{Proposed Framework}
\label{sec:framework}

\subsection{Overview of the Proposed Framework}
\label{sec:framework_overview}

The key idea for our proposed framework, \textbf{RecPIE} (\textbf{Rec}ommendation with \textbf{P}rediction-\textbf{I}nformed \textbf{E}xplanations), is to jointly train an Explanation Component and a Recommendation Component toward a shared downstream objective, where predictions guide the generation of explanations, and explanations, in turn, improve predictions, creating a mutually reinforcing learning loop.

In the \textit{Explanation Component} (\textbf{prediction-informed explanations}), we train a personalized LLM to generate contrastive explanations (reasons for why a consumer may or may not want to visit a place) by optimizing a \emph{usefulness} reward. The input to the LLM fine-tuning includes the consumer's visitation history and a learnable consumer embedding (i.e., \emph{soft prompt}) to enable personalized explanations. The usefulness reward, measuring how much the explanations improve the downstream recommendation accuracy\endnote{In the rest of the paper, we use ``recommendation accuracy'', ``prediction accuracy'' and ``recommendation performance'' interchangeably.}, ensures the generated explanations are aligned with the recommendation task. 

In the \textit{Recommendation Component} (\textbf{explanation-informed predictions}), explanations generated from the Explanation Component are converted into embeddings using a pretrained text encoder. These explanation embeddings are then concatenated with consumer, place, and sequence embeddings and fed into a deep neural network (DNN) to generate the predicted outcomes (i.e., rating) for each $(consumer, place)$ pair. Through backpropagation, the Recommendation Component learns how much weight to assign to positive and negative explanations when making predictions.

\begin{figure}[hbtp!]
    \centering
    \includegraphics[width=1.0\linewidth]{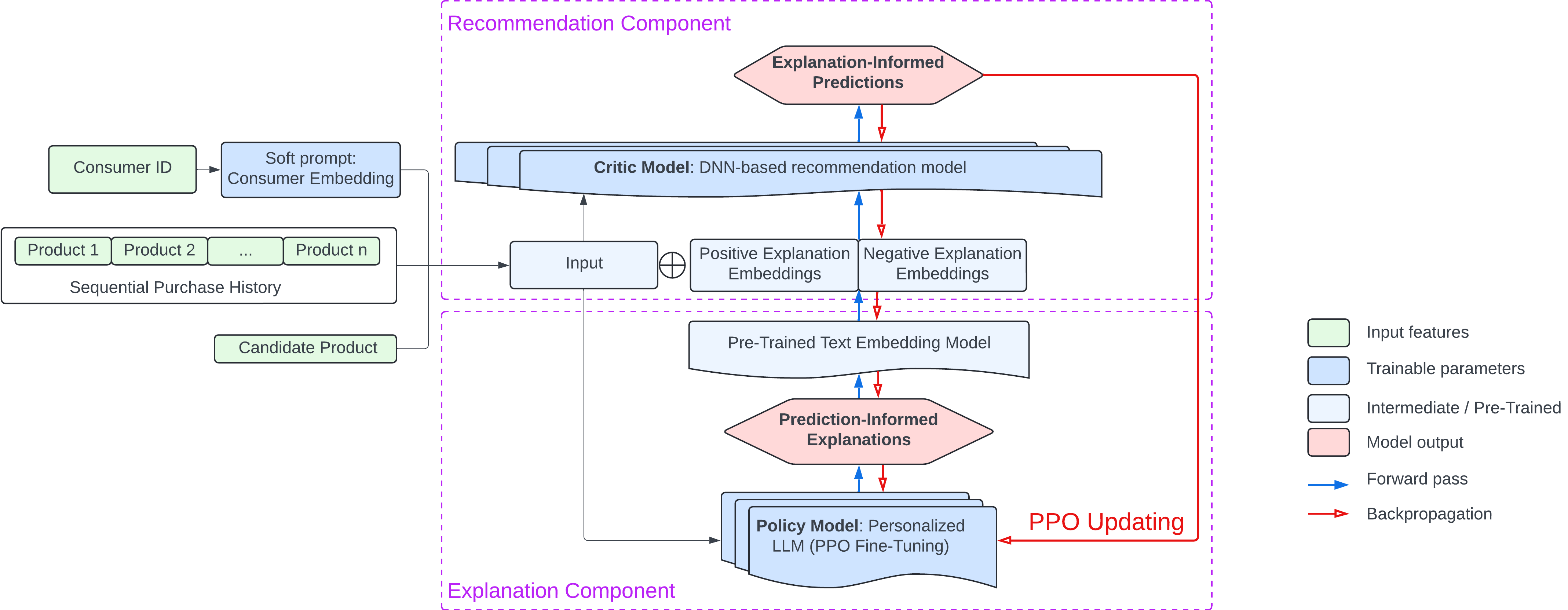}
    \caption{(Color online) Detailed architecture of RecPIE.}\label{fig:llm_recsys_diagram}
\end{figure}

These two components are trained iteratively in an alternating fashion until both tasks converge, exploiting the mutual reinforcement between explanation and prediction. The parameters of the recommender (for predictions) and the LLM (for explanations) are updated in alternating steps, as we describe in Section~\ref{sec:framework_training}. Next, we describe each component (Figure~\ref{fig:llm_recsys_diagram}) in detail. 

\subsection{Explanation Component: Prediction-Informed Explanations}
\label{sec:framework_explanation_generator}

The Explanation Component has two goals: (i) generate \emph{personalized} explanations for why a consumer may or may not like a place, and (ii) ensure that these explanations are \emph{useful} for improving recommendation accuracy. We achieve these goals through a learnable, personalized \emph{soft prompt}, and reinforcement-learning-based fine-tuning that guides explanation generation using recommendation accuracy.   

\subsubsection{Learnable soft prompt for personalization.}
\label{sec:framework_soft_prompt}
We introduce a \emph{learnable soft prompt} $\ve_i$ for each consumer $i$, which is a trainable consumer-specific embedding vector that serves as the input to the LLM during fine-tuning \citep{lester2021power}. We adopt the soft prompt for two reasons. First, consumer preferences are complex and high-dimensional. Therefore, rather than pre-specifying preference attributes in textual prompts, we learn a compact representation that flexibly captures each consumer's latent preferences. Second, the soft prompt is shared with the Recommendation Component and jointly trained to optimize recommendation accuracy, therefore allowing the explanation generator to leverage \emph{collaborative} signals learned \emph{across} users (i.e., information about the preferences and behaviors of other users in addition to themselves) when generating personalized explanations for a given user. 

The learnable soft prompt $\ve_i$ is concatenated with the consumer's visitation history and candidate place as input to the LLM for training the personalized explanation generator. The full explanation prompt $\mathbf{X}^{(k)}$ for observation $k$, which involves consumer $i$, is:
\begin{quote}
\emph{
$\mathbf{X}^{(k)} \coloneqq$ ``$\ve_i$ $\oplus$ E(The consumer's visitation history includes \{history\}.
Please generate a reason why \{candidate place\} would (or would not) be a good recommendation for this consumer. Answer in at most two sentences.)''}
\end{quote}
where $\oplus$ denotes vector concatenation, $\ve_i$ is the soft prompt, and $E(\cdot)$ denotes the tokenizer of the LLM. In Section~\ref{sec:ablation}, we show that RecPIE is robust to the wording or language used in the explanation prompt.

\subsubsection{Usefulness reward for RL-based LLM fine-tuning.}
\label{sec:framework_ppo}
We fine-tune the LLM to select explanations that improve recommendation accuracy among all plausible explanations. In the absence of ground-truth explanations, we define a \emph{usefulness reward} measuring how much the generated explanations improve the downstream recommendation accuracy, and adopt a reinforcement learning (RL)-based fine-tuning approach using \emph{Proximal Policy Optimization} (PPO) \citep{schulman2017proximal} to optimize that reward. 

The usefulness reward has two advantages. First, it addresses the missing ground-truth issue by using recommendation accuracy as a proxy reward for explanation quality. Second, it encourages learned explanations to improve the recommendation performance instead of trading off with it. As a result, our fine-tuning approach differs from the classical supervised fine-tuning paradigm \citep{brown2020language} that maximizes the likelihood of ground-truth tokens, which are unavailable in our setting (see Appendix~\ref{appen:sft_background} for details). We instead optimize the usefulness reward. Because the generated explanations are guided by how much they improve downstream prediction accuracy, we call them \emph{prediction-informed explanations}. 

\phantomsection
\label{para:usefulness_reward}
\textbf{Usefulness reward.}
Let $\pi_{\phi}$ denote the explanation generation policy induced by the LLM, parameterized by the trainable parameters $\phi$. Each training instance $k$ corresponds to a consumer with a given visitation history and a candidate place, from which we construct the prompt $\mathbf{X}^{(k)}$ as described in Section~\ref{sec:framework_soft_prompt}. Given $\mathbf{X}^{(k)}$, the policy generates a distribution $\pi_{\phi}(\mZ^{\mathrm{pos}}, \mZ^{\mathrm{neg}} \mid \mathbf{X}^{(k)})$ over a positive explanation $\mZ^{\mathrm{pos}}$ and a negative explanation $\mZ^{\mathrm{neg}}$, and the Recommendation Component (Section~\ref{sec:framework_dnn}) produces a predicted outcome $\hat{y}_{k}$. We define the \emph{usefulness reward} for the explanations as a function of prediction accuracy:
\begin{equation}
\label{eq:reward}
R_{k} = \frac{1}{1 + \lvert \hat{y}_{k} - y_{k} \rvert},
\end{equation}
where $y_{k}$ is the observed outcome (e.g., actual rating). Because the generated explanations $\mZ^{\mathrm{pos}}$ and $\mZ^{\mathrm{neg}}$ are incorporated as inputs to the Recommendation Component, they directly affect $\hat{y}_{k}$ and, consequently, the reward $R_{k}$. As a result, the reward provides a direct learning signal for explanation quality, guiding the explanation generation process toward improving recommendation accuracy.

\textbf{PPO objective.}
We update the explanation-generation policy by minimizing the sum of empirical PPO losses for positive and negative explanations:
\begin{equation}
\label{eq:ppo_total}
\hat{\mathcal{L}}_{\mathrm{PIE}}(\phi)
=
\hat{\mathcal{L}}_{\mathrm{PPO}}^{\mathrm{pos}}(\phi)
+
\hat{\mathcal{L}}_{\mathrm{PPO}}^{\mathrm{neg}}(\phi)
=
-
\frac{1}{N}
\sum_{k=1}^{N}
\sum_{t=1}^{T_k}
\Bigl[
\mathrm{clip}\!\left(
r^{\mathrm{pos},(k)}_{t}(\phi),\,
\epsilon
\right)
+
\mathrm{clip}\!\left(
r^{\mathrm{neg},(k)}_{t}(\phi),\,
\epsilon
\right)
\Bigr]
R_{k},
\end{equation}
where ``PIE'' stands for \emph{prediction-informed explanations}, $N$ is the number of training observations, $T_k$ is the number of tokens in the generated explanation, and $R_k$ is the reward defined in Equation~(\ref{eq:reward}). The likelihood ratio is $r^{s,(k)}_{t}(\phi) = \pi_{\phi}\bigl(z^{s,(k)}_{t} \mid \mathbf{X}^{(k)}, z^{s,(k)}_{<t}\bigr) \big/ \pi_{\mathrm{ref}}\bigl(z^{s,(k)}_{t} \mid \mathbf{X}^{(k)}, z^{s,(k)}_{<t}\bigr)$, where $z^{s,(k)}_{t}$ denotes the $t$-th token of the explanation of type $s \in \{\mathrm{pos}, \mathrm{neg}\}$ for the $k$-th training instance. Here, $\pi_{\mathrm{ref}}(\cdot)$ denotes the original, untuned language model, and $\epsilon$ is a small hyperparameter controlling the trust region. The clipping operation $\mathrm{clip}(\cdot)$ prevents large policy updates and stabilizes training.\endnote{The clipping function is defined as
$ \mathrm{clip}(x,\epsilon) = \max\!\bigl(\min(x,\,1+\epsilon),\,1-\epsilon\bigr) $.} Following standard RLHF practice \citep{ouyang2022training}, the reward $R_k$ is assigned uniformly across all tokens in the generated explanation. 


\textbf{Parameter updates.}
The parameters $\phi$ are updated by gradient descent on the empirical PPO objective, $\phi^{\mathrm{new}} = \phi - \eta_{\phi} \nabla_{\phi} \hat{\mathcal{L}}_{\mathrm{PIE}}(\phi)$, where $\eta_{\phi}$ is the learning rate. We use Low-Rank Adaptation (LoRA) \citep{hu2022lora} for parameter-efficient fine-tuning, so that only a small fraction of LLM parameters are updated (see Appendix~\ref{appen:lora} for mathematical details). Each PPO iteration consists of generating positive and negative explanations using the current policy $\pi_{\phi}$, computing the usefulness rewards based on the downstream recommendation accuracy, and updating $\phi$ accordingly.

\subsection{Recommendation Component: Explanation-Informed Predictions}
\label{sec:framework_dnn}  

The Recommendation Component takes Explanation Component's output, combined with the other information about the consumer, the visitation history and the context as the input, and passes them through a DNN to generate predictions for each $(consumer, candidate place)$ pair, which determine the final ranking.  

\subsubsection{Input.} 
\label{sec:framework_dnn_cons_embed}  

As illustrated in the \textit{Recommendation Component} of Figure~\ref{fig:llm_recsys_diagram}, the DNN recommender takes as input the concatenated embeddings of the generated explanations, consumer, candidate place, and sequential visitation history. We describe the embedding construction for each component below.

 
\textbf{Explanation embedding.} To convert the natural language explanations generated by the Explanation Component into embeddings, we can either use the embeddings produced by the PPO-tuned LLM above or adopt a separate, smaller pretrained text encoder. We adopt the latter approach using DistilBERT \citep{sanh2019distilbert}, since the former is a decoder-only model optimized for next-token prediction, whereas encoder-only models (such as DistilBERT) are built on bidirectional attention and are better suited for producing sentence-level semantic embeddings \citep{reimers2019sentence,lyu2024llm}. As a result, we denote the embeddings of the positive and negative explanations for the $k$-th training instance as $\mathbf{\vz}^{(\mathrm{pos})}_k$ and $\mathbf{\vz}^{(\mathrm{neg})}_k$.

\textbf{Consumer and place embedding.} 
We construct two sets of embeddings, $\mathbf{\ve}^{(c)}_i$ and $\mathbf{\ve}^{(p)}_j$, which serve as compact representations of consumer $i$ and place $j$, respectively. Appendix~\ref{appen:consumer_place_embed} describes details of the construction. An important design choice is that the consumer embedding $\mathbf{\ve}^{(c)}_i$ is shared with the Explanation Component, i.e., the personalized soft prompt described in Section~\ref{sec:framework_soft_prompt}, ensuring that the same consumer's latent preference is used for both explanation generation and recommendation prediction. 

\textbf{Sequential history embedding.} 
We map the visitation history, which is a sequence of previously visited places indexed by $\{j_{i_{1}}, j_{i_{2}}, \ldots, j_{i_{n}}\}$,  into a \emph{sequence embedding}, a learned vector representation that summarizes the consumer's past history and can be jointly optimized with the rest of the model. An effective sequence embedding should capture both the \emph{local and temporal dependencies} (e.g., consecutive visits to the same restaurant) and \emph{global dependencies} that may span across a long horizon (e.g., monthly visits to a car wash). We combine two sequence modeling techniques to capture both. We first apply a \emph{self-attention} layer, as used in the Transformer \citep{vaswani2017attention}, to model global relationships across the sequence. This is followed by a \emph{gated recurrent unit} (GRU) layer \citep{cho2014learning}, which is designed to capture local and temporal dynamics. The output of this stacked architecture serves as the final sequence embedding.\endnote{A related hybrid approach is used by \citet{li2020purs}. However, their focus is on item-level heterogeneity, not on jointly modeling local and global dependencies.} The details are provided in Appendix~\ref{appen:seq_embed}. The final sequence embedding is denoted as $\ve^{(\mathrm{seq})}_k$.

\subsubsection{Model architecture and loss function.}
\label{sec:framework_dnn_loss}
Putting everything together, the final input layer is
\begin{equation}
\label{eq:input_layer}
X_{\text{input}}
=
\big[
\vz^{(\mathrm{pos})}_k,\,
\vz^{(\mathrm{neg})}_k,\,
\mathbf{\ve}^{(c)}_i,\,
\mathbf{\ve}^{(p)}_j,\,
\ve^{(\mathrm{seq})}_k
\big],
\end{equation}
consisting of five elements: the positive and negative explanation embeddings $\vz^{(\mathrm{pos})}_k$ and $\vz^{(\mathrm{neg})}_k$; the consumer and place embeddings $\mathbf{\ve}^{(c)}_i$ and $\mathbf{\ve}^{(p)}_j$ and the sequence embedding $\ve^{(\mathrm{seq})}_k$. To model interactions among these inputs, we apply an additional self-attention layer \citep{vaswani2017attention} to the concatenated input, allowing the components to interact with one another. Let $I = \{\mathrm{pos}, \mathrm{neg}, c, p, \mathrm{seq}\}$ index the elements of $X_{\text{input}}$. The output is computed as a weighted combination of all elements in $I$ \citep{vaswani2017attention}:
\begin{equation}
\label{eq:self-attn}
X_{\text{self-attn}} =
\big[
\sum_{i \in I} \alpha_{\mathrm{pos}, i} \vv_i,\,
\sum_{i \in I} \alpha_{\mathrm{neg}, i} \vv_i,\,
\sum_{i \in I} \alpha_{c, i} \vv_i,\,
\sum_{i \in I} \alpha_{p, i} \vv_i,\,
\sum_{i \in I} \alpha_{\mathrm{seq}, i} \vv_i
\big],
\end{equation}
where $\alpha_{u,i}$ denotes the attention weight from element $u$ to element $i$, and $\vv_i$ is the learned value vector associated with element $i$. For example, $\alpha_{\mathrm{pos}, c}$ captures how much the positive explanation attends to the consumer embedding. Details of the attention weight computation are provided in Appendix~\ref{appen:self_attn_2}.

This self-attention layer enables the recommender to explicitly model interactions among heterogeneous inputs, including explanations, consumer preferences, place attributes, and sequential context. In particular, it allows the model to learn how positive and negative explanations should be weighted differently across consumers and places. The output of the self-attention layer $X_{\text{self-attn}}$ is passed through a feedforward neural network $f_{\mathrm{MLP}}(\cdot)$, a multi-layer perceptron (MLP), to generate the final prediction $\hat{y}_{k} = f_{\mathrm{MLP}}\!\left(\mathrm{Flatten}(X_{\text{self-attn}})\right)$. 

The recommendation loss uses mean squared error for continuous outcomes (e.g. numerical ratings), and binary cross-entropy for binary outcomes (e.g. thumbs-up or thumbs-down), in which case the final layer applies a sigmoid function $\sigma(z) = 1/(1+e^{-z})$ so that $\hat{y}_k$ is a probability:
\begin{equation}
\label{eq:rec_loss}
\hat{\mathcal{L}}_{\mathrm{Rec}}(w) =
\begin{cases}
\dfrac{1}{N} \sum_{k=1}^{N} (\hat{y}_k - y_k)^2, & \text{continuous outcomes},\\[6pt]
-\dfrac{1}{N} \sum_{k=1}^{N} \bigl[ y_k \log \hat{y}_k + (1 - y_k) \log (1 - \hat{y}_k) \bigr], & \text{binary outcomes}.
\end{cases}
\end{equation}
Here $y_k$ is the observed outcome for instance $k$. The parameter vector $w$ collects all trainable parameters in the Recommendation Component, including consumer embeddings, place embeddings, sequence embeddings, and the weights of the self-attention and feedforward layers.

\subsection{Alternating Training Process for RecPIE}
\label{sec:framework_training}  

Putting everything together, RecPIE consists of two components: the \emph{Explanation Component}, which optimizes the loss $\hat{\mathcal{L}}_{\mathrm{PIE}}$ to generate \emph{prediction-informed explanations}; and the \emph{Recommendation Component}, which optimizes the loss $\hat{\mathcal{L}}_{\mathrm{Rec}}$ to generate \emph{explanation-informed predictions}. The training objective is
\begin{equation}
\label{eq:recpie_loss}
\hat{\mathcal{L}}_{\mathrm{RecPIE}}(\phi, w) = \hat{\mathcal{L}}_{\mathrm{Rec}}(w) + \hat{\mathcal{L}}_{\mathrm{PIE}}(\phi),
\end{equation}
where $\phi$ and $w$ denote the parameters of the LLM Explanation Component and DNN Recommendation Component, respectively. Training alternates between updating $\phi$ by minimizing $\hat{\mathcal{L}}{\mathrm{PIE}}(\phi)$ with $w$ fixed (odd iterations) and updating $w$ by minimizing $\hat{\mathcal{L}}{\mathrm{Rec}}(w)$ with $\phi$ fixed (even iterations), until both converge.


The complete RecPIE framework is illustrated in Figure~\ref{fig:llm_recsys_diagram}. The trainable parameters are summarized in Appendix~\ref{appen:trainable}. Pseudo-code is provided in Algorithm~\ref{algo:le_recsys} of Appendix~\ref{appen:algorithm}. Implementation details, including architectures and hyperparameters, are reported in Appendix~\ref{appen:model_architecture_dnn}. 

\phantomsection
\label{remark:diff_ExpCTR}
\noindent \paragraph{\textbf{Remark 2.}} Closest to RecPIE's alternating training process is ExpCTR \citep{yu2025expctr}, which integrates explanation generation into CTR prediction. The main differences are twofold. First, RecPIE aligns explanation generation with recommendation accuracy through a usefulness reward derived directly from prediction loss, allowing both components to optimize a \emph{shared} downstream objective. In contrast, ExpCTR optimizes separate LC- and IC-alignment objectives that are not directly tied to prediction accuracy. Second, while ExpCTR represents users through textual histories of liked and disliked items, RecPIE uses learnable consumer soft prompts \emph{shared} with the recommender, enabling explanation generation to leverage collaborative signals across users.

\section{Empirical Context and Experiment Setup}
\label{sec:context}
 Our empirical context is \emph{map-based POI recommendation} on Google Maps, where each consumer has an associated sequence of historically visited places, and the goal is to predict the consumer's preference for each candidate place and rank the candidates accordingly. We use the Google Maps dataset\endnote{\url{https://mcauleylab.ucsd.edu/public_datasets/gdrive/googlelocal/}} released by \citet{li2022uctopic, yan2023personalized}, which contains user ratings and reviews, along with business metadata (e.g., address, description, category) in the United States. We focus on a subset in Mountain View, California, by filtering zip codes from 94035 to 94043. In the end, the sampled dataset contains 7,546,912 ratings (1 to 5 stars) from 47,377 users over 133,599 locations.

\subsection{Experiment Setup}
\label{sec:context_exp_setup}

\subsubsection{Baselines.}
\label{sec:context_baselines} We compare RecPIE against 21 state-of-the-art baselines spanning LLM-based, aspect-based, sequential, interpretable, and POI recommender systems. We categorize them as \emph{explanation-focused} or \emph{recommendation-focused} based on whether their primary objective is to improve explanation quality or recommendation accuracy. Appendix~\ref{appen:baselines} provides detailed descriptions of these baselines.

For both our RecPIE framework and all LLM-based baselines, we use Llama~3.1-8B \citep{dubey2024llama}, a strong, recent open-source LLM released by Meta. We focus on open-source models because, in real-world POI recommendation settings, user location histories and visited places are sensitive information, making it impractical to rely on closed-source or externally hosted models. Using open-source LLMs therefore better reflects realistic deployment constraints for privacy-preserving recommendation systems.

We split the data into training and test sets using an 80/20 split at the user-temporal level. To ensure a fair comparison, we perform grid search \citep{bergstra2011algorithms} and allocate equal computational resources (training time and memory) to tune hyperparameters for RecPIE and all baselines. Selected hyperparameter values are reported in Appendix~\ref{appen:hyper_param}. We run each method 10 times and report the mean performance with standard deviations across runs. Our RecPIE model was trained using 4 NVIDIA H200 GPUs at a high-performance computing (HPC) center, and each training run took roughly 20 hours to complete.\endnote{The use of PPO provides stability through bounded policy updates. In practice, we observe stable training behavior and consistent convergence across datasets and random seeds.}

\subsubsection{Evaluation metrics.}
\label{sec:context_eval_method}

RecPIE is evaluated on two tasks: recommendation and explanation. 

For the recommendation task, we measure \emph{recommendation accuracy} using rating prediction. Treating ratings as continuous outcomes, we report RMSE and MAE. We also construct a binary classification task by labeling ratings $\ge 4$ as positive and $<4$ as negative, and report AUC accordingly. 

For the explanation task, we evaluate \emph{explanation quality} using both human evaluations and standard text metrics. Details of the human evaluation are provided in Section~\ref{sec:human_evaluation}. For standard text metrics, following \citet{celikyilmaz2020evaluation}, we report: (1) \textbf{BLEURT} \citep{sellam2020bleurt}, which measures semantic similarity between generated explanations and ground-truth explanations; (2) \textbf{Coverage}, defined as the proportion of aspect terms in generated explanations relative to those extracted from ground-truth explanations \citep{ding2024boosting}\endnote{Aspect terms are extracted using the Double Propagation algorithm \citep{qiu2011opinion}, initialized with the sentiment lexicon from \citet{hu2004mining} and expanded iteratively, following \citet{bauman2017aspect}.}; (3) \textbf{Informativeness}, measuring relevance using an LLM-based evaluator \citep{yuan2021bartscore}; and (4) \textbf{Fluency}, measuring grammatical quality, also using the evaluator of \citet{yuan2021bartscore}.



\section{Main Results}
\label{sec:results}


\subsection{Main Results I: Recommendation Accuracy}
\label{sec:main_results_recs}

Table~\ref{tab:googlemap_results} reports recommendation results, where RecPIE consistently outperforms both groups of baselines\endnote{Some recommendation-focused baselines such as TallRec \citep{bao2023tallrec} are LLM-based recommenders, so they are naturally able to generate explanations. However, these methods focus on improving recommendation accuracy instead of explanation quality, so we put them under ``recommendation-focused baselines''.} across all three metrics. Compared with the best-performing baseline, RecPIE achieves relative improvements of \textbf{3.28\%} in RMSE, \textbf{3.56\%} in MAE, and \textbf{1.67\%} in AUC. We highlight a few comparisons below.

First, RecPIE outperforms LLM-based recommenders such as TallRec and RecSAVER, which rely on LLMs for either direct prediction or standalone reasoning. This highlights the benefit of combining LLM reasoning with a discriminative DNN recommender, rather than replacing the recommender with an LLM.

Second, RecPIE outperforms strong sequential recommenders such as BERT4Rec and UniSRec, indicating that the gains do not come from improved sequence modeling (Section~\ref{sec:framework_dnn}) alone, but also from the additional preference information captured by personalized explanations. 

Third, RecPIE outperforms ExpCTR, which also incorporates explanations into recommendation but uses a different reward for explanation generation. This confirms the value of RecPIE's \emph{usefulness} reward, which directly aligns explanation generation with downstream recommendation accuracy.
 
Together, these results demonstrate the value of jointly optimizing explanations and recommendations guided by the \emph{usefulness} reward in RecPIE. 

\begin{table}[htbp]
\centering
\footnotesize
\setlength\extrarowheight{2pt}
\begin{tabular}{l ccc}
\toprule
\textbf{Method} & RMSE $\downarrow$ & MAE $\downarrow$ & AUC $\uparrow$ \\ 
\midrule

\textbf{RecPIE (Ours)}
& \textbf{0.2976***} (0.0011) & \textbf{0.1976***} (0.0010) & \textbf{0.6576***} (0.0021) \\

\textit{\% Improved}
& \textit{+3.28\%} & \textit{+3.56\%} & \textit{+1.67\%} \\

\midrule
\multicolumn{4}{l}{\textit{Recommendation-focused baselines}} \\

RecSAVER 
& 0.3173 (0.0015) & 0.2104 (0.0012) & 0.6402 (0.0027) \\ 

LLMRG 
& 0.3177 (0.0015) & 0.2101 (0.0012) & 0.6402 (0.0027) \\ 

SASRec 
& 0.3118 (0.0011) & 0.2058 (0.0010) & 0.6388 (0.0017) \\ 

DIN 
& 0.3134 (0.0011) & 0.2070 (0.0010) & 0.6366 (0.0017) \\ 

TallRec
& 0.3118 (0.0015) & 0.2072 (0.0012) & \underline{0.6468} (0.0027) \\

BERT4Rec
& 0.3110 (0.0011) & 0.2054 (0.0010) & 0.6394 (0.0017) \\

UniSRec 
& 0.3144 (0.0011) & 0.2075 (0.0010) & 0.6349 (0.0017) \\ 

LLM4POI 
& 0.3134 (0.0013) & 0.2076 (0.0012) & 0.6366 (0.0026) \\ 

ExpCTR
& \underline{0.3077} (0.0015) & \underline{0.2049} (0.0011) & 0.6408 (0.0026) \\

\midrule
\multicolumn{4}{l}{\textit{Explanation-focused baselines}} \\

LLM2ER-EQR
& 0.3089 (0.0013) & 0.2056 (0.0011) & 0.6420 (0.0023) \\

RXR
& 0.3197 (0.0011) & 0.2139 (0.0007) & 0.6349 (0.0017) \\

A3NCF 
& 0.3238 (0.0035) & 0.2124 (0.0021) & 0.6238 (0.0037) \\ 

SULM 
& 0.3244 (0.0034) & 0.2150 (0.0021) & 0.6226 (0.0037) \\ 

AARM 
& 0.3238 (0.0035) & 0.2124 (0.0021) & 0.6236 (0.0037) \\ 

MMALFM 
& 0.3255 (0.0036) & 0.2136 (0.0021) & 0.6210 (0.0037) \\ 

ANR 
& 0.3220 (0.0035) & 0.2118 (0.0021) & 0.6249 (0.0037) \\ 

MTER 
& 0.3218 (0.0035) & 0.2116 (0.0021) & 0.6255 (0.0037) \\ 

AMCF 
& 0.3148 (0.0021) & 0.2076 (0.0014) & 0.6352 (0.0030) \\ 

PETER 
& 0.3126 (0.0021) & 0.2059 (0.0013) & 0.6382 (0.0029) \\ 

UCEPic 
& 0.3148 (0.0021) & 0.2078 (0.0014) & 0.6349 (0.0030) \\ 

PARSRec 
& 0.3166 (0.0021) & 0.2089 (0.0014) & 0.6313 (0.0030) \\ 

\bottomrule
\end{tabular}
\caption{Recommendation accuracy results on Google Maps. ``\% Improved'' reports RecPIE's gains over the best baseline in each metric (underlined); parenthetical values are standard deviations. ***p$<$0.01}
\label{tab:googlemap_results}
\end{table}

\paragraph{\textbf{Potential managerial value.}}\phantomsection\label{para:economic_impact} While the percentage improvements in Table~\ref{tab:googlemap_results} may appear modest, evidence from large-scale online experiments suggests that small improvements in recommendation accuracy can generate substantial business value when deployed at scale \citep{zhang2023empowering, chen2024shopping, zhou2018deep, wang2024limaml, gunawardana2022evaluating}. For example, \citet{li2024variety} showed that a 3\% AUC improvement led to a 2.29\% increase in actual click-through rates (CTR) on a major video streaming platform, translating to approximately \$30 million in additional annual revenue; \citet{wang2025recommending} showed that a 0.5\% increase in conversion rate translated to a \$1.5 million weekly increase in revenue for Uber Eats. Netflix has noted that even a 0.1\% improvement in recommendation performance yields sizable economic value \citep{gomez2015netflix}. These findings suggest that RecPIE's 1.7--3.6\% improvements in predictive accuracy could generate meaningful economic value when deployed at scale. Appendix~\ref{appen:economic_value} provides an illustrative back-of-the-envelope calculation with explicit assumptions and limitations, as translating offline accuracy gains into actual business outcomes is context-dependent and would require field experiments.

Beyond potential revenue gains, RecPIE offers two additional managerial benefits. First, its personalized natural-language explanations improve transparency and interpretability of AI systems, which are increasingly important in AI governance \citep{johnson2023privacy}, addressing a barrier for AI adoption \citep{bauer2023expl}. Second, RecPIE supports efficient online deployment: explanations and their embeddings can be precomputed and cached, allowing the recommender to use cached representations without real-time LLM inference. This avoids costly real-time LLM inference and substantially reduces online serving costs, addressing another practical barrier to deploying LLM-based recommender systems at scale \citep{lin2025efficient}.

\subsection{Main Results II: Explanation Quality}
\label{sec:main_results_exp} 

\subsubsection{Linguistic quality metrics.}
\label{sec:exp_linguistic}

We evaluate RecPIE on four standard text evaluation metrics: BLEURT, Coverage, Informativeness, and Fluency as described in Section~\ref{sec:context_eval_method}, treating user reviews as ground-truth explanations.\endnote{We also evaluate RecPIE against elicited ground-truth reasons on a different benchmark dataset (REASONER in Appendix~\ref{sec:faithfulness_user}).} Most recommendation-focused baselines are excluded here as they do not generate natural-language explanations.\endnote{\label{endnote:tallrec} TallRec, although based on LLMs, is considered a recommendation-focused baseline because its LLM was fine-tuned to generate recommendations and not explanations. However, since it is an LLM and can naturally generate text, we prompt TallRec to generate explanations and treat them as TallRec's explanations.} Results are summarized in Table~\ref{tab:googlemap_reason}.

RecPIE consistently outperforms all baselines across all four metrics. It achieves the highest BLEURT score, indicating a strong semantic alignment with user-written reviews, and the highest Coverage score, suggesting its explanations capture a larger fraction of relevant aspect terms than baselines. Improvements in Informativeness score confirm that RecPIE's explanations are more relevant. For Fluency, RecPIE also achieves the highest score, although we note that RecPIE is optimized for explanation usefulness and not for linguistic fluency, so the goal is to ensure that fluency is not degraded relative to existing methods, not to maximize it explicitly. Given that RecPIE does not rely on ground-truth explanations during training, these gains explanation quality emerges from joint optimization with the recommendation task.

\begin{table}[htbp]
\centering
\footnotesize
\setlength\extrarowheight{2pt}
\begin{tabular}{lcccc}
\toprule
 & BLEURT $\uparrow$ & Coverage $\uparrow$ & Informativeness $\uparrow$ & Fluency $\uparrow$ \\
\midrule
\textbf{RecPIE (Ours)} & \textbf{0.3998***} (0.0003) & \textbf{0.3138***} (0.0011) & \textbf{0.4997***} (0.0011) & \textbf{0.3836***} (0.0009) \\
TallRec & \underline{0.3980} (0.0003) & \underline{0.3099} (0.0013) & \underline{0.4930} (0.0011) & \underline{0.3818} (0.0009) \\
ExpCTR & 0.3826 (0.0039) & 0.3086 (0.0013) & 0.4701 (0.0017) &  0.3511 (0.0023) \\
LLM2ER-EQR & 0.3773 (0.0037) & 0.3045 (0.0013) &  0.4538 (0.0016) &  0.3338 (0.0026) \\
RXR & 0.3038 (0.0044) & 0.2498 (0.0011) & 0.3789 (0.0016) & 0.2384 (0.0019) \\
A3NCF & 0.2298 (0.0632) & 0.2137 (0.0238) & 0.2622 (0.0809) & 0.2523 (0.0704) \\
SULM & 0.2501 (0.0579) & 0.2463 (0.0117) & 0.2501 (0.0657) & 0.2713 (0.0581) \\
AARM & 0.2509 (0.0574) & 0.2461 (0.0117) & 0.2496 (0.0651) & 0.2707 (0.0579) \\
MMALFM & 0.2528 (0.0533) & 0.2489 (0.0103) & 0.2526 (0.0633) & 0.2726 (0.0589) \\
ANR & 0.2523 (0.0566) & 0.2637 (0.0133) & 0.2726 (0.0788) & 0.2711 (0.0601) \\
MTER & 0.2496 (0.0523) & 0.2585 (0.0105) & 0.2484 (0.0738) & 0.2731 (0.0609) \\
AMCF & 0.2517 (0.0611) & 0.2644 (0.0138) & 0.2483 (0.0765) & 0.2698 (0.0611) \\
PETER & 0.3792 (0.0004) & 0.3074 (0.0013) & 0.4525 (0.0012) & 0.3597 (0.0009) \\
UCEPic & 0.3583 (0.0029) & 0.2883 (0.0033) & 0.4127 (0.0019) & 0.3384 (0.0029) \\
PARSRec & 0.2069 (0.0937) & 0.1886 (0.0337) & 0.2428 (0.0938) & 0.2510 (0.0771) \\
\bottomrule
\end{tabular}
\caption{Linguistic quality of generated explanations. The best baseline in each metric is underlined; parenthetical values are standard deviations. ***p$<$0.01}
\label{tab:googlemap_reason}
\vspace{-0.1in}
\end{table}


\subsubsection{Human evaluation.}
\label{sec:human_evaluation}
Note that a Google Maps review describes the experience \emph{after} a visit, not the reasoning that led to it, so agreement with review content is an imperfect, ex-post proxy for explanation quality. To complement the automated linguistic metrics, we conduct a human-subject study to assess the perceived quality of RecPIE's explanations and validate the importance of jointly learning explanations and predictions. We compare RecPIE against four alternatives: \textbf{Explanation-only}, an ablated variant that uses the LLM component without gradient updates from the recommendation task; \textbf{Llama~3} (8B), direct zero-shot prompting of the LLM; \textbf{PETER}, a strong interpretable recommender system; and \textbf{TallRec}, the best-performing LLM-based recommender system.

\paragraph{\textbf{Experimental design.}}
The study is conducted on Prolific and approved by our IRB. Participants are recruited from California, United States, to ensure familiarity with the locations in the study, and are required to be fluent English speakers with at least an undergraduate degree. Each participant evaluates six recommendation scenarios: in each, they are shown a list of ten locations to imagine having previously visited, then presented with a new candidate location and asked to engage with its explanation.


Providing participants with a common preference history follows established practice in consumer-behavior studies \citep{morales2017keeping,haubl2000consumer} and allows explanations from different methods to be evaluated against the same history. Because participants adopt a provided profile rather than evaluate their own behavioral histories, the study measures the perceived plausibility and relevance of an explanation given a history, rather than whether it recovers participants' own underlying motivations.

The six scenarios are divided into two groups. In the first three (\textbf{multiple-choice}), participants select the single most informative explanation from five candidates generated by the methods above, presented in randomized order. In the remaining three (\textbf{free-form}), participants write their own explanation for why they might visit the candidate location given their history. To ensure authenticity, we use Prolific's built-in tools to discourage LLM use and prevent copy-and-paste responses. We recruited 566 qualifying participants and collected 3,220 valid responses.\endnote{All six scenarios were presented to each participant; however, some participants skipped individual scenarios. After removing empty responses, the number of valid responses per scenario is 550, 547, 544, 535, 526, and 518.} The full questionnaire and protocol are provided in Appendix~\ref{appen:human_eval}.



\paragraph{\textbf{Multiple-choice responses.}}
RecPIE explanations are overwhelmingly preferred across all three scenarios. The average selection rate for RecPIE is \textbf{61.5\%}, compared with 16.6\% for PETER, 10.2\% for Llama~3, 9.1\% for Explanation-only, and 2.6\% for TallRec. RecPIE's preference share is stable across scenarios, ranging from 58\% to 64\%. We visualize the aggregated results in Figure~\ref{fig:bar_plot_human_eval}, while providing per-scenario breakdowns in Appendix~\ref{appendix:bar_plot}. Chi-square goodness-of-fit tests confirm that preferences deviate significantly from random choice (20\% expected under uniform selection) in all three scenarios ($p \ll 0.01$). Pairwise binomial proportion tests further confirm that RecPIE significantly outperforms each alternative in every scenario ($p \ll 0.01$). The large effect sizes (Cram\'{e}r's $V > 0.5$ \citep{cramer1999mathematical}) indicate a strong and systematic preference, not random variation.

\begin{figure}[hbtp!]
    \centering
    \includegraphics[width=0.5\linewidth]{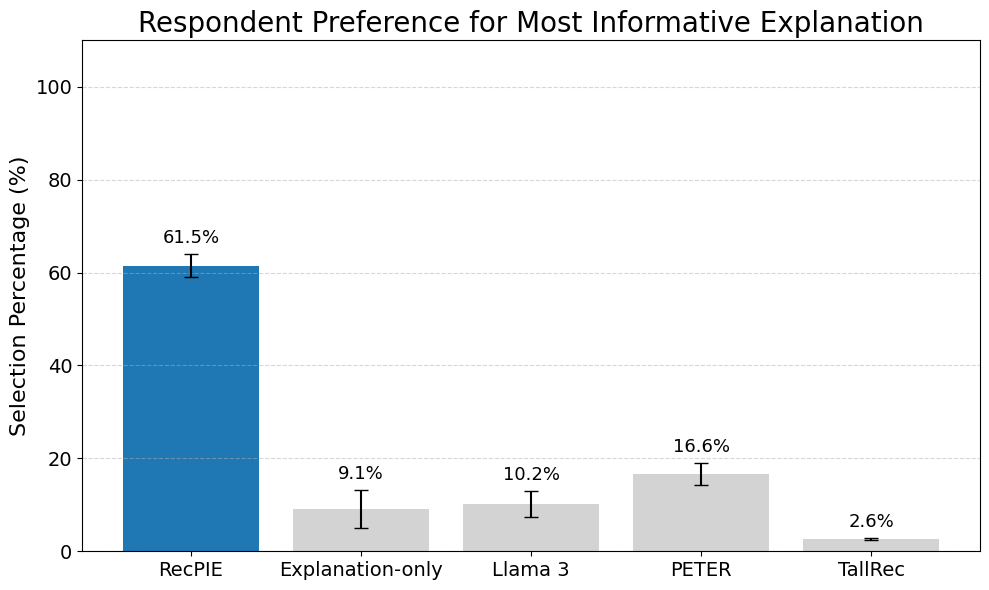}
    \caption{(Color online) Human evaluation: respondent preference for the most informative explanation, averaged across three multiple-choice scenarios (error bars indicate standard deviation).}
    \label{fig:bar_plot_human_eval}
\end{figure}

\paragraph{\textbf{Free-form responses.}}
\label{para:free_form} 

We compare model-generated explanations against user-written explanations as ground truth using the same four linguistic metrics as in Table~\ref{tab:googlemap_reason}. This comparison is a stronger test than evaluation against platform reviews, as the ground truth here directly reflects how real users reason about recommendations. Since scoring an explanation against these free-form responses requires no additional participation, we also include ExpCTR in this comparison, although it was not among the methods shown to participants in the multiple-choice scenarios. Results in Table~\ref{user_study_reason} show that RecPIE achieves the highest scores across all four metrics, indicating that its explanations are both semantically closer to human reasoning and better at capturing the decision-relevant aspects that users actually consider.

\begin{table}[htbp]
\centering
\footnotesize
\setlength\extrarowheight{2pt}
\begin{tabular}{lcccc}
\toprule
Method & BLEURT $\uparrow$ & Coverage $\uparrow$ & Informativeness $\uparrow$ & Fluency $\uparrow$ \\
\midrule
\textbf{RecPIE (Ours)} & \textbf{0.5384} & \textbf{0.3932} & \textbf{0.6368} & \textbf{0.4432} \\
ExpCTR & 0.4636 & 0.3815 & 0.6027 & 0.4193 \\
Explanation-only & 0.5175 & 0.3827 & 0.6139 & 0.4275 \\
Llama~3 & 0.4427 & 0.3650 & 0.6035 & 0.4179 \\
TallRec & 0.4698 & 0.3803 & 0.5988 & 0.4170 \\
PETER & 0.4892 & 0.3688 & 0.5872 & 0.4133 \\
\bottomrule
\end{tabular}
\caption{%
Linguistic evaluation comparing model-generated explanations with user-written explanations collected in the free-form responses. Higher values indicate closer alignment with human reasoning.}
\label{user_study_reason}
\end{table}

Taken together, the multiple-choice and free-form results provide evidence that RecPIE generates explanations that are more aligned with human reasoning. The gap between RecPIE and Explanation-only validates the value of joint optimization to generate \emph{prediction-informed} explanations.

\subsubsection{Evidence on explanation faithfulness.}
\label{sec:faithfulness_evidence}
Although RecPIE optimizes explanations for predictive \emph{usefulness}, we also evaluate their faithfulness, which is a desired quality for explanations in the XAI literature \citep{lyu2024faithful}. We conduct additional evaluations of both \emph{model-side faithfulness} and \emph{user-side faithfulness}, where the former measures whether the explanation accurately reflects how the model arrives at its prediction, and the latter measures whether the explanation captures the true reasons behind the consumer’s choice. The results suggest that RecPIE's explanations contain information that is relevant to the model's predictions and align with consumers' reported reasons, which indicates that optimizing usefulness does not come at the cost of faithfulness. Detailed evaluations are reported in Appendix~\ref{appen:faithfulness_evidence}. 

\phantomsection
\label{remark:faithfulness}
\noindent \paragraph{\textbf{Remark 3.}} RecPIE achieves comparable or even better performance on the faithfulness metrics, even though its explanations are optimized for \emph{usefulness} in downstream prediction rather than faithfulness directly. We propose a possible mechanism. Intuitively, both model-side faithfulness and user-side faithfulness will reward explanations that capture the true underlying data-generating process (DGP). Black-box recommendation models have no guarantee of capturing such DGPs. RecPIE's explanations are trained to be \emph{useful}, in the sense of being informative about consumer behavior to improve predictions of future behavior. However, consumer preferences and decision-making contexts are constantly evolving, meaning that future consumer behavior may frequently fall outside the distribution of observed data. In such settings, models that rely on unfaithful explanations may struggle to generalize to out-of-distribution observations \citep{wang2024going}. Thus, by optimizing explanations for their usefulness in predicting future consumer behavior, RecPIE may implicitly encourage them to capture the underlying data-generating process. This provides a plausible explanation for RecPIE's strong performance on the evaluated faithfulness metrics.

That being said, RecPIE does not provide any theoretical guarantee of recovering the underlying DGP or ensuring faithfulness. An important direction for future research is to characterize the assumptions under which predictive usefulness can imply, or provide bounds on, a specified measure of faithfulness.

\subsubsection{What do the explanations capture?}
\label{sec:what_do_explanations_capture}

 Table~\ref{tab:case_box} illustrates how RecPIE's explanations compare with ExpCTR's, the strongest baseline. Given a consumer's history dominated by restaurants and specialty beverage shops, RecPIE's positive explanation identifies an affinity for Red Rock Coffee's community atmosphere (attention = 0.92), while its negative explanation highlights potential mismatches with the consumer's preference for polished environments (attention = 0.11).\endnote{We explain how the attention value is computed in Appendix~\ref{appen:attention_analysis}.} RecPIE predicts a rating of 4.83, close to the ground-truth rating of 5, compared with ExpCTR's 4.09. Moreover, RecPIE's explanation captures three of the four aspects in the ground-truth review (Coverage = 0.75), compared with only one for ExpCTR (Coverage = 0.25), illustrating its ability to identify decision-relevant preference information.

\phantomsection
\label{para:illustrative_example}
\begin{table}[htbp]
\begin{tcolorbox}[halign=left, boxrule=0.4pt, colback=gray!5]
{\footnotesize
\setlength{\parskip}{0.2em}
\linespread{0.9}\selectfont

\textbf{Consumer History}: Tea Era, La Salsa, Cucina Venti Restaurant, Jersey Mike's Subs, Mamacitas! Mexican Grill, Chop \& Pub, Cost Plus World Market, Alexander's Patisserie, Sweetgreen, QBB - Quality Bourbons \& Barbecue

\smallskip
\textbf{Destination}: ``Red Rock Coffee: An independently-owned coffee shop known for its community-oriented atmosphere and live events. It's a multi-level space that serves as a hub for locals to work and socialize.''

\medskip\hrule\medskip

\textbf{RecPIE (Ours)}

\textbf{Positive Explanation} [\emph{Attention = 0.92}]: ``Given the history of independent and atmosphere-rich spots like Alexander's Patisserie and QBB, the consumer was likely drawn to Red Rock Coffee for its authentic local character and `third place' vibe. The shop's emphasis on community events perfectly complements the appreciation for the distinct, gathering-focused energy found at places like Tea Era and Chop \& Pub.''
[\emph{Aspect Terms: Unique Vibe, Independent, Community Atmosphere, Socializing; \ul{Coverage = 0.75}}]

\textbf{Negative Explanation} [\emph{Attention = 0.11}]: ``Considering the patronage of the refined, polished environment at Alexander's Patisserie and the streamlined efficiency of Sweetgreen, the consumer might find Red Rock Coffee's busy, event-heavy atmosphere too chaotic and noisy.''

\textbf{Predicted Rating}: 4.83

\medskip\hrule\medskip

\textbf{ExpCTR}: ``Red Rock Coffee is a strong match for this consumer, who has previously visited Alexander's Patisserie, QBB, and Tea Era. The shop serves quality espresso drinks in a relaxed setting with plenty of seating, and its regular events give it a welcoming, community feel that visitors with similar tastes tend to enjoy.''
[\emph{Aspect Terms: Community Atmosphere; \ul{Coverage = 0.25}}]

\textbf{Predicted Rating}: 4.09

\medskip\hrule\medskip

\textbf{Ground-Truth}: ``Great place to sit and work, but I'm also a fan of meeting people here. Nice local coffee shop that gives off the vibes of Silicon Valley. Overall, love the ambiance and atmosphere of this place. Upstairs particularly is great for working.''
[\emph{Aspect Terms: Working, Socializing, Community Atmosphere, Unique Vibe; \ul{Coverage = 1.00}}]

\textbf{Ground-Truth Rating}: 5

}
\end{tcolorbox}
\captionof{table}{Illustrative example comparing RecPIE and ExpCTR for a randomly selected Google Maps consumer. Attention values indicate how much each explanation influences the final predicted rating.}
\label{tab:case_box}
\end{table}

Beyond this case study, a broader qualitative analysis by two researchers identified three recurring patterns in RecPIE's explanations: (i) inferring \emph{subtle user traits}, such as a consumer's ``curiosity to explore new and offbeat destinations''; (ii) \emph{bridging categories} through non-obvious reasoning, such as linking a party-store visit to a sports-bar through an inferred interest in social gatherings; and (iii) generating \emph{personalized reasons for the same place} for different consumers. These patterns correspond nicely to the three POI recommendation challenges in Section~\ref{sec:formulation_poi}: cross-category generalization, heterogeneous and unobserved motivations, and long-tailed interactions. Together, they illustrate how prediction-informed explanations can make higher-level consumer preferences explicit, potentially contributing to the accuracy gains reported in Section~\ref{sec:main_results_recs}. Appendix~\ref{appen:case_study} provides additional examples and a detailed discussion of these patterns.

\section{Understanding the Improvements}
\label{sec:understanding}

In this section, we present additional analyses to decompose and better understand the performance gains observed in the previous section.

\subsection{Heterogeneity Analysis: Larger Gains for Harder Examples and Niche Tastes}
\label{sec:harder_examples} 

If RecPIE's explanations capture the decision-relevant preferences that conventional recommender systems struggle to extract, we would expect larger accuracy gains for harder-to-predict examples. To examine this, we construct two measures of hardness-to-predict. First, \textbf{prediction uncertainty} is measured by the variance of predictions across RecPIE and all baselines for the same consumer–place pair. If different models disagree more, this indicates a potentially harder example. Second, \textbf{taste uniqueness} measures how much a consumer's ratings deviate from population-level preferences, computed as the average absolute deviation from each place's mean rating.\endnote{Formally, $\text{Taste Uniqueness}_i = \frac{1}{N_i} \sum_{j} | r_{ij} - \bar{r}_j |$, where $N_i$ is the number of places rated by consumer $i$ and $\bar{r}_j$ is the average rating of place $j$ across consumers.} Consumers with higher taste uniqueness have more idiosyncratic preferences that may be harder to predict from purely data-driven patterns.

\begin{figure}[hbtp!]
    \begin{subfigure}[b]{0.48\textwidth}
        \centering
        \includegraphics[width=\textwidth]{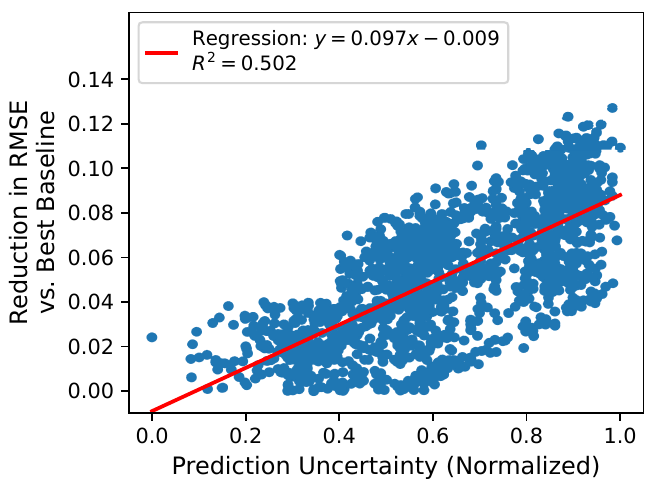}
        \caption{Performance gain vs. prediction uncertainty.}
        \label{fig:uncertainty:googlemap}
    \end{subfigure}
     \centering
     \hfill
    \begin{subfigure}[b]{0.48\textwidth}
        \centering
        \includegraphics[width=\textwidth]{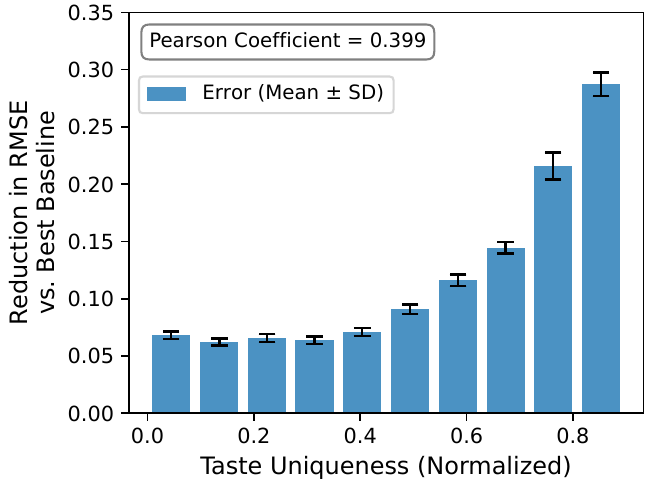}
        \caption{Performance gain vs. taste uniqueness.}
        \label{fig:niche:googlemap}
    \end{subfigure}
   \caption{Performance gains of RecPIE are larger for harder-to-predict examples.}
  \label{fig:harder}
\end{figure}

We plot RecPIE's accuracy gains, measured by absolute improvement in RMSE over the best-performing baseline, against these two measures of hardness-to-predict in Figure~\ref{fig:harder}. We see a very clear and statistically significant positive trend in both plots, confirming that RecPIE's gains are larger for harder-to-predict examples, consistent with our hypothesis that prediction-informed explanations are particularly valuable when conventional behavioral patterns provide less reliable signals. The fact that RecPIE's gains are larger for users with niche tastes is also managerially meaningful, suggesting that RecPIE may help platforms better serve heterogeneous consumer preferences, as platforms are increasingly characterized by their long-tail demand \citep{brynjolfsson2011goodbye}.

\subsection{Data Efficiency Gains}
\label{sec:learning_effiency}

Based on the conceptual motivation in Section~\ref{sec:conceptual_motivation}, which is formalized in Appendix~\ref{appen:theory}, incorporating prediction-informed explanations should improve the recommender system's \emph{data efficiency}, i.e., achieving the same performance with less training data. To validate this, we train RecPIE on random subsets of the Google Maps dataset (15\%, 25\%, and 50\% of training data) while keeping the same test set.

As shown in Table~\ref{tab:googlemaps_efficiency}, RecPIE trained on only \textbf{25\%} of the data already surpasses the strongest full-data baseline on all three metrics. Even with \textbf{15\%} of the data, its performance is not statistically significantly different from the strongest full-data baseline (ExpCTR) on any of the three metrics. These findings validate RecPIE's strong empirical data efficiency, consistent with our hypothesis that prediction-informed explanations help extract more information from limited training data.

\begin{table}[htbp]
\centering
\footnotesize
\setlength\extrarowheight{2pt}
\begin{tabular}{lccc}
\toprule
Method & RMSE $\downarrow$ & MAE $\downarrow$ & AUC $\uparrow$ \\
\midrule
RecPIE (100\%) & 0.2976$^{***}$ (0.0011) & 0.1976$^{***}$ (0.0010) & 0.6576$^{***}$ (0.0021) \\
RecPIE (50\%)  & 0.2998$^{***}$ (0.0014) & 0.1985$^{***}$ (0.0014) & 0.6528$^{***}$ (0.0029) \\
RecPIE (25\%)  & 0.3037$^{***}$ (0.0027) & 0.2009$^{***}$ (0.0023) & 0.6498$^{***}$ (0.0041) \\
RecPIE (15\%)  & 0.3089 (0.0029) & 0.2055 (0.0027) & 0.6406 (0.0047) \\
\midrule
ExpCTR (100\%) & 0.3077 (0.0015) & 0.2049 (0.0011) & 0.6408 (0.0026) \\
TallRec (100\%) & 0.3118 (0.0015) & 0.2072 (0.0012) & 0.6468 (0.0027) \\
PETER (100\%)  & 0.3126 (0.0021) & 0.2059 (0.0013) & 0.6382 (0.0029) \\
\bottomrule
\end{tabular}
\caption{\textbf{Data efficiency of RecPIE.} Parenthetical values are standard deviations. ***p$<$0.01}
\label{tab:googlemaps_efficiency}
\end{table}

\paragraph{\textbf{Faster representation learning.}} To further illustrate RecPIE's learning efficiency, we examine how its learned place embeddings evolve during training. We picked three representative place pairs: similar (In-N-Out Burger and Five Guys), related-but-distinct (a Chinese restaurant and an Asian supermarket), and unrelated. For all cases, RecPIE converges to semantically meaningful embeddings up to \textbf{four} times faster than BERT4Rec and SASRec, which are state-of-the-art sequential representation learning baselines. In addition, it learns to increase the embedding similarity for similar places, correctly separate the related-but-distinct places, and decrease the embedding similarity for unrelated places, all within the first few training epochs and even with only 25\% of the training data. The full results are presented in Appendix~\ref{appen:similarity}. These results provide further evidence of the improved learning efficiency of RecPIE.

\subsection{Potential Mechanism: Evidence Consistent with the Role of LLM Reasoning}  
\label{res_understanding_reasoning}

The conceptual motivation in Section~\ref{sec:conceptual_motivation} suggests that RecPIE's gains may stem from the LLM's ability to infer decision-relevant variables, so LLMs with stronger reasoning capabilities should yield larger gains. Next, we present some additional experiments in Table~\ref{tab:googlemaps_llm} to investigate the potential role of LLM reasoning and alternative explanations for RecPIE's gains.

\textbf{Stronger reasoning capabilities yield larger gains.} RecPIE with Llama~3.1 achieves the best performance, followed by Llama~3, Mixtral-8$\times$7B, Qwen2-7B, Vicuna-7B-v1.5, and GPT-2. This performance order is aligned with the order of the reasoning capabilities of the LLMs, according to the ZebraLogic reasoning leaderboard.\endnote{\url{https://huggingface.co/spaces/allenai/ZebraLogic}} While not definitive, this suggests that reasoning capability plays an important role in RecPIE's performance gains.

\textbf{Direct LLM recommendation performs substantially worse.} Simply replacing the discriminative recommender with a generative LLM (``Llama~3.1 Direct Recommendation'' in Table~\ref{tab:googlemaps_llm})\endnote{\label{endnote:direct_llama}``Llama~3.1 Direct Recommendation'' refers to prompting the base Llama~3.1 model directly to predict the consumer's rating of the candidate place, given a textual description of the consumer's visit history and the candidate. It uses no downstream DNN recommender and no prediction-informed explanation; it therefore measures how well the base LLM performs the recommendation task on its own, and serves as a reference point for the value added by the rest of the RecPIE framework.} performs far worse, suggesting the gains arise specifically from using LLMs to generate prediction-informed explanations that support a discriminative recommender, not from the predictive capabilities of the LLM alone.

\textbf{External knowledge and memorization play limited roles.} Augmenting inputs with LLM-generated location profiles yields only marginal and statistically insignificant improvements (``RecPIE with Profile Augmentation'' in Table~\ref{tab:googlemaps_llm}), ruling out the LLM's external knowledge as the main driver. In addition, RecPIE with GPT-2 (pretrained on data from 2019, predating the 2021 Google Maps dataset) still outperforms all non-RecPIE baselines (``RecPIE with GPT-2'' in Table~\ref{tab:googlemaps_llm}), ruling out dataset memorization.

\textbf{Summarization prompts perform worse than reasoning prompts.} Replacing reasoning-oriented prompts with summarization prompts (presented in Appendix~\ref{appen:prompts}) that ask the LLM to summarize past behavior performs worse (``Consumption History Summarization'' in Table~\ref{tab:googlemaps_llm}), confirming that the LLM's reasoning capability, and not its summarization skill, is the key driver of RecPIE's performance gains.

Together, these findings are consistent with our proposed mechanism: LLM reasoning can generate informative explanations that improve downstream recommendation when integrated with a discriminative model. Specifically, RecPIE's performance gains are closely associated with the \emph{reasoning} capabilities of LLMs, not with their direct prediction capability, external dataset knowledge or summarization skills.

\begin{table}[htbp]
\centering
\footnotesize
\setlength\extrarowheight{2pt}
\begin{tabular}{lccc}
\toprule
Method & RMSE $\downarrow$ & MAE $\downarrow$ & AUC $\uparrow$ \\
\midrule
\textbf{RecPIE with Llama~3.1} & \textbf{0.2976} (0.0011) & \textbf{0.1976} (0.0010) & \textbf{0.6576} (0.0021) \\
RecPIE with Llama~3 & 0.2999 (0.0011) & 0.1995 (0.0010) & 0.6553 (0.0021) \\
RecPIE with Mixtral-8$\times$7B & 0.3001 (0.0011) & 0.1996 (0.0010) & 0.6550 (0.0021) \\
RecPIE with Qwen2-7B & 0.3003 (0.0011) & 0.2003 (0.0011) & 0.6545 (0.0023) \\
RecPIE with Vicuna-7B-v1.5 & 0.3010 (0.0011) & 0.2005 (0.0011) & 0.6531 (0.0023) \\
RecPIE with GPT-2 & 0.3045 (0.0013) & 0.2031 (0.0013) & 0.6498 (0.0026) \\
\midrule
Llama~3.1 Direct Recommendation & 0.3477 (0.0041) & 0.2295 (0.0047) & 0.6103 (0.0069) \\
RecPIE with Profile Augmentation & 0.2971 (0.0011) & 0.1973 (0.0011) & 0.6580 (0.0023) \\
Consumption History Summarization & 0.3008 (0.0011) & 0.1994 (0.0011) & 0.6542 (0.0023) \\
\bottomrule
\end{tabular}
\caption{
\textbf{Additional experiments with different LLM backbones and RecPIE variants.}
}
\label{tab:googlemaps_llm}
\end{table}

\subsection{Role of Contrastive Explanations}
\label{sec:role_pos_neg}

A unique design of RecPIE is that it generates both a positive explanation and a negative explanation, and lets the Recommendation Component learn the weight to put on each explanation for each observation. Thanks to the attention mechanism used in the Recommendation Component, we can numerically compute the attention weights that it puts on different parts of the input, including positive and negative explanations, consumer, place and sequential history embeddings. 

\begin{figure}[hbtp!]
    \begin{subfigure}[b]{0.32\textwidth}
        \centering
        \includegraphics[width=\textwidth]{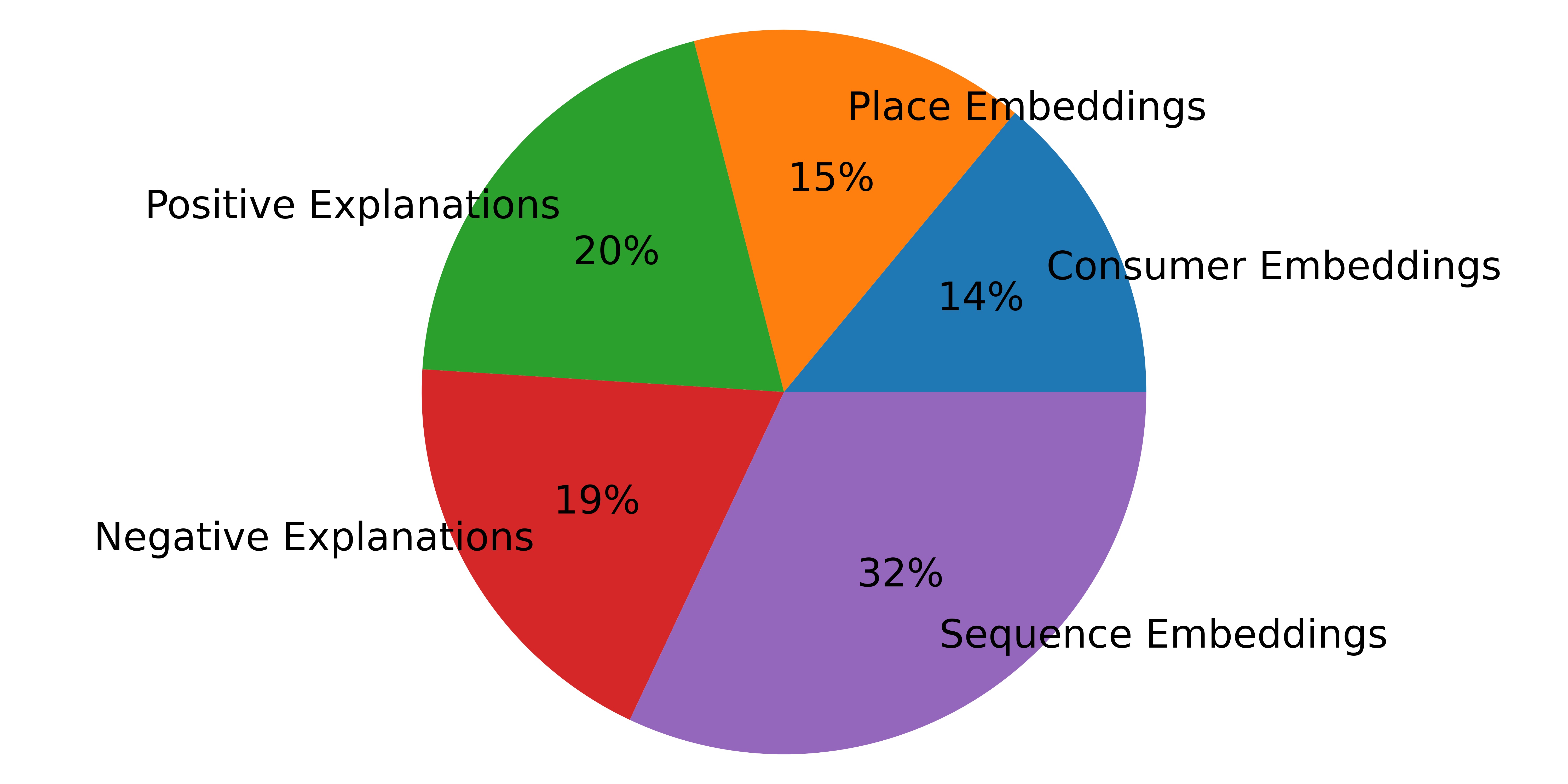}
        \caption{Attention value distribution.}
        \label{fig:pie_googlemap}
    \end{subfigure}
    \begin{subfigure}[b]{0.33\textwidth}
        \centering
        \includegraphics[width=\textwidth]{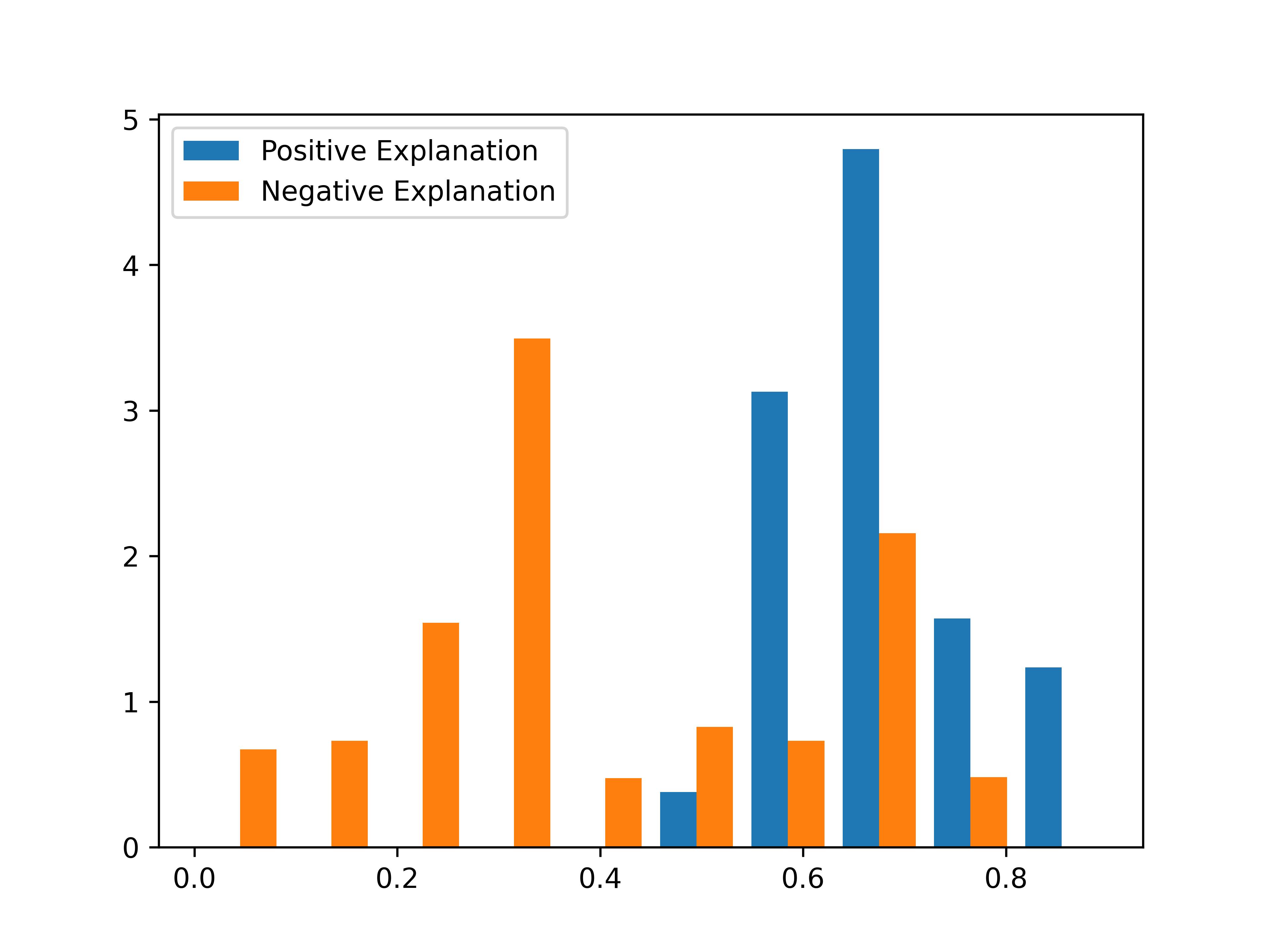}
        \caption{Positive examples.}
        \label{fig:positive:google}
    \end{subfigure}
     \centering
    \begin{subfigure}[b]{0.33\textwidth}
        \centering
        \includegraphics[width=\textwidth]{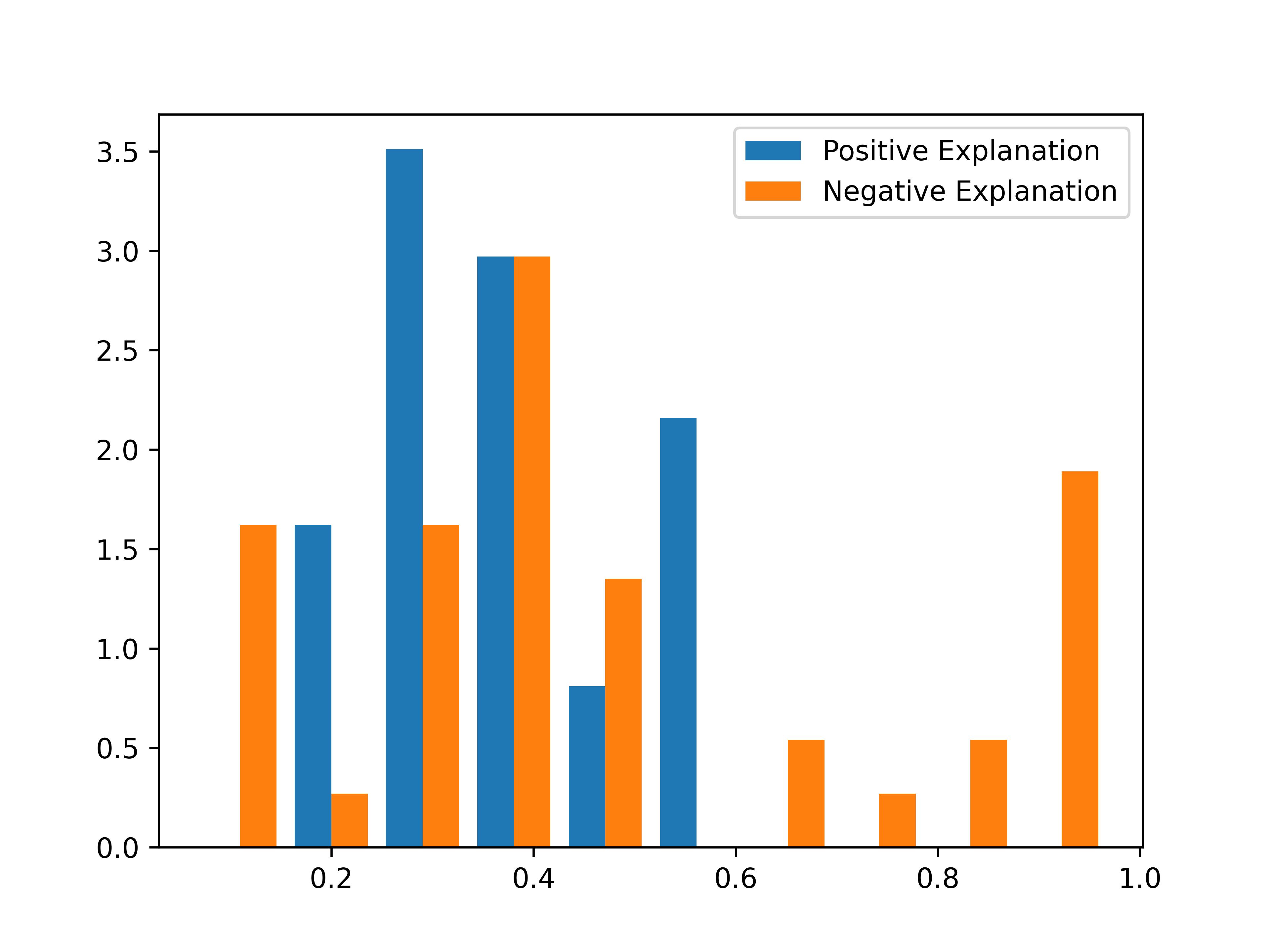}
        \caption{Negative examples.}
        \label{fig:negative:google}
    \end{subfigure}
   \caption{(Color online) Role of contrastive explanations in RecPIE. (a) Prediction-informed explanations account for a substantial share of total attention.
    (b) For positively rated items, the model assigns greater attention to positive explanations.
    (c) For negatively rated items, attention shifts toward negative explanations.
   } 
  \label{fig:attention}
\end{figure}

An attention analysis\endnote{We explain how the attention value is computed in Appendix~\ref{appen:attention_analysis}.} reveals that the positive and negative explanations together account for approximately 39\% of the total attention in the Recommendation Component (Figure~\ref{fig:pie_googlemap}), complementing the consumer, place, and sequential history embeddings. Moreover, for positively rated items, RecPIE assigns significantly more attention to positive explanations (Figure~\ref{fig:positive:google}), while for negatively rated items, attention weights shift toward negative explanations (Figure~\ref{fig:negative:google}). This confirms that RecPIE makes effective use of the explanations, as it adaptively determines which signal (positive or negative explanation) is most informative for each prediction. Full details for the attention weights computation are provided in Appendix~\ref{appen:dual_explanation}.

In Appendix~\ref{appen:dual_explanation}, we also conduct ablation studies where we ablate the contrastive explanation design by keeping only the positive explanation, by switching to a single non-contrastive explanation, or by using aspect terms alone without natural-language explanations. All perform significantly worse than RecPIE, confirming the value of our contrastive explanation design.

\subsection{Ablation Studies, Robustness Checks and Generalizability}
\label{sec:ablation}

\subsubsection{Ablation studies and robustness checks.}
\label{sec:ablation_ablation}

We conducted ablation experiments to further examine the value of RecPIE's \emph{prediction-informed explanations}, showing that its gains cannot be replicated by simply incorporating LLM-generated semantic information or additional model capacity. In addition, replacing RecPIE's \emph{usefulness} reward with alternative explanation-quality objectives commonly used in the XAI literature leads to degraded recommendation accuracy (Appendix~\ref{appen:explanation_value}). 

Additional robustness checks show that RecPIE's performance is robust to prompt wording, neural network architecture, and evaluation framework. The results are summarized in Appendix~\ref{appen:robusness}. 

\phantomsection
\label{para:wyoming}
To examine whether RecPIE's gains depend on the LLM's familiarity with Mountain View, which may be well represented in pretraining data due to its proximity to Silicon Valley, we replicate the analysis in Wyoming using the same Google Maps dataset. RecPIE achieves even larger relative improvements in both recommendation accuracy and explanation quality (Appendix~\ref{appen:wyoming}), suggesting that its benefits are not limited to geographically familiar settings.

\subsubsection{Generalizability.}
\label{sec:generalizability} 

We conducted additional experiments to validate RecPIE's generalizability. 

\paragraph{Generalizability to unseen consumers, places, and regions.} In Appendix~\ref{appen:coldstart}, we evaluate RecPIE on previously unseen consumers and places, as well as cross-region transfer. RecPIE consistently outperforms all baselines across all three metrics, with larger relative gains than in the original setting, suggesting stronger generalization to data-sparse and out-of-distribution settings. These results also validate the insights from Section~\ref{sec:harder_examples} that prediction-informed explanations provide larger gains for harder-to-predict examples.

\paragraph{Generalizability to three additional benchmark datasets.} To test RecPIE's generalizability beyond the POI setting, we evaluate RecPIE on three additional benchmarks: TripAdvisor \citep{li2023personalized} (hotel recommendation), Yelp\endnote{\url{https://www.yelp.com/dataset}} (restaurant recommendation), and Amazon Movie\endnote{\url{https://nijianmo.github.io/amazon/index.html}} (movie recommendation). RecPIE achieves consistent and significant improvements across all three applications. Full results are reported in Appendix~\ref{appen:generalizability_three_datasets}.


\section{Practical Deployment Considerations: Privacy and Latency}
\label{sec:exp_additional_privacy_latency}

We design RecPIE with deployment constraints in mind, particularly privacy and latency, which are central considerations for real-world POI recommendation systems.

\paragraph{\textbf{Privacy.}} RecPIE uses open-source LLMs for two reasons. First, fine-tuning with a customized reward signal is generally infeasible with closed-source APIs. Second, POI recommendation involves sensitive location histories, and open-source models enable on-premise deployment and full control over data handling, making RecPIE compatible with privacy-preserving requirements in large-scale systems.

\paragraph{\textbf{Latency.}} Directly calling LLMs at serving time is impractical due to latency constraints. For example, generating 50 words takes about 1{,}250 milliseconds with GPT-5 \citep{openrouter_gpt5}, while production recommendation systems typically require end-to-end response times within dozens of milliseconds \citep{nitropack2023webperformance}, making real-time LLM calls infeasible. RecPIE addresses this challenge by decoupling explanation generation from real-time inference. Explanations can be generated offline during training or periodically updated after user interactions. At serving time, the system performs only lightweight embedding lookups and standard neural network inference, making it feasible on large-scale online platforms.

\section{Conclusion and Managerial Implications}
\label{sec:discussion}

Explanations play a critical role in the adoption of AI systems by improving trust, transparency, and user understanding \citep{gregor1999explanations, bauer2023expl}. However, the trade-off between interpretability and accuracy \citep{mohammadi2025regulating, zhang2023comprehensible} serves as a major barrier for the adoption of explainable AI. In this work, we show that this trade-off can be mitigated by design: explanations can be incorporated as an \emph{integral} component of AI systems and, when aligned with predictions, can improve \emph{both} interpretability and accuracy. Our key idea is to \emph{learn which explanations are useful}: a useful explanation of consumer behavior is one that improves the prediction of future behavior. This contrasts with the existing XAI literature, which typically optimizes explanation quality independently of downstream prediction and therefore does not guarantee that the resulting explanations improve accuracy. Building on this principle, we propose RecPIE, a framework that jointly learns \emph{prediction-informed explanations} and \emph{explanation-informed predictions}. In a large-scale Google Maps application, RecPIE significantly improves recommendation accuracy, outperforming the strongest full-data baseline with only 25\% of the training data, while generating explanations that are more informative and better aligned with human judgments. 

\phantomsection
\label{para:design_implications}
Our work has direct implications for the design of digital platforms. By integrating explanations into the learning process, platforms can provide meaningful explanations \emph{without sacrificing performance}, addressing a key barrier to AI adoption. The benefits of prediction-informed explanations apply to multiple stakeholders on digital platforms. For consumers, contrastive explanations demystify black-box recommendations by clarifying why they may or may not prefer a product, increasing consumer trust. For sellers and content creators, the framework offers understandings on which product attributes drive consumer interest, offering guidance for product design. For platforms, aggregating explanation signals yields actionable insights \citep{cheng2025llms} into which features attract which consumer segments, enabling more effective targeting and marketplace design. These are especially beneficial in POI recommendation settings, where consumer preferences typically span multiple place categories that may seem unrelated (e.g. a garage and a museum), requiring cross-category reasoning that LLM-generated explanations are uniquely positioned to provide. RecPIE is easy to deploy at scale, by allowing explanations to be precomputed and cached, avoiding costly real-time LLM inference; it also helps protect consumer privacy by allowing platforms to retain control over sensitive consumer data, including location histories, using on-premise deployment with open-source LLMs. 

The benefits of prediction-informed explanations are not uniform across all settings. RecPIE's performance gains are largest for consumers with niche tastes and sparse histories, where prediction uncertainty is high and traditional models struggle most. In these settings, explanations provide additional structure that helps the system generalize beyond limited data. In addition, RecPIE substantially improves data efficiency, outperforming the best full-data baseline using only 25\% of the training data, and converging to semantically meaningful item representations up to four times faster. This is especially valuable in settings where high-quality data are costly or scarce, and even synthetic data generated by LLMs can be unreliable \citep{gui2023challenge}. 

More broadly, our work highlights the importance of incorporating behavioral insights into AI system design. Black-box AI systems today often struggle to extrapolate to new consumers, products, or environments because they rely primarily on correlations in observed data and fail to capture the underlying data-generating process. Our results suggest that, by enabling AI systems to reason explicitly about the drivers of user behavior, RecPIE improves both predictive accuracy and generalization in uncertain settings. This suggests a new paradigm in how AI systems should be designed: from purely data-driven prediction toward systems that combine statistical learning with structured reasoning \emph{jointly}. As such, RecPIE represents a step toward AI systems that are not only more accurate, but also more interpretable, data-efficient, and aligned with human decision-making.


\section{Funding and Competing Interests Declarations}
\label{sec:declarations}
Google recognizes a potential need to disclose certain confidential information, and to protect such information from unauthorized use and disclosure. Google had the right to remove its intellectual property or trade secrets, subject to the following stipulations: 
\begin{enumerate}
    \item Removing Google confidential information from the paper, including data, code, and statistics. 
    \item Compliance with Google's obligations as it relates to applicable laws, including the Data Protection Law, security laws, confidentiality requirements, and contractual commitments.
\end{enumerate}



\theendnotes

\bibliographystyle{informs2014} 
\bibliography{a_reference} 

\newpage
\begin{APPENDIX}{Can Explanations Improve Recommendations? Evidence from Prediction-Informed Explanations}
 \section{Additional Related Work}
\label{appen:additional_related_work}

\subsection{XAI in Recommender Systems}
\label{appen:related_work_expl_recsys}
Recommender systems have been a central topic in information systems research \citep{adamopoulos2014unexpectedness, chen2024background, wang2025probing}, with direct implications for user behavior and firm outcomes \citep{bi2024consumer, ghose2014examining, simester2020targeting, gabel2022product, ghose2012designing, lee2019recommender, wang2024quality, wan2024product}. With the growing demand for transparency, several explainable recommender systems have been proposed using post-hoc or supervised explanation training methods. For example, \citet{wang2018reinforcement} introduced a reinforcement learning framework that generates explanations for an \emph{existing} recommender, and \citet{li2021personalized} used sequence models and review text to produce personalized explanations. While these approaches focus on generating coherent explanations, they do \textbf{not} improve the predictive performance of the underlying recommender.

Another line of work constructs aspect-based recommendations, recommending items based on salient aspects extracted from user reviews \citep{bauman2017aspect, cheng20183ncf, cheng2019mmalfm, chin2018anr, guan2019attentive, le2021explainable, li2023prompt} or structured knowledge graphs \citep{lee2018explainable, wang2019explainable}. However, these methods rely on explicit text or curated knowledge bases, which may be unavailable or incomplete, and the generated explanations are inherently constrained by the provided content. In contrast, our approach does not require review text or external knowledge and, importantly, \emph{jointly} trains the explanation and recommendation tasks rather than treating them as separate tasks. This joint optimization enables explanations to directly \emph{improve} predictive accuracy.


\section{Motivating Examples}
\subsection{Examples of Point-of-Interest Recommender Systems}
\label{appen:poi_interfaces}

Personalized location recommendations are a central feature of services such as Airbnb, Yelp, TripAdvisor, and Google Maps, which draw on a consumer's prior activity and current context to suggest destinations, restaurants, and attractions. Figure~\ref{fig:screenshots_trip_google} shows representative interfaces from TripAdvisor and Google Maps.

\begin{figure}[hbtp!]
    \begin{subfigure}[b]{0.4\textwidth}
        \centering
        \includegraphics[width=\textwidth]{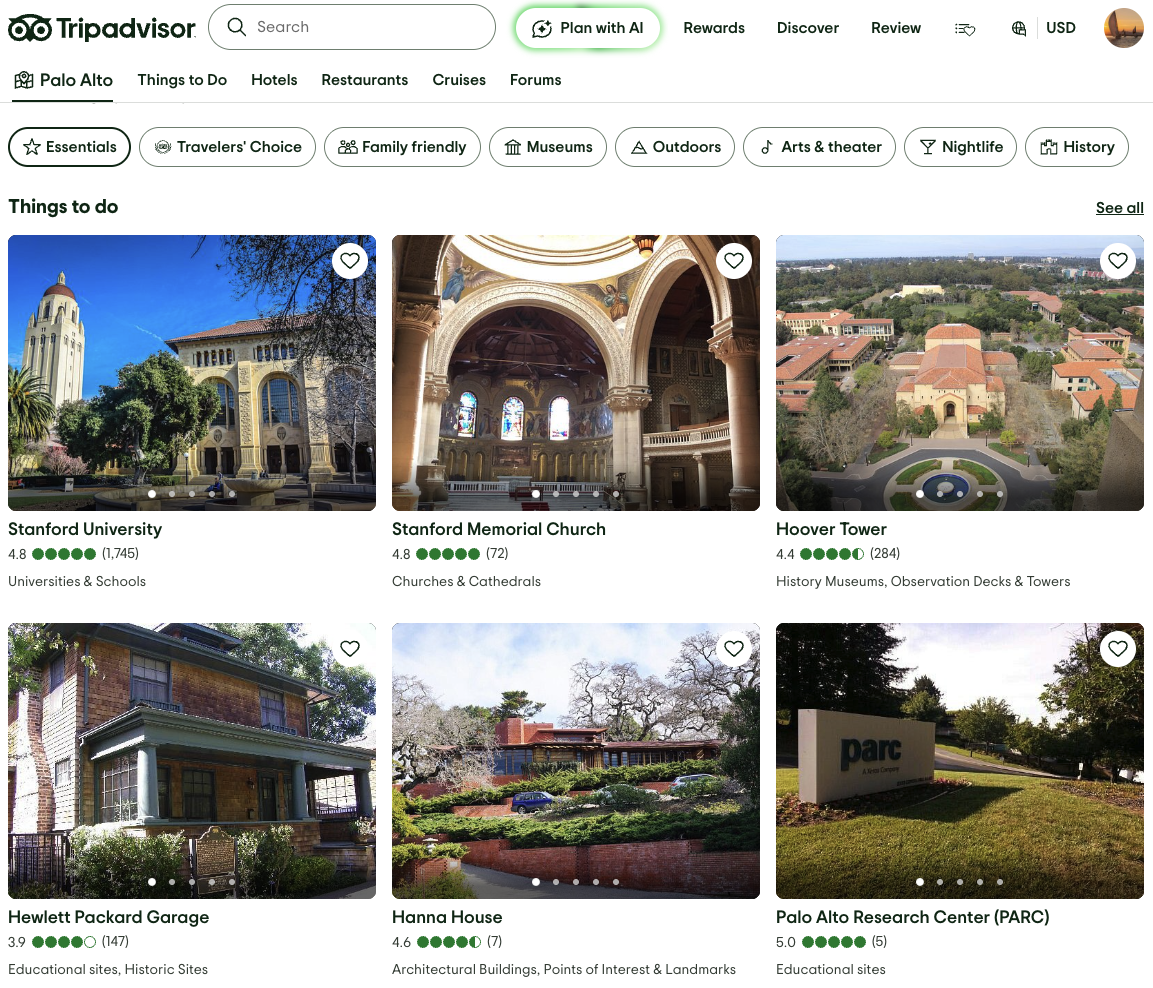}
        \caption{TripAdvisor.}
        \label{fig:tripadvisor_screenshot}
    \end{subfigure}
    \hspace{12mm}
     \centering
    \begin{subfigure}[b]{0.2\textwidth}
        \centering
        \includegraphics[width=\textwidth]{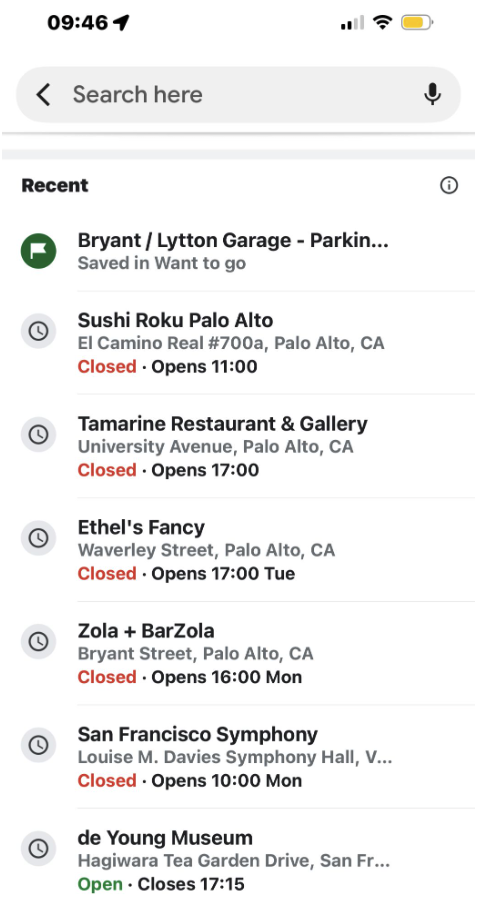}
        \caption{Google Maps.}
        \label{fig:googlemaps_screenshot}
    \end{subfigure}
   \caption{(Color online) Example applications of point-of-interest (POI) recommender systems.}
  \label{fig:screenshots_trip_google}
\end{figure}

\subsection{LLM Reasoning over Relevant and Irrelevant Attributes}
\label{appen:llm_examples}

We present two examples of the reasoning a pretrained LLM brings to a recommendation setting. In the first, we asked GPT-4o \citep{achiam2023gpt} to explain why a consumer might or might not purchase orange juice given her prior purchase history. We also included potentially irrelevant details for the product, such as packaging and logo, to see how the model separates decision-relevant from incidental information. As shown in Figure~\ref{fig:example_explanations}, GPT-4o generated plausible explanations for both outcomes. In the positive case it inferred that the consumer ``regularly buys fruits including oranges, and the product is an orange juice'', identifying the semantic connection between oranges and orange juice (Figure~\ref{fig:pos_explanation}). In the negative case it reasoned that ``the consumer may prefer whole fruits over processed juices'', drawing the distinction between whole-fruit preference and juice consumption (Figure~\ref{fig:neg_explanation}).

\begin{figure}[hbtp!]
    \begin{subfigure}[b]{0.45\textwidth}
        \centering
        \includegraphics[width=\textwidth]{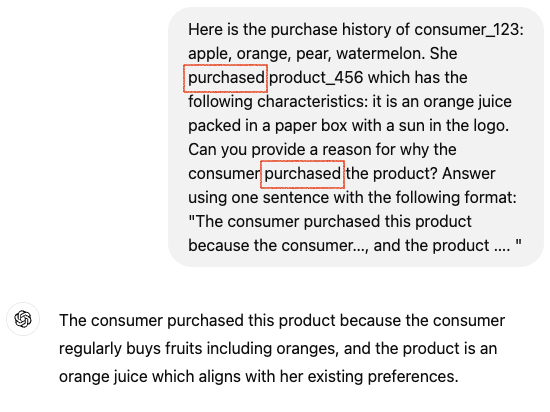}
        \caption{Positive explanation connecting orange juice to past fruit purchases.}
        \label{fig:pos_explanation}
    \end{subfigure}
    \hfill
     \centering
    \begin{subfigure}[b]{0.45\textwidth}
        \centering
        \includegraphics[width=\textwidth]{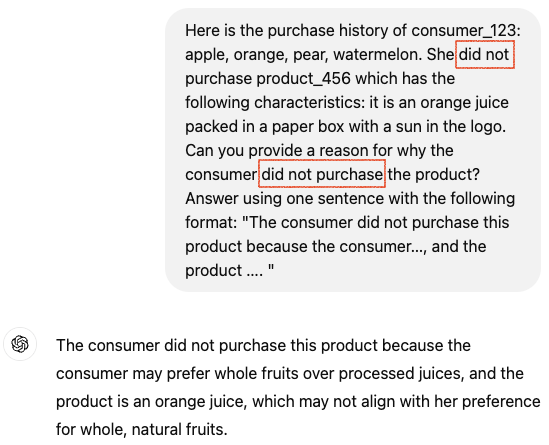}
        \caption{Negative explanation contrasting whole-fruit preference with processed juice.}
        \label{fig:neg_explanation}
    \end{subfigure}
    \caption{(Color online) GPT-4o's explanations of positive and negative outcomes.}
  \label{fig:example_explanations}
\end{figure}

A similar opportunity arises in POI recommendation, where relevant preference patterns often span categories. Given a visitation history that includes a garage, upscale restaurants, and a symphony venue, we asked Claude Opus~4.8 \citep{anthropic2026claudeopus48} to explain why the consumer might visit a museum. The model connected these otherwise distinct places through possible preferences for fine art, cultural experiences, and architectural design (Figure~\ref{fig:claude_example}), pointing to cross-category preference dimensions that are difficult to learn from item categories or sparse interaction data alone.

\begin{figure}[hbtp!]
    \centering
    \includegraphics[width=0.5\linewidth]{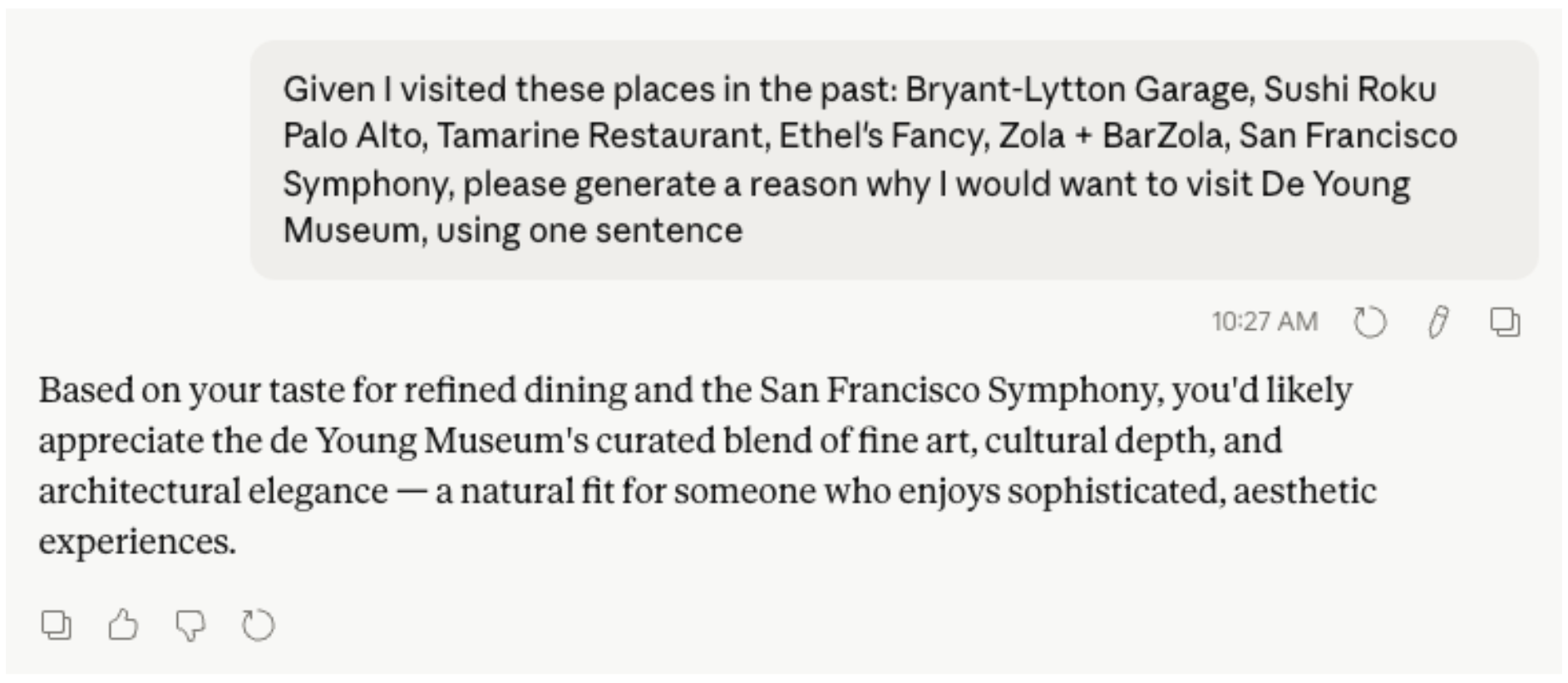}
    \caption{(Color online) Explanations for POI recommendations generated directly by Claude Opus 4.8.}\label{fig:claude_example}
\end{figure}

\subsection{LLMs Can Generate Multiple Different but Plausible Explanations}
\label{appen:llm_hallucination}

Figure~\ref{fig:claude_hallucination} illustrates a central challenge in using LLM-generated explanations for recommendation. Given the same POI recommendation scenario and visitation history, Claude Opus~4.8 generates several distinct explanations, each of which appears plausible in isolation. However, the observed data do not reveal which explanation, if any, reflects the consumer's \emph{actual} motivation or is most useful for predicting future behavior. This example motivates the need for a task-based learning signal, as in RecPIE, that evaluates explanations according to their contribution to downstream recommendation accuracy.
\begin{figure}[hbtp!]
    \centering
    \includegraphics[width=0.65\linewidth]{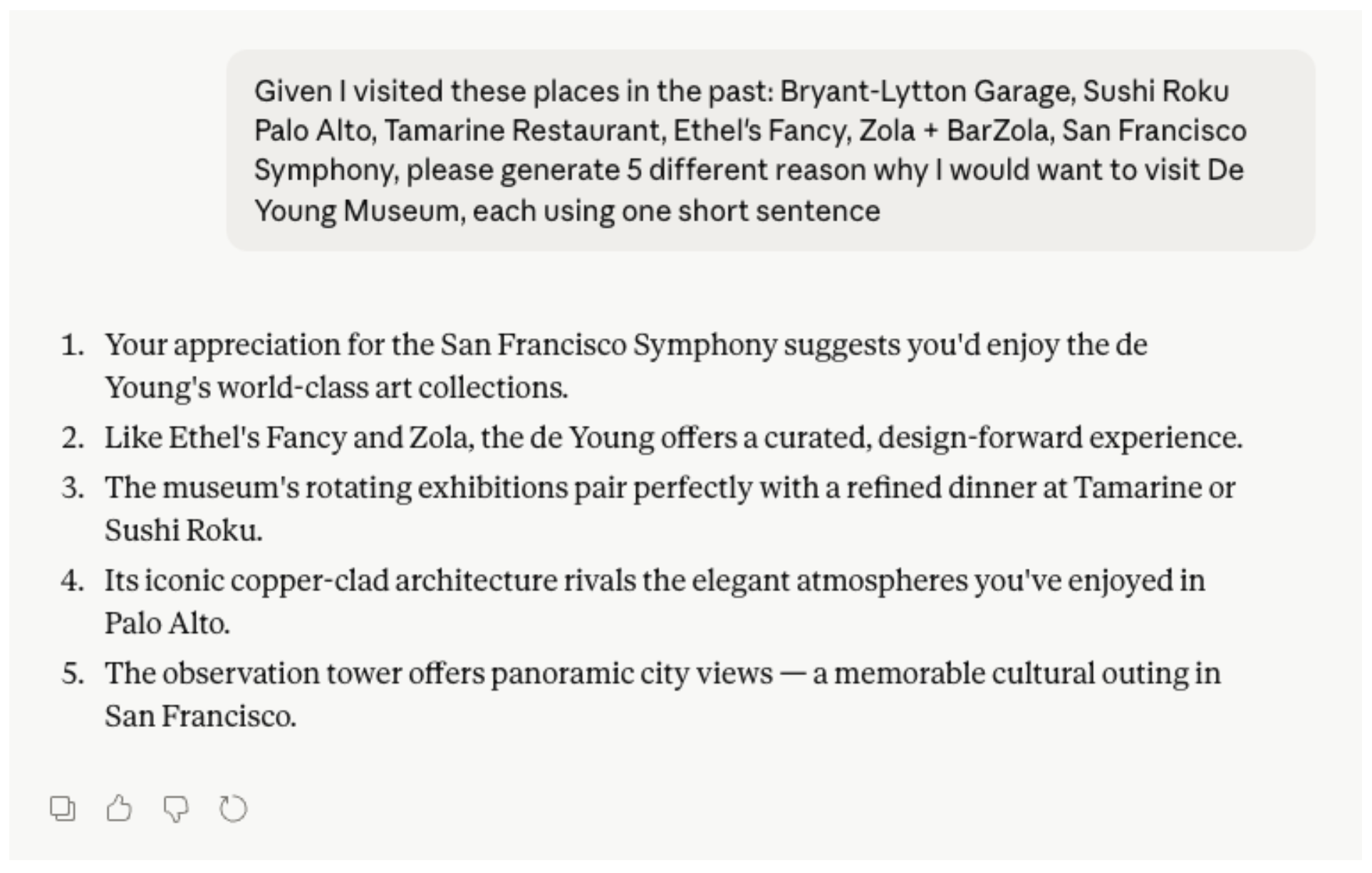}
    \caption{Multiple explanations generated by Claude Opus~4.8 for the same POI recommendation scenario.}\label{fig:claude_hallucination}
\end{figure}


\section{Conceptual Motivation from High-Dimensional Statistical Learning}
\label{appen:theory}

In this section, we provide intuition for why jointly modeling explanations and predictions can improve the performance for both tasks. Consumer decision-making can be viewed as a high-dimensional prediction problem. Using orange juice as an example, the product may be described by many attributes such as ingredients, packaging, brand, and size, forming a high-dimensional input $X$. The goal of the recommender system is to predict an outcome $Y$ (e.g., whether the consumer makes a purchase) based on $X$.

Figure~\ref{fig:illustration_orange} shows how a conventional recommender approaches this problem: consumers and products are mapped into a shared latent embedding space, and products lying close to a consumer's representation are recommended. For example, such a system can learn that ``oranges'' and ``orange juice'' belong together, but only after observing many consumers who bought both, since nothing in the representation encodes \emph{why} the two are related.

\begin{figure}[hbtp!]
    \centering
    \includegraphics[width=0.85\linewidth]{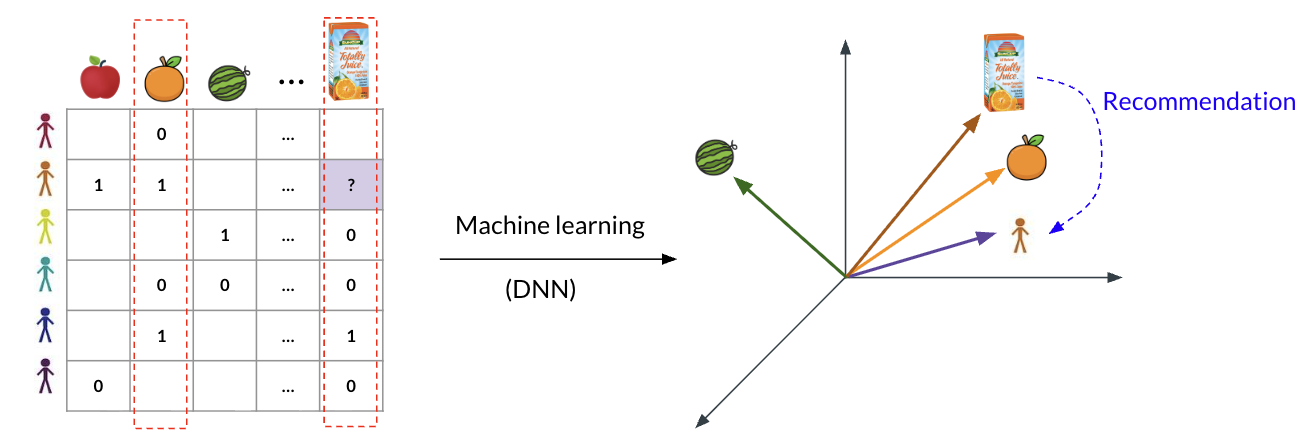}
    \caption{(Color online) Illustration of modern recommender systems via latent embeddings.}\label{fig:illustration_orange}
\end{figure}

When prompting LLMs to generate explanations for consumer preferences, we effectively ask the model to identify the key factors driving the observed outcome, which can be interpreted as a low-dimensional representation of the underlying data-generating process. In the orange juice example, although $X$ may contain many attributes, the LLM highlights a small subset of relevant features (related to ingredients) among these high-dimensional inputs (Figure~\ref{fig:pos_explanation}). We denote this subset of important variables by $S^*$.

Under this perspective, the prediction task corresponds to estimating $Y$ given $X$, while the explanation task corresponds to identifying the subset of features in $X$ that most strongly influence $Y$. When an LLM explains why an input $\vx$ leads to an outcome $y$, it effectively provides a candidate approximation of the relevant feature set $S^*$. For instance, in Figure~\ref{fig:example_explanations}, the LLM-generated explanations highlight salient factors (e.g., \texttt{hasBoughtOrange}, \texttt{isOnlyWholeFruitBuyer}) from a high-dimensional representation of consumer behavior and product attributes.

Intuitively, identifying the relevant subset $S^*$ reduces the effective dimensionality of the problem, allowing models to focus on the most informative signals. Consider a prediction problem
\[
Y=f^*(\vx_{S^*})+\epsilon,
\]
where $\vx\in\mathbb{R}^p$ contains consumer, item, and contextual information, $S^*\subseteq\{1,\ldots,p\}$ indexes the relevant variables, and $\mathbb{E}[\epsilon\mid\vx_{S^*}]=0$. Suppose that $s^*=|S^*|\ll p$. An explanation may provide information about this lower-dimensional structure by highlighting factors relevant to consumer preferences. For example, the orange juice example in Section~\ref{sec:llm_promise_challenges} connects a consumer's purchase history to preferences for orange-based or unprocessed products. Such explanations provide candidate information about relevant factors, although they need not identify $S^*$ accurately or completely.

We next present two statistical insights. First, LLMs can provide more accurate estimates of $S^*$ than models trained in a single environment with limited data (\emph{prediction-informed explanations}). Second, incorporating such estimates improves learning efficiency within a given environment (\emph{explanation-informed predictions}).  


\subsection{Prediction-Informed Explanations: LLMs Encode Knowledge of $S^*$}
\label{appen:theory_better_knowledge}
 
How do LLMs acquire information about $S^*$, the set of important variables? We draw on insights from \emph{multi-environment learning} \citep{peters2016causal}. In this setting, the goal is to predict an outcome $y$ from inputs $\vx$ using data from multiple environments, where each environment corresponds to a distinct context with potentially different input distributions but a shared underlying data-generating mechanism.

LLMs are trained on large and diverse corpora spanning many domains, contexts, and linguistic structures \citep{dubey2024llama, achiam2023gpt}. This training process can be viewed as analogous to learning from multiple environments: while the distribution of $\vx$ varies across contexts, the mapping from $\vx$ to $y$, which reflects the underlying decision-making process, remains stable. For example, in the orange juice example (Section~\ref{sec:llm_promise_challenges}), purchasing behavior may differ across contexts (e.g., grocery stores, online platforms, or farmers' markets), but the key decision factors (e.g., whether the product is orange-based or processed) remain invariant.

This intuition suggests that models trained across many environments are better able to recover the set of relevant variables $S^*$. We summarize this insight in Lemma~\ref{lem:var_selection_ellis}.

\begin{lem}
\label{lem:var_selection_ellis}
Consider a multi-environment linear model with a shared coefficient vector $\vbeta^*$ and support $S^*$. Under the invariance, heterogeneity, regularity, sample-size, and tuning conditions of Theorem~4.5 in \citet{fan2023environment}, the $\ell_0$-regularized environment-invariant least-squares estimator $\hat{\vbeta}_L$ satisfies
\[
\Pr\!\left(\mathrm{supp}(\hat{\vbeta}_L)=S^*\right)\to 1
\]
as sample size and dimensionality grow while these conditions hold.
\end{lem}

Lemma~\ref{lem:var_selection_ellis} implies that access to diverse environments facilitates accurate identification of the relevant variables $S^*$. Given the scale and diversity of data used in LLM pretraining, it is plausible that LLMs encode informative estimates of $S^*$. In contrast, models trained on a single environment with limited data may struggle to identify these variables. This provides intuition for why prediction-informed explanations can improve learning efficiency by providing additional structure about which features are most relevant.

\subsection{Explanation-Informed Predictions: Knowledge of $S^*$ Improves Prediction}
\label{appen:theory_better_performance}

The number of truly relevant variables $s^* = |S^*|$ is typically much smaller than the total number of variables $p$. Intuitively, knowledge of the support set $S^*$ allows the model to focus on relevant variables, thereby improving learning efficiency. We formalize this intuition in Lemma~\ref{lem:explanation_informed_predictions}.
 
\begin{lem}
\label{lem:explanation_informed_predictions}
Consider the model $Y=\vx^\top\vbeta^*+\epsilon$ with $n$ independent observations, $p$ predictors, and support $S^*$ of size $s^*$. Assume mean-zero sub-Gaussian errors independent of the predictors, bounded noise variance, normalized predictors, and suitable restricted-eigenvalue conditions with constants bounded away from zero. An appropriately tuned Lasso estimator satisfies
\[
\|\hat{\vbeta}_{\mathrm{Lasso}}-\vbeta^*\|_2
=O_P\!\left(\sqrt{\frac{s^*\log p}{n}}\right).
\]
If $S^*$ is known, least squares restricted to $S^*$ satisfies
\[
\|\hat{\vbeta}_{\mathrm{Orc}}-\vbeta^*\|_2
=O_P\!\left(\sqrt{\frac{s^*}{n}}\right).
\]
Thus, knowledge of the relevant variables removes the logarithmic dimensionality factor from these estimation-error bounds.
\end{lem}

The Lasso bound follows from \citet{bickel2009simultaneous}; the oracle bound follows from the standard least-squares variance calculation on the well-conditioned design restricted to $S^*$. These bounds illustrate how knowledge of relevant structure can reduce the estimation burden. They provide an oracle benchmark rather than a guarantee that imperfect explanations achieve the same improvement.

Identifying relevant factors also differs from estimating their predictive effects. An explanation may suggest that a consumer values unprocessed ingredients without determining how strongly this preference affects the outcome for a particular item. This distinction motivates using explanations as inputs to a discriminative recommender that learns their predictive relationships from observed behavior.

The rate improvements in Lemma~\ref{lem:explanation_informed_predictions} are significant because $p$ is usually very large in modern recommender systems.

\subsection{Implications}
\label{appen:theory_implications}

The statistical insights above directly inform the design of our proposed framework, RecPIE in Section~\ref{sec:framework}, through two complementary components:
\begin{itemize}
    \item \textbf{Prediction-informed explanations}: LLM-generated explanations, guided by prediction outcomes, provide informative estimates of the relevant variables $S^*$. Because LLMs are trained on diverse environments, they can capture generalizable decision-making factors beyond what is available in the training data.
    \item \textbf{Explanation-informed predictions}: Incorporating these explanations into the prediction model improves learning efficiency by focusing on the most relevant variables. This reduces the effective dimensionality of the problem and lowers data requirements. Note that while statistical theory suggests discarding variables outside $S^*$, in practice RecPIE retains low-dimensional, environment-specific representations (e.g., consumer and place embeddings) to capture idiosyncratic variation across environments. This design balances invariant structure with environment-specific effects and improves empirical performance.
    
\end{itemize}

Note that these statistical insights serve as motivations for the design of the proposed RecPIE framework, not as theoretical guarantees for its performance. Establishing such guarantees for LLM-based learning remains an important open problem and is beyond the scope of this paper. It is worth noting that the analogy is limited in the following three respects. First, the results above concern idealized estimators, environment-invariant least squares and an oracle least-squares estimator, under high-dimensional assumptions that a fine-tuned LLM does not satisfy. Second, the corpora on which LLMs are pretrained are not ``environments'' in the technical sense that the multi-environment results require. Third, the step from ``such estimators can recover the relevant subset'' to ``LLMs recover this subset'' is intuition and not a derivation. A rigorous theory of LLM representation learning is not available without imposing assumptions we regard as unrealistic, so we present this section as a conceptual motivation. In the empirical evaluation (Sections~\ref{sec:results}--\ref{sec:understanding}), we show that RecPIE consistently outperforms both prediction-only and explanation-only approaches while substantially reducing data requirements (beating the best-performing baseline using only 25\% of the training data), with findings that are consistent with the statistical insights presented here.


\section{Technical Details of the RecPIE Framework}

\subsection{Background: Supervised Fine-Tuning of LLMs}
\label{appen:sft_background}

Section~\ref{sec:framework_ppo} contrasts our reinforcement-learning fine-tuning with the standard supervised paradigm. This subsection states that paradigm for reference.

Traditional supervised fine-tuning adapts a pretrained LLM to a downstream task by improving the accuracy of next-token prediction. Let $f_{\theta}$ denote an autoregressive language model with parameters $\theta$. Given a dataset $\mathcal{D}=\{(\mathbf{X}^{(k)}, \mathbf{Z}^{(k)})\}_{k=1}^{N}$,
where $\mathbf{X}^{(k)}=(x^{(k)}_{1},\dots,x^{(k)}_{n_k})$ is the input token sequence (e.g., an explanation prompt) and $\mathbf{Z}^{(k)}=(z^{(k)}_{1},\dots,z^{(k)}_{T_k})$ is the target output sequence (e.g., a desired explanation), the empirical supervised fine-tuning objective is
\begin{equation}
\label{eqn:sft}
\hat{\mathcal{L}}_{\mathrm{SFT}}(\theta)
= -
\frac{1}{N}
\sum_{k=1}^{N}
\sum_{t=1}^{T_k}
\log
p_{\theta}
\!\left(
z^{(k)}_{t}
\mid
\mathbf{X}^{(k)}, z^{(k)}_{<t}
\right),
\end{equation}
where $\theta$ stands for the parameters of the LLM and $p_{\theta}\!\left(z^{(k)}_{t}\mid\mathbf{X}^{(k)}, z^{(k)}_{<t}\right)$ refers to the probability of the target token $z^{(k)}_{t}$ given the prompt $\mathbf{X}^{(k)}$ and preceding tokens $z^{(k)}_{<t}$. This objective corresponds to minimizing the token-level cross-entropy loss between the empirical data distribution and the model distribution, thereby maximizing the likelihood of the observed tokens.

\subsection{LoRA for Parameter-Efficient Fine-Tuning}
\label{appen:lora}

In practice, updating all parameters $\theta$ is computationally expensive, so a common approach is \emph{Low-Rank Adaptation} (LoRA) \citep{hu2022lora}, which freezes the base model and updates only a small number of parameters. For a linear projection matrix
$\mathbf{W} \in \mathbb{R}^{d_{\text{out}} \times d_{\text{in}}}$,
LoRA parameterizes the update as
\[
\mathbf{W}' = \mathbf{W} + \Delta \mathbf{W},
\qquad
\Delta \mathbf{W} = \frac{\alpha}{r}\mathbf{B}\mathbf{A},
\]
where $\mathbf{A} \in \mathbb{R}^{r \times d_{\text{in}}}$ and
$\mathbf{B} \in \mathbb{R}^{d_{\text{out}} \times r}$ are trainable low-rank matrices with
$r \ll \min(d_{\text{in}}, d_{\text{out}})$, and $\alpha$ is a scaling hyperparameter. Let $\phi$ denote the collection of all LoRA parameters, yielding effective model parameters
$\theta'(\phi)$. The corresponding LoRA-based supervised fine-tuning objective is
\begin{equation}
\label{eqn:lora}
\hat{\mathcal{L}}_{\mathrm{LoRA}}(\phi)
=
-
\frac{1}{N}
\sum_{k=1}^{N}
\sum_{t=1}^{T_k}
\log
p_{\theta'(\phi)}
\!\left(
z^{(k)}_{t}
\mid
\mathbf{X}^{(k)}, z^{(k)}_{<t}
\right),
\end{equation}
where optimization updates only the LoRA parameters $\phi$. RecPIE retains this LoRA parameterization but replaces the supervised objective, which requires ground-truth target explanations, with the reward defined in Section~\ref{sec:framework_ppo}.

\subsection{Construction of Consumer and Place Embeddings}
\label{appen:consumer_place_embed}

We create two embedding tables: a consumer embedding table $\mathbf{E}_{c}$ and a place embedding table $\mathbf{E}_{p}$. Each row of $\mathbf{E}_{c}$ corresponds to a unique consumer ID, and each row of $\mathbf{E}_{p}$ corresponds to a unique place ID.

For an interaction between consumer $i$ and place $j$, the corresponding embeddings are retrieved as
\begin{equation}
\label{eq:embed}
\mathbf{\ve}^{(c)}_i = \mathrm{Lookup}(\mathbf{E}_{c}, i), \qquad
\mathbf{\ve}^{(p)}_j = \mathrm{Lookup}(\mathbf{E}_{p}, j),
\end{equation}
where $\mathrm{Lookup}(\mathbf{E}, i)$ returns the $i$-th row of the embedding matrix $\mathbf{E}$. The embeddings $\mathbf{\ve}^{(c)}_i$ and $\mathbf{\ve}^{(p)}_j$ serve as compact representations of consumer $i$ and place $j$ and are used as inputs to the downstream network.

Notably, the consumer embedding $\mathbf{\ve}^{(c)}_i$ is shared with the Explanation Component and corresponds to the personalized soft prompt described in Section~\ref{sec:framework_soft_prompt}. This shared embedding ensures that the same consumer representation is used consistently for both explanation generation and recommendation prediction.

During training, these embeddings are first randomly initialized and are then updated through back-propagation to minimize the recommendation loss function $\mathcal{L}$:
\begin{equation}
\label{eq:embedding_update}
\mathbf{E}_{c} \leftarrow \mathbf{E}_{c} - \eta \frac{\partial \mathcal{L}}{\partial \mathbf{E}_{c}}, \quad \mathbf{E}_{p} \leftarrow \mathbf{E}_{p} - \eta \frac{\partial \mathcal{L}}{\partial \mathbf{E}_{p}},
\end{equation}
where $\eta$ is the learning rate, and the loss $\mathcal{L}$ measures the deviation between the predicted outcome (e.g., predicted rating) and the actual outcome (e.g., actual rating) as detailed in Section~\ref{sec:framework_dnn_loss}. $\frac{\partial \mathcal{L}}{\partial \mathbf{E}_{c}}$ and $\frac{\partial \mathcal{L}}{\partial \mathbf{E}_{p}}$ are the gradients of the loss function with respect to the consumer/place embedding table, which indicate the direction in which the embeddings should be adjusted to achieve the largest decrease in prediction error. This process is repeated iteratively during training to improve the embeddings and to capture latent consumer/place characteristics.  


\subsection{How RecPIE Differs from LLM-Based and Explanation-Only Recommenders}
\label{appen:recpie_contrast}

RecPIE differs fundamentally from existing LLM-based recommender systems (e.g., \citealp{bao2023tallrec}) that use LLMs to generate recommendations directly. Pure LLM-based approaches generally underperform state-of-the-art DNN recommenders, because recommendation is inherently a \emph{discriminative} task (selecting which item to recommend), whereas LLMs are primarily \emph{generative} models and often struggle when applied directly to discriminative prediction problems \citep{ye2025lola}. RecPIE also differs from prior work that uses LLMs only for explanation generation: existing explanations are typically ad hoc and do \emph{not} improve recommendation accuracy. In contrast, we integrate LLMs into the recommendation model itself and explicitly learn \emph{which} explanations are most useful for improving predictive performance.

\subsection{Details of the Sequence Embedding in the Recommendation Component}
\label{appen:seq_embed}

We build a self-attention layer followed by a GRU layer to formulate the sequence embedding that captures global, local and temporal dependencies in the consumption history. We describe them below. 

\subsubsection*{Self-attention layer.}
The self-attention mechanism enables each item in a sequence to attend to all other items, effectively capturing global dependencies regardless of their position within the sequence. This is achieved by computing attention weights that quantify the influence of each element.

For consumer $i$'s consumption history $\{j_{i_1}, j_{i_2}, \ldots, j_{i_n}\}$, we transform this sequence into a series of embeddings. Specifically, we retrieve each place's embedding from the embedding table described in Appendix~\ref{appen:consumer_place_embed}. The consumption history is represented as $\mathbf{S} = [\vs_1, \vs_2, \dots, \vs_n]$, where each $\vs_l$ is the embedding of the $l$-th place in the sequence. $\vs_l$ is obtained through the lookup operation: $\vs_{l} = \text{Lookup}(\mathbf{E}_{p}, j_{i_l}), \forall l = 1,\dots,n$. For simplicity, we omit the consumer index $i$ in this notation.

For each input \(\mathbf{s}_l\), we compute \textit{query}, \textit{key}, and \textit{value} using learned weight matrices \(\mathbf{W}^Q\), \(\mathbf{W}^K\), and \(\mathbf{W}^V\):
\begin{equation}
\label{eq:qkv}
\mathbf{Q} = \mathbf{S} \mathbf{W}^Q, \quad \mathbf{K} = \mathbf{S} \mathbf{W}^K, \quad \mathbf{V} = \mathbf{S} \mathbf{W}^V,
\end{equation}
where $\mathbf{Q} = [\mathbf{q}_1, \mathbf{q}_2, \dots, \mathbf{q}_n],  \mathbf{K} = [\mathbf{k}_1, \mathbf{k}_2, \dots, \mathbf{k}_n], \mathbf{V} = [\mathbf{v}_1, \mathbf{v}_2, \dots, \mathbf{v}_n]$. Then for each query vector \(\mathbf{q}_l\), we compute the \emph{attention score} with each key vector \(\mathbf{k}_j\):
\begin{equation}
\label{eq:score_lj}
\text{score}_{lj} = \frac{\mathbf{q}_l \cdot \mathbf{k}_j^T}{\sqrt{d_k}}
\end{equation}
where $d_k$ is the dimensionality of the key vectors. This formulation is also known as ``scaled dot-product attention'' \citep{vaswani2017attention}, where the scaling factor $\sqrt{d_{k}}$ is used to counteract the effect of large dot product values, which can push the softmax function into regions with very small gradients and make learning less stable. The value of $\text{score}_{lj}$ can be viewed as the degree of ``attention'' that the $l$-th item in the sequence should give to the $j$-th item. A softmax function is applied to normalize the attention scores into attention weights so that they sum up to 1:
\begin{equation}
\label{eq:alpha_lj}
\alpha_{lj} = \frac{\exp(\text{score}_{lj})}{\sum_{j'=1}^{n} \exp(\text{score}_{lj'})}.
\end{equation}
The output for each query vector is the weighted sum of the value vectors:
\begin{equation}
\label{eq:z_l}
\mathbf{z}_l = \sum_{j=1}^{n} \alpha_{lj} \mathbf{v}_j
\end{equation}
Thus, the output of the self-attention layer is $\mathbf{Z} = [\mathbf{z}_1, \mathbf{z}_2, \dots, \mathbf{z}_n]$, which is a matrix that represents the consumer's sequential consumption history, accounting for global dependencies among the consumptions.

\subsubsection*{GRU layer.} The Gated Recurrent Unit (GRU) models sequences by processing them one step at a time, which makes it naturally suited for capturing local temporal dependencies and sequential order, and effective in tasks that require maintaining memory over time. The GRU deploys two gating mechanisms that are detailed below: the update gate $\mathbf{u}_t$, which controls how much of the past information is retained at time $t$, and the reset gate $\mathbf{r}_t$, which determines how much of the previous state is forgotten at time $t$.

The output $\mathbf{Z} = [\mathbf{z}_1, \mathbf{z}_2, \dots, \mathbf{z}_n]$ from the self-attention layer is passed through the GRU, which processes the sequence step-by-step to compute the hidden states at each time step $t$. Let the first output of the GRU layer be the first vector from the self-attention layer: $\mathbf{h}_1 = \mathbf{z}_1$. Then starting from $t = 2$, the output of the GRU layer $\rvh_t$ depends on $\rvh_{t-1}$ and $\rvz_t$ through a reset gate $\mathbf{r}_t$ and an update gate $\mathbf{u}_t$:
\begin{equation}
\label{eq:reset_gate}
\mathbf{r}_t = \sigma(\mathbf{W}_r \mathbf{z}_t + \mathbf{U}_r \mathbf{h}_{t-1} + \mathbf{b}_r), 
\end{equation}
\begin{equation}
\label{eq:update_gate}
\mathbf{u}_t = \sigma(\mathbf{W}_u \mathbf{z}_t + \mathbf{U}_u \mathbf{h}_{t-1} + \mathbf{b}_u),
\end{equation}
where $\mathbf{W}_r$, $\mathbf{U}_r$, $\mathbf{W}_u$, and $\mathbf{U}_u$ are learnable weight matrices and $\mathbf{b}_r$ and $\mathbf{b}_u$ are learnable vectors. $\mathbf{r}_t$ controls how much of the previous hidden state \(\mathbf{h}_{t-1}\) to forget; $\mathbf{u}_t$ determines how much of the previous hidden state should be carried forward, and $\sigma(\cdot)$ denotes the sigmoid activation function as $\sigma(z)=\frac{1}{1+e^{-z}}$. The candidate hidden state $\tilde{\mathbf{h}}_t$ is then calculated as 
\begin{equation}
\label{eq:cand_hid_state}
\tilde{\mathbf{h}}_t = \tanh(\mathbf{W}_h \mathbf{z}_t + \mathbf{U}_h (\mathbf{r}_t \odot \mathbf{h}_{t-1}) + \mathbf{b}_h),
\end{equation}
which incorporates the reset gate to adjust the influence of the past state. Here $\odot$ denotes the element-wise product, and $\mathbf{W}_h$, $\mathbf{U}_h$ and $\mathbf{b}_h$ represent learnable weight matrices and a learnable bias vector, respectively. $\tanh$ is the hyperbolic tangent activation function, which squashes the output to a range of $[-1, 1]$ for regulating the information flow within the network. Finally, the new hidden state \(\mathbf{h}_t\) is updated by interpolating between the previous hidden state \(\mathbf{h}_{t-1}\) and the candidate hidden state $\tilde{\mathbf{h}}_t$ using the update gate $\mathbf{u}_t$: 
\begin{equation}
\label{eq:new_hid_state}
\mathbf{h}_t = \mathbf{u}_t \odot \mathbf{h}_{t-1} + (1 - \mathbf{u}_t) \odot \tilde{\mathbf{h}}_t.
\end{equation}
After processing the entire sequence in \(\mathbf{Z}\) one by one, the output of the GRU layer is the sequence of hidden states $\mathbf{H} = [\mathbf{h}_1, \mathbf{h}_2, \dots, \mathbf{h}_n]$. We use the last hidden state, $\mathbf{h}_n$, as the final embedding for the consumer's sequential consumption history, i.e., $\ve_k^{(\mathrm{seq})}$, which encapsulates \emph{both} global dependencies from the self-attention layer and temporal dependencies captured by the GRU.

\subsection{Details of the Self-Attention Layer in the Recommendation Component}
\label{appen:self_attn_2}

Each element in the input layer $X_{\text{input}} = [\vz^{(\mathrm{pos})}_k, \vz^{(\mathrm{neg})}_k, \mathbf{\ve}^{(c)}_i, \mathbf{\ve}^{(p)}_j, \ve_k^{(\mathrm{seq})}]$ is an 8-dimensional vector, which is viewed as one item in the sequence for a standard self-attention layer. For the context features, we first apply a multi-layer perceptron (MLP) to transform them into an 8-dimensional vector. We construct three learned matrices for \emph{query}, \emph{key} and \emph{value}: \(\mathbf{W}^Q\), \(\mathbf{W}^K\), and \(\mathbf{W}^V\):
\begin{equation}
\label{eq:qkv_2}
\mathbf{Q} = X_{\text{input}} \mathbf{W}^Q, \quad \mathbf{K} = X_{\text{input}} \mathbf{W}^K, \quad \mathbf{V} = X_{\text{input}} \mathbf{W}^V.
\end{equation}
Specifically, 
\begin{equation}
\begin{aligned}
\label{eq:qkv_2_expanded}
\mathbf{Q} &= [\vq_{\text{pos}}, \vq_{\text{neg}}, \vq_c, \vq_p, \vq_{\text{seq}}, \vq_{\text{context}}], \\
\mathbf{K} &= [\vk_{\text{pos}}, \vk_{\text{neg}}, \vk_c, \vk_p, \vk_{\text{seq}}, \vk_{\text{context}}], \\
\mathbf{V} &= [\vv_{\text{pos}}, \vv_{\text{neg}}, \vv_c, \vv_p, \vv_{\text{seq}}, \vv_{\text{context}}].
\end{aligned}
\end{equation}


Then for any $i, j \in I = \{\mathrm{pos}, \mathrm{neg}, c, p, \mathrm{seq}\}$, the attention score $\text{score}_{ij}$, the normalized attention weight $\alpha_{ij}$, and the output for each element $i$, $\vz_i$, are computed using the same formulas in Eqs.~(\ref{eq:score_lj})--(\ref{eq:z_l}), with the dimensionality $d_k = 8$. The final output of the self-attention layer is then 
\begin{equation}
\begin{aligned}
X_{\text{self-attn}} &= [\vz_{\text{pos}}, \vz_{\text{neg}}, \vz_c, \vz_p, \vz_{\text{seq}}] \\
&= [\sum_{i \in I} \alpha_{\text{pos}, i} \vv_i, \sum_{i \in I} \alpha_{\text{neg}, i} \vv_i, \sum_{i \in I} \alpha_{c, i} \vv_i, \sum_{i \in I} \alpha_{p, i} \vv_i, \\
& \quad \sum_{i \in I} \alpha_{\text{seq}, i} \vv_i],
\end{aligned}
\end{equation}
which accounts for the interactions among the input elements.

\subsection{Model Architecture and Hyperparameters of the DNN Recommendation Component}
\label{appen:model_architecture_dnn}

After the input layer and the self-attention layer described in Appendix~\ref{appen:self_attn_2} above, the model passes through an MLP with layers of size [64, 8, 1]. The DNN recommendation component was trained using a learning rate of 0.01, a batch size of 256, and for 10 epochs.


\subsection{Trainable Parameters of Each Component}
\label{appen:trainable}

Table~\ref{tab:trainable} summarizes the trainable parameters of the two components of RecPIE and the role each plays.

\begin{table}[hbtp!]
  \centering
  \footnotesize
  \setlength\extrarowheight{4pt}
  \setlength{\tabcolsep}{8pt}
  \begin{tabular}{llp{6.2cm}}
    \toprule
    \textbf{Component} & \textbf{Trainable Parameters} & \textbf{Role} \\
    \midrule
    Explanation Component ($\phi$) & LLM LoRA parameters & Produces positive and negative explanations. \\
    \midrule
    \multirow{4}{*}{Recommendation Component ($w$)}
      & Consumer/Place embedding tables & Latent representations of consumers/places. \\
      & Self-attentive GRU layer & Encodes sequential consumption history. \\
      & Self-attention layer &  Models interactions among explanations, preferences, places, and sequence context. \\
      & Multi-layer perceptron (MLP) layers & Produces final predictions. \\
    \bottomrule
  \end{tabular}
  \caption{Trainable parameters within each component of the RecPIE framework.}
  \label{tab:trainable}
\end{table}

\subsection{Training Procedure}
\label{appen:algorithm}

Algorithm~\ref{algo:le_recsys} gives the pseudocode for the alternating training procedure described in Section~\ref{sec:framework_training}.

\begin{algorithm}[hbtp!]
\caption{Recommendation with Prediction-Informed Explanations (RecPIE)}
\label{algo:le_recsys}
\small
\begin{algorithmic}[1]
\REQUIRE Dataset $\mathcal{D}=\{(\vx_k,y_k)\}_{k=1}^N$, pretrained LLM, pretrained text embedding model, learning rates $\eta_{\phi},\eta_{w}$, number of epochs $E$, batch size $B$
\ENSURE Ranked list of recommended places
\STATE \textbf{Initialize} LoRA parameters $\phi$ (Explanation Component) and DNN parameters $w$ (Recommendation Component: embedding tables $\mathbf{E}_c, \mathbf{E}_p$, sequence encoder, self-attention layer, MLP layers)
\REPEAT
\STATE \textbf{\emph{Explanation Component (prediction-informed explanations):}}
\FOR{each minibatch $\mathcal{B} \subset \mathcal{D}$}
    \STATE Construct explanation prompts using the soft prompt $\mathbf{\ve}^{(c)}_i$, sequential history, and candidate place
    \STATE Generate positive and negative explanations using the LLM policy $\pi_{\phi}$
    \STATE Encode explanations into embeddings $\vz_k^{(\mathrm{pos})}$ and $\vz_k^{(\mathrm{neg})}$ using the text embedding model
    \STATE \textbf{PPO update:} fix $w$, set $\phi \leftarrow \phi - \eta_{\phi}\nabla_{\phi}\hat{\mathcal{L}}_{\mathrm{PIE}}(\phi)$ (Eq.~(\ref{eq:ppo_total}))
\ENDFOR
\STATE \textbf{\emph{Recommendation Component (explanation-informed predictions):}}
\FOR{epoch $=1$ to $E$}
    \STATE Shuffle dataset $\mathcal{D}$
    \FOR{each minibatch $\mathcal{B}_b$ of size $B$}
        \STATE Form inputs $\vx^{\text{input}}_k = [\vz^{(\mathrm{pos})}_k, \vz^{(\mathrm{neg})}_k, \mathbf{\ve}^{(c)}_i, \mathbf{\ve}^{(p)}_j, \ve^{(\mathrm{seq})}_k]$ and compute predictions $\hat{y}_k$
        \STATE \textbf{Backpropagation:} fix $\phi$, set $w \leftarrow w - \eta_w \nabla_w \hat{\mathcal{L}}_{\mathrm{Rec}}(w)$ (Eq.~(\ref{eq:rec_loss}))
    \ENDFOR
\ENDFOR
\UNTIL{both $\phi$ and $w$ converge}
\RETURN Places ranked by predicted outcomes $\hat{y}_k$
\end{algorithmic}
\end{algorithm}

\section{Additional Experimental Details}

\subsection{Baselines}
\label{appen:baselines}

We compare RecPIE against 21 state-of-the-art baselines, grouped into five categories: LLM-based, aspect-based, sequential, interpretable, and POI recommender systems.

\begin{enumerate}
    \item \textbf{LLM-Based Recommender Systems}: These models use LLMs as the core of the recommender.
    \begin{itemize}
        \item \textbf{RecSAVER} \citep{tsai2024leveraging} automatically assesses the quality of LLM reasoning responses through a self-verification process, and then uses them for the rating prediction task.
        \item \textbf{LLMRG} \citep{wang2023enhancing} uses LLMs to construct personalized reasoning graphs, representing the user's interests in an interpretable way.
        \item \textbf{TallRec} \citep{bao2023tallrec} fine-tunes LLMs with a large set of recommendation data to align LLMs with the recommendation tasks. 
    \end{itemize}
    \item \textbf{Aspect-Based Recommender Systems}: These models extract aspect terms (typically from reviews) and use them for preference reasoning.
    \begin{itemize}
        \item \textbf{A3NCF} \citep{cheng20183ncf} constructs a topic model to extract user preferences and item characteristics from reviews, and captures user attention on specific item aspects via an attention network.
        \item \textbf{SULM} \citep{bauman2017aspect} predicts the sentiment of a user about an item's aspects, identifies the most valuable aspects of their potential experience, and recommends items based on these aspects.
        \item \textbf{AARM} \citep{guan2019attentive} models interactions between similar aspects to enrich aspect connections between users and products, using an attention network to focus on aspect-level importance.
        \item \textbf{MMALFM} \citep{cheng2019mmalfm} applies a multi-modal aspect-aware topic model to estimate aspect importance and predict overall ratings as a weighted linear combination of aspect ratings.
        \item \textbf{ANR} \citep{chin2018anr} learns aspect-based user and item representations through an attention mechanism and models multi-faceted recommendations using a neural co-attention framework.
        \item \textbf{MTER} \citep{wang2018explainable} jointly factorizes tensors of consumers, items, aspects, and opinions extracted from reviews through multi-task learning, predicting ratings and generating aspect-level explanations.
    \end{itemize}

    \item \textbf{Sequential Recommender Systems}: These models predict future choices from past sequences using neural sequence architectures.
    \begin{itemize}
        \item \textbf{SASRec} \citep{kang2018self} utilizes self-attention to capture long-term semantics in user actions and identify relevant items in a user's history.
        \item \textbf{DIN} \citep{zhou2018deep} adopts a local activation unit to adaptively learn user interest representations from historical behaviors and predict preferences for candidate items.
        \item \textbf{BERT4Rec} \citep{sun2019bert4rec} adopts bidirectional self-attention and the Cloze objective to model user behavior sequences and avoid information leakage, enhancing recommendation efficiency.
        \item \textbf{UniSRec} \citep{hou2022towards} uses contrastive pre-training to learn universal sequence representations of user preferences, improving recommendation accuracy.
    \end{itemize}

    \item \textbf{Interpretable Recommender Systems}: These models produce recommendations together with explicit natural language or aspect-based explanations.
    \begin{itemize}
        \item \textbf{AMCF} \citep{pan2021explainable} maps uninterpretable general features to interpretable aspect features, optimizing for both recommendation accuracy and explanation clarity through dual-loss minimization.
        \item \textbf{PETER} \citep{li2021personalized} predicts words in target explanations using IDs, endowing them with linguistic meaning to generate personalized recommendations.
        \item \textbf{UCEPic} \citep{li2023ucepic} combines aspect planning and lexical constraints to produce personalized explanations through insertion-based generation, improving recommendation performance.
        \item \textbf{PARSRec} \citep{gholami2022parsrec} exploits common and individual behavior patterns via an attention mechanism to tailor recommendations and generate explanations based on these patterns.
        \item \textbf{ExpCTR} \citep{yu2025expctr} fine-tunes an LLM with reinforcement learning to generate explanations for click-through-rate prediction, using an LC-alignment reward that scores whether a generated explanation is consistent with the user's observed behavior.
        \item \textbf{LLM2ER-EQR} \citep{yang2024llm2er} fine-tunes an LLM with an explanation-quality reward to produce explanations for an already-trained recommender, optimizing the quality of the explanation text alone.
        \item \textbf{RXR} \citep{wang2018reinforcement} uses reinforcement learning to select review sentences as explanations for an already-trained recommender, rewarding selections from which a second agent can recover the recommender's predicted rating and that have good readability and consistency.
    \end{itemize}

    \item \textbf{POI Recommender Systems}: These models are designed specifically for location and POI recommendations.
    \begin{itemize}
        \item \textbf{LLM4POI} \citep{li2024large} presents a framework that uses pretrained large language models to preserve heterogeneous location-based social network data and to capture the inherent meaning of contextual information to produce POI recommendations.
    \end{itemize}
    
\end{enumerate}

Among these, ``LLM-Based Recommender Systems'', ``Sequential Recommender Systems'' and ``POI Recommender Systems'' are considered \textbf{recommendation-focused baselines}, as they are designed mainly to improve recommendation accuracy, while ``Aspect-Based Recommender Systems'' and ``Interpretable Recommender Systems'' are considered \textbf{explanation-focused baselines}, as they are designed mainly to improve explanation quality.

Table~\ref{tab:related_work_paper} summarizes how RecPIE differs from the three most closely related methods discussed in Section~\ref{sec:related_llm_rec}, along the dimensions that separate them.

\begin{table}[htbp]
\centering
\footnotesize
\setlength\tabcolsep{5pt}
\begin{tabular}{p{3.0cm} cccc c ccc}
\toprule
Method & \shortstack{Usefulness\\reward} & \shortstack{Shared\\objective} & \shortstack{Explanation\\feedback} & \shortstack{Contrastive\\pos./neg.} & \shortstack{Downstream\\recommender} & \shortstack{Improves\\both} & \shortstack{Human\\study} & \shortstack{Benefit\\analysis} \\
\midrule
RecPIE (Ours)     & \cmark & \cmark & \cmark & \cmark & DNN seq.\ & \cmark  & \cmark & \cmark \\
\citet{yu2025expctr}     & \cmark & \xmark & \cmark & \xmark & DeepFM    & Partial & \xmark & \xmark \\
\citet{yang2024llm2er} & \xmark & \xmark & \xmark & \xmark & \shortstack{GraphNN\\+ MLP} & \xmark  & \xmark & \xmark \\
\citet{wang2018reinforcement} & \xmark & \xmark & \xmark & \xmark & Any       & \xmark  & \cmark & \xmark \\
\bottomrule
\end{tabular}
\caption{Comparison of RecPIE with the most closely related methods (\cmark~= yes, \xmark~= no). ``Usefulness reward'' = reward derived from downstream recommendation accuracy; ``Shared objective'' = explanation and prediction share the same downstream objective; ``Explanation feedback'' = generated explanations are fed back to improve the recommender; ``Contrastive pos./neg.'' = contrastive positive and negative explanations; ``Improves both'' = improves recommendation accuracy \emph{and} explanation quality; ``Benefit analysis'' = analysis of data efficiency and niche consumers.}
\label{tab:related_work_paper}
\end{table}

\subsection{Hyperparameter Settings}
\label{appen:hyper_param}
To ensure a fair comparison in our experiments, we use grid search \citep{bergstra2011algorithms} and allocate an equal amount of effort, in terms of training time and memory usage, to identify the optimal hyperparameters for both our proposed approach and all baseline models. Consequently, we determine the following experimental settings for our proposed RecPIE framework:
\begin{itemize}
\item \textbf{Large Language Model}: We utilize the open-source Llama~3.1 \citep{dubey2024llama} as the LLM in our experiments. To demonstrate the flexibility of our approach, we also evaluate its performance with other LLMs in Section~\ref{res_understanding_reasoning}.

\item \textbf{Recommendation Model}: The recommendation model predicts the outcome (either the actual rating for a regression task or high vs. low rating for a classification task) for a candidate item using inputs that include consumer ID, place ID, consumer sequential behaviors, and LLM-generated explanations. Consumer and place IDs are mapped to 8-dimensional latent vectors through an initialized embedding table, which are learned and updated during training. Sequential behaviors are aggregated using a Self-Attentive GRU model (Section~\ref{sec:framework_dnn_cons_embed}) with one hidden layer of size 8. 

\item \textbf{Overall Architecture}: The inputs, consisting of 8-dimensional embeddings for positive explanations, negative explanations, consumer ID, place ID, and sequential behaviors, are concatenated via the attention layer described in Section~\ref{sec:framework_dnn_loss} and passed through three fully connected layers with hidden sizes [64, 8, 1] to produce the predicted outcome. For regression tasks (i.e., predicting actual ratings), the model is optimized by minimizing the mean squared error (MSE) between predicted and ground-truth ratings. For classification tasks (i.e., predicting high vs. low ratings), the model is optimized by minimizing the binary cross-entropy loss between predicted and ground-truth ratings. We use a learning rate of 0.01, a batch size of 256, and train the model for 10 epochs.

\end{itemize}


\section{Details of Human Evaluation}
\label{appen:human_eval}

\subsection{Experimental Protocol}

\textbf{Overview and Recruitment.}
The study was conducted using the Qualtrics survey platform, and participants were identified via Prolific IDs. The objective of the study was to evaluate explanations generated by RecPIE and four alternatives within the context of recommender systems.

\textbf{Participant Instructions.}
Participants were presented with six scenarios. In each, they were asked to imagine having visited and liked a sequence of 10 specific locations in the Mountain View, California area. If a place appeared more than once, it indicated multiple visits. Following the presentation of this history, participants were shown a new place they had purportedly visited and liked, and asked to provide feedback on it.

\textbf{Tasks.}
The experiment involved two distinct tasks:
\begin{enumerate}
    \item \textit{Multiple-choice responses:} Participants selected the ``most informative explanation'' from five candidates (generated by the five methods described in Section~\ref{sec:human_evaluation}, presented in random order).
    \item \textit{Free-form responses:} Participants wrote their own explanation (1--2 sentences) for why they would visit the new place given their history.
\end{enumerate}

\textbf{Informed Consent.}
Prior to beginning the tasks, participants were provided with a consent form stating that participation was voluntary and they could withdraw at any time. Participants were required to explicitly select ``I consent to participate in this research'' to proceed.

\subsection{Questionnaire Content and Scenarios}

\newcommand{\scenario}[4]{%
  \subsubsection*{#1}\mbox{}
  \noindent\textbf{History:} #2

  \smallskip
  \noindent\textbf{Target:} \textit{#3}

  \smallskip
  \noindent\textbf{#4}
  \smallskip\hrule\smallskip
}

\subsubsection*{Scenario 1 (Selection Task)}\hfill\break
\noindent\textbf{History:} New York Pizza, Srasa Kitchen, Charleston Plaza, Carl's Jr., Best Buy, PetSmart, Trader Joe's, Starbucks, Hobee's, Shorebird Egret Rookery

\smallskip
\noindent\textbf{Target:} \textit{In-N-Out Burger} --- A popular West Coast fast-food chain famous for its simple menu of burgers, fries, and shakes, known for fresh ingredients and ``secret menu'' items.

\smallskip
\noindent\textbf{Options (select one):}
\begin{enumerate}[label=(\alph*)]
    \item Since you've already been to Carl's Jr., you have to try the legendary West Coast burger experience at In-N-Out.
    \item You went to Carl's Jr., which is a fast-food place that sells burgers. In-N-Out is also a fast-food place that sells burgers, so you could go there too.
    \item Based on your visits to Carl's Jr. for casual meals and Trader Joe's for specialty items, In-N-Out Burger is a recommendation. It is a casual place where you can get specialty drinks and food.
    \item Your visit to Carl's Jr. shows you enjoy a classic burger, and In-N-Out is an iconic West Coast institution famous for perfecting that meal with fresh ingredients. Just as you visited Hobee's for its beloved coffeecake, you'll appreciate In-N-Out's reputation for being a local favorite with a simple menu done right.
    \item The user's history includes visits to fast-food chains that serve burgers, such as Carl's Jr. In-N-Out Burger is another establishment in the same category, aligning with this demonstrated preference for quick-service burger restaurants.
\end{enumerate}

\smallskip\hrule\smallskip

\subsubsection*{Scenario 2 (Selection Task)}\hfill\break
\noindent\textbf{History:} Diddams Party \& Toy Store, San Antonio Center, Best Buy, Trader Joe's, In-N-Out Burger, Orchard Supply Hardware, Bed Bath \& Beyond, Charleston Plaza, JOANN Fabrics and Crafts, Computer History Museum

\smallskip
\noindent\textbf{Target:} \textit{Ross Dress for Less} --- An off-price department store selling brand-name clothing, accessories, and home goods at a discount.

\smallskip
\noindent\textbf{Options (select one):}
\begin{enumerate}[label=(\alph*)]
    \item Given your visit to the Computer History Museum to see unique items and In-N-Out Burger for good value, you might like Ross. It is a place to find unique items at a good value.
    \item The user's history demonstrates a pattern of visiting large-format retail stores, such as Bed Bath \& Beyond and Best Buy. Ross Dress for Less is a retail establishment of a similar category and scale, aligning with this established consumer behavior.
    \item Your visits to large shopping centers and multi-category stores like Bed Bath \& Beyond show you enjoy browsing a wide variety of merchandise. Ross fits this pattern perfectly, offering a diverse selection of clothing and home goods with the added appeal of finding brand names at a discount.
    \item You have been to big stores like Bed Bath \& Beyond that sell many different things. Ross is another big store that sells different things for your house and for you to wear.
    \item You're already a shopping pro, hitting up major hubs like San Antonio Center and stores like Best Buy. Think of Ross as your next great treasure hunt.
\end{enumerate}

\smallskip\hrule\smallskip

\subsubsection*{Scenario 3 (Selection Task)}\hfill\break
\noindent\textbf{History:} Taco Bell, TacoMania, San Antonio Center, O'Reilly Auto Parts, McDonald's, Stierlin Auto Wash, Charleston Plaza, Rincon Sabroso Restaurant, Diddams Party \& Toy Store, Bailey Park Plaza Shopping Center

\smallskip
\noindent\textbf{Target:} \textit{Baskin-Robbins} --- An international chain of ice cream parlors famous for its ``31 flavors'' concept, offering a wide variety of ice cream flavors, cakes, and frozen treats.

\smallskip
\noindent\textbf{Options (select one):}
\begin{enumerate}[label=(\alph*)]
    \item Your history of visiting practical places like O'Reilly Auto Parts and fun places like Diddams Party Store suggests an interest in Baskin-Robbins. It can be a fun stop during a day of practical errands.
    \item After running errands at places like Charleston Plaza and O'Reilly's, you totally deserve a treat, and Baskin-Robbins is the perfect reward.
    \item The user's history shows a pattern of visiting quick-service restaurants like McDonald's for convenient meals. Baskin-Robbins is an establishment of the same service model that specializes in the dessert category, aligning with this preference.
    \item You have been to fast-food places like Taco Bell and McDonald's to eat a meal. Baskin-Robbins is a place where you can get a quick dessert like ice cream.
    \item Your visits to fast-food spots like McDonald's and Taco Bell show you appreciate a quick, classic treat, making Baskin-Robbins a perfect dessert equivalent. It's also a convenient and rewarding stop to add to your shopping trips at plazas like San Antonio Center.
\end{enumerate}

\smallskip\hrule\smallskip

\subsubsection*{Scenario 4 (Generation Task)}\hfill\break
\noindent\textbf{History:} Jack in the Box ($\times$3), Century Cinema 16, Little Caesars Pizza, Cheztakos!!!, Walmart, McDonald's, Safeway, Mountain View Farmers' Market

\smallskip
\noindent\textbf{Target:} \textit{Sylvan Park} --- A city park with recreational facilities including sports fields, a playground, and picnic areas, serving as a community space for outdoor activities.

\smallskip
\noindent\textbf{Prompt:} Please write your own explanation (1--2 sentences) for why you, as the user who liked the 10 places, might have visited and liked this new place.

\smallskip\hrule\smallskip

\subsubsection*{Scenario 5 (Generation Task)}\hfill\break
\noindent\textbf{History:} Sylvan Park, ICICLES, Extended Stay America, Pokeworks, Walmart, Kohl's, JOANN Fabrics and Crafts, Safeway, Don Giovanni, Sweet Tomatoes

\smallskip
\noindent\textbf{Target:} \textit{La Fontaine} --- A restaurant serving fine French and European cuisine, known for its elegant atmosphere and popular for special occasions.

\smallskip
\noindent\textbf{Prompt:} Please write your own explanation (1--2 sentences) for why you, as the user who liked the 10 places, might have visited and liked this new place.

\smallskip\hrule\smallskip

\subsubsection*{Scenario 6 (Generation Task)}\hfill\break
\noindent\textbf{History:} Tea Era, La Salsa, Cucina Venti Restaurant, Jersey Mike's Subs, Mamacitas! Mexican Grill, Chop \& Pub, Cost Plus World Market, Alexander's Patisserie, Sweetgreen, QBB - Quality Bourbons \& Barbecue

\smallskip
\noindent\textbf{Target:} \textit{Red Rock Coffee} --- An independently-owned coffee shop known for its community-oriented atmosphere and live events, serving as a hub for locals to work and socialize.

\smallskip
\noindent\textbf{Prompt:} Please write your own explanation (1--2 sentences) for why you, as the user who liked the 10 places, might have visited and liked this new place.

\subsection{Explanation Length and Human Preference}
\label{appen:length}

Section~\ref{sec:human_evaluation} reports that RecPIE's explanations are preferred 61.5\% of the time. This subsection examines whether this preference is driven by explanation length.



To test this, we regenerate every method's explanations under a common length constraint of 30--40 words. For the methods that generate from an instruction prompt, we append ``Write between 30 and 40 words, double-check to confirm.'' to the prompt in Section~\ref{sec:framework_soft_prompt}, while for PETER, which generates with a sequence model and does not follow instructions, we control the length at decoding. We then repeat the recommendation evaluation of Section~\ref{sec:main_results_recs}, and we score the regenerated explanations against the free-form explanations written by our human-study participants, since it is that comparison, reported in Section~\ref{sec:human_evaluation}, on which the length concern bears. Tables~\ref{tab:length_rec_paper} and~\ref{tab:length_expl_paper} report the results. With length held roughly constant across methods, RecPIE remains the best performer on recommendation accuracy and on all four explanation-quality metrics, and its recommendation margins are of a similar magnitude to those in Table~\ref{tab:googlemap_results}. Explanation-only and Llama~3 do not produce recommendations, so they are omitted from the accuracy table.

\begin{table}[htbp]
\centering
\footnotesize
\setlength\extrarowheight{2pt}
\begin{tabular}{l c ccc}
\toprule
Method & Mean length (words) & RMSE $\downarrow$ & MAE $\downarrow$ & AUC $\uparrow$ \\
\midrule
\textbf{RecPIE (Ours)} & 35.9 & \textbf{0.2995}$^{***}$ & \textbf{0.1994}$^{***}$ & \textbf{0.6541}$^{***}$ \\
ExpCTR           & 34.7 & 0.3101 & 0.2055 & 0.6425 \\
LLM2ER-EQR       & 34.3 & 0.3112 & 0.2059 & 0.6413 \\
PETER            & 35.2 & 0.3117 & 0.2048 & 0.6395 \\
TallRec          & 34.3 & 0.3130 & 0.2079 & 0.6472 \\
\bottomrule
\end{tabular}
\caption{Recommendation accuracy on Google Maps with all methods' explanations length-controlled to 30--40 words. $^{***}$ denotes significance at $p<0.01$ relative to the best baseline.}
\label{tab:length_rec_paper}
\end{table}

\begin{table}[htbp]
\centering
\footnotesize
\setlength\extrarowheight{2pt}
\begin{tabular}{l c cccc}
\toprule
Method & Mean length (words) & BLEURT $\uparrow$ & Coverage $\uparrow$ & Informativeness $\uparrow$ & Fluency $\uparrow$ \\
\midrule
\textbf{RecPIE (Ours)} & 35.9 & \textbf{0.5318}$^{***}$ & \textbf{0.3886}$^{***}$ & \textbf{0.6294}$^{***}$ & \textbf{0.4401}$^{***}$ \\
ExpCTR           & 34.7 & 0.4593 & 0.3774 & 0.5979 & 0.4188 \\
LLM2ER-EQR       & 34.3 & 0.4532 & 0.3739 & 0.5798 & 0.4018 \\
PETER            & 35.2 & 0.4901 & 0.3684 & 0.5860 & 0.4138 \\
TallRec          & 34.3 & 0.4703 & 0.3795 & 0.5971 & 0.4174 \\
Explanation-only & 34.8 & 0.5162 & 0.3823 & 0.6145 & 0.4278 \\
Llama 3          & 35.9 & 0.4529 & 0.3726 & 0.6098 & 0.4217 \\
\bottomrule
\end{tabular}
\caption{Explanation quality on Google Maps with all methods' explanations length-controlled to 30--40 words, measured against the free-form explanations written by human-study participants. $^{***}$ denotes significance at $p<0.01$ relative to the best baseline.}
\label{tab:length_expl_paper}
\end{table}

\subsection{Bar Plots for Individual Scenarios in the Human Evaluation}
\label{appendix:bar_plot}
Figure~\ref{fig:three_scenarios} presents a separate bar plot for each of the three multiple-choice scenarios described in Section~\ref{sec:human_evaluation}, and RecPIE is the most preferred method in all three.

\begin{figure}[hbtp!]
    \begin{subfigure}[b]{0.4\textwidth}
        \centering
        \includegraphics[width=\textwidth]{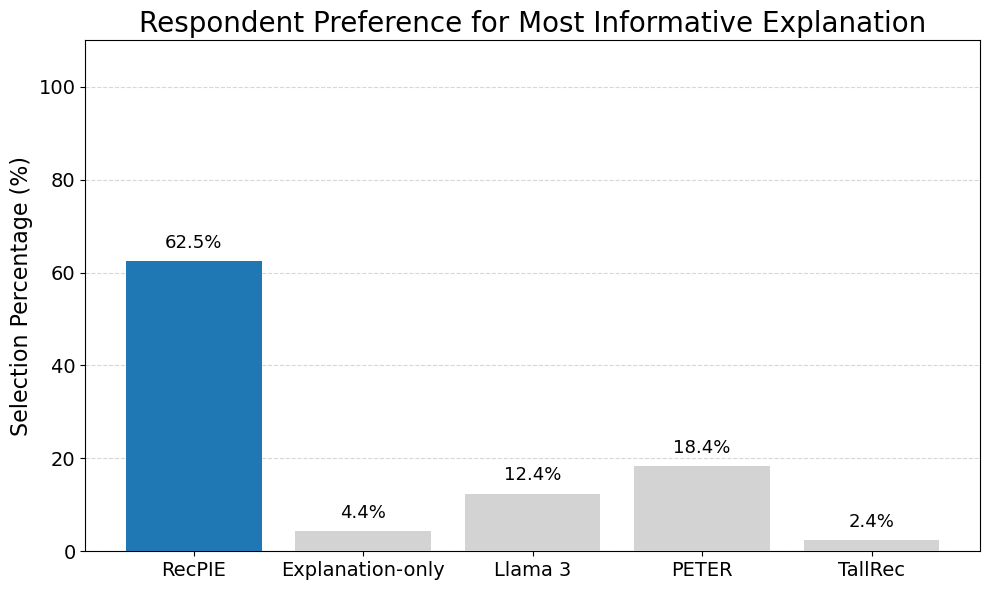}
        \caption{Scenario 1.}
    \end{subfigure}
    \hspace{0.5mm}
     \centering
    \begin{subfigure}[b]{0.4\textwidth}
        \centering
        \includegraphics[width=\textwidth]{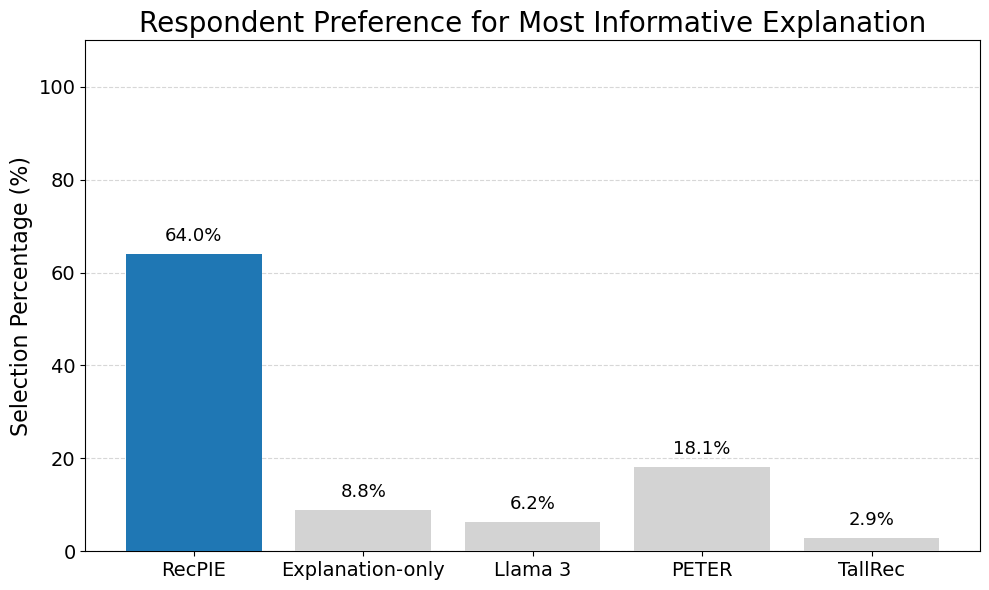}
        \caption{Scenario 2.}
    \end{subfigure}
    \begin{subfigure}[b]{0.4\textwidth}
        \centering
        \includegraphics[width=\textwidth]{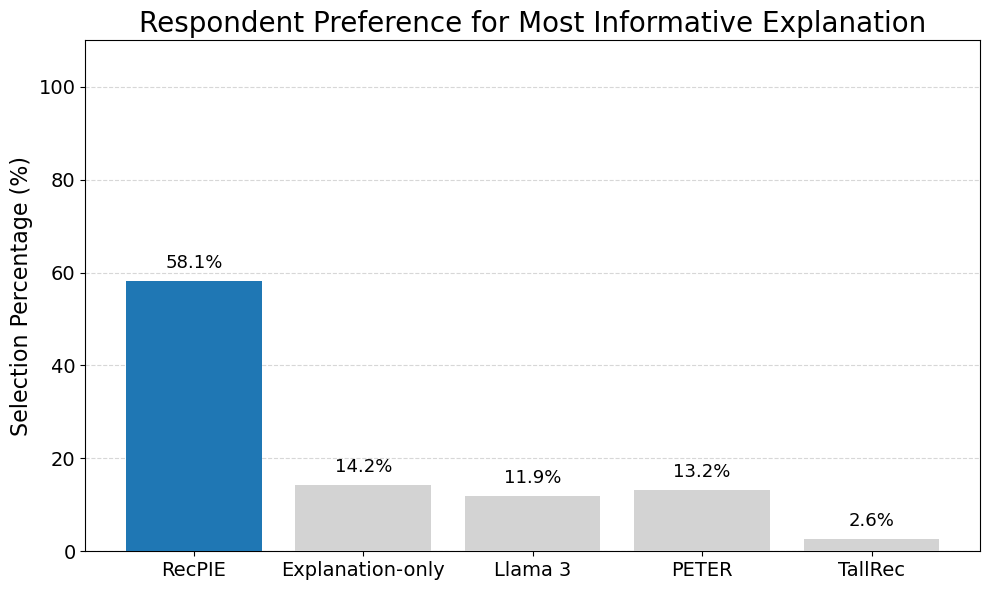}
        \caption{Scenario 3.}
    \end{subfigure}
   \caption{Human evaluation: respondent preference for the most informative explanation across three different scenarios.}
  \label{fig:three_scenarios}
\end{figure}

\section{Understanding the Improvements}

\subsection{What Explanations Capture Beyond Conventional Recommender Systems}
\label{appen:case_study}

Two researchers manually examined a random selection of 1,425 explanations generated by RecPIE and identified three systematic patterns of these explanations that a conventional recommender system struggles to capture explicitly. First, explanations infer subtle consumer traits from observed behavior. Second, they bridge across place categories through non-obvious reasoning, enabling recommendations that go beyond surface-level similarity, for example, linking a party store in the visit history to a sports bar based on an inferred interest in social gatherings. This cross-category reasoning is a key advantage of RecPIE in the POI recommendation setting, where consumer preferences regularly span diverse place categories without concentrating within a single one. Third, explanations are highly personalized: even for the same destination, different consumers receive different explanations reflecting distinct underlying motivations.

The three subsections below give additional evidence for each pattern in turn, showing how RecPIE's explanations capture higher-order dimensions of consumer preference that conventional recommender systems do not represent explicitly.

\subsection*{1. Inferring Subtle User Traits from Observed Behavior}

\noindent \emph{\textbf{Example A1: Restaurants and bars $\rightarrow$ Escape Room}}

\begin{tcolorbox}
\textbf{Visit history:} Scratch; The Voya Restaurant; Steins Beer Garden \& Restaurant; Starbucks; Sports Page 

\textbf{Place of interest:} Ryptic Room Escape Mountain View

\textbf{Positive explanation:}
\emph{
Based on the user's interest in unique and local experiences, such as visiting ``Steins Beer Garden \& Restaurant'' and ``The Voya Restaurant'', Ryptic Room Escape Mountain View is a good recommendation because it offers an immersive escape room experience that combines puzzle-solving with a fun, social atmosphere. The venue's unusual concept aligns well with the user's curiosity to explore new and offbeat destinations.
}

\textbf{Negative explanation:}
\emph{
Based on the user's past visits, it appears they tend to frequent casual eateries and coffee shops, rather than escape rooms or unique entertainment venues like ``Ryptic Room Escape Mountain View''. This suggests that the user might not find an escape room appealing as their next visit choice.
}
\end{tcolorbox}

The consumer only visited restaurants and bars, so a category-based system would be unlikely to recommend an escape room. The explanation, however, identifies the consumer as someone with ``curiosity to explore new and offbeat destinations'', which is an inference about personality that goes beyond matching features.

\noindent \emph{\textbf{Example A2: Googleplex + restaurants $\rightarrow$ Computer History Museum}}

\begin{tcolorbox}
\textbf{Visit history:} Kirin Restaurant; Google Merchandise Store; Googleplex; Target; In-N-Out Burger

\textbf{Place of interest:} Computer History Museum

\textbf{Positive explanation:}
\emph{
The ``Google Merchandise Store'' and ``Googleplex'' in the user's history suggest an interest in Google, which aligns with Computer History Museum as it is dedicated to preserving and presenting the history of computing and technology, including Google's early years. Therefore, recommending Computer History Museum based on their interest in Google could lead to a relevant and engaging experience for the user.
}

\textbf{Negative explanation:}
\emph{
The ``Computer History Museum'' is unlikely to be a good fit because it doesn't align with the user's interests in food, shopping, or entertainment that are reflected in their previous visits. The museum falls outside of these categories and may not appeal to someone who has shown preferences for more lighthearted activities.
}
\end{tcolorbox}

The explanation connects Googleplex and Google Merchandise Store to an ``interest in Google'' and extends it to ``the history of computing and technology, including Google's early years''. The explanation thus reasons at the level of the consumer's identity, as visits to Google's workplace suggest an interest in technology that extends naturally to a museum of computing.

\subsection*{2. Bridging Categories via Non-Obvious Reasoning}

\noindent \emph{\textbf{Example B1: Delis + Party Store $\rightarrow$ Sports Bar}}

\begin{tcolorbox}
\textbf{Visit history:} Roger's Deli \& Donuts; La Espuela Mexican Restaurant \& Cantina; Asian Box; Diddams Party \& Toy Store; Diddams Party \& Toy Store

\textbf{Place of interest:} Sports Page

\textbf{Positive explanation:}
\emph{
The reason is that both ``Roger's Deli \& Donuts'' and ``La Espuela Mexican Restaurant \& Cantina'' are food establishments, suggesting the user enjoys dining out, which makes it likely they'll appreciate other casual eateries like Sports Page. Additionally, the presence of party store in their history indicates an interest in social gatherings or celebrations, making sports bars like Sports Page a fitting suggestion.
}

\textbf{Negative explanation:}
\emph{
The reason is that ``Sports Page'' does not appear in the user's history, and there may be other recommendations based on their preferences and tastes from their previous visits to different types of establishments such as restaurants (``Roger's Deli \& Donuts'', ``La Espuela Mexican Restaurant \& Cantina'', ``Asian Box'') and stores like toy store (Diddams Party \& Toy Store).
}
\end{tcolorbox}

The explanation interprets the party store in the history as an interest in social gatherings, and carries that inference across to a sports bar. Connecting party supplies to socializing is a behavioral inference that no feature vector encodes. It parallels a classic retail analytics insight: people who buy diapers also buy beer, not because diapers relate to beer, but because both reflect the same latent lifestyle (new parents running errands). These examples show our prediction-informed explanations discovering similar hidden connections between POIs.

\noindent \emph{\textbf{Example B2: Niche Restaurants $\rightarrow$ Museum}}

\begin{tcolorbox}
\textbf{Visit history:} Kappo Nami Nami; Nijiya Market; Sports Page; Oren's Hummus; In-N-Out Burger

\textbf{Place of interest:} Computer History Museum

\textbf{Positive explanation:}
\emph{
Based on the user's interest in unique and local spots, such as Kappo Nami Nami (a Japanese restaurant) and Oren's Hummus (an Israeli-Mediterranean eatery), the Computer History Museum could be recommended because it is an unusual attraction that celebrates technology and innovation, reflecting the user's appreciation for niche interests. Additionally, its location in Mountain View, California, near many tech companies like Google, might appeal to someone who values proximity to innovative environments.
}

\textbf{Negative explanation:}
\emph{
The reason is that there is no evidence in the user's history of visiting cultural institutions like museums, whereas their visits are mostly to restaurants and markets, indicating they might prefer recommendations related to food or entertainment. Therefore, suggesting ``Computer History Museum'' would be an outlier from the user's typical interests.
}
\end{tcolorbox}

The explanation transfers preferences for niche food experiences into niche cultural interests.


\subsection*{3. Personalized Reasons for the Same Place}

The explanations generated by RecPIE carry strong consumer-specific signals. Even when the recommended POI is identical (so that item identity and description are held constant), the explanations differ systematically across consumers. These differences reflect distinct underlying motivations and correctly differentiate which consumers are more or less likely to visit the same place.

\noindent \emph{\textbf{Example C1: McDonald's rated 5 by a convenience-seeker, rated 1 by a niche-seeker}}

\begin{tcolorbox}
\textbf{Consumer A (rating = 5)}

\textbf{Visit history:} Fiesta Del Mar Too; Walmart; Maldonado's Pizzeria; Maldonado's Pizzeria; Marshalls \& HomeGoods

\textbf{Place of interest:} McDonald's

\textbf{Positive explanation:}
\emph{
The reason why McDonald's could be a good recommendation is that it appears in many people's daily routines, and users with similar interests or preferences tend to visit multiple fast-food restaurants like Walmart (a common destination). This pattern suggests an affinity for convenience stores and quick bites. Final Answer: The final answer is that McDonald's could be a good recommendation because of its widespread presence and appeal to those who frequent convenience stores and enjoy quick meals.
}

\textbf{Negative explanation:}
\emph{
The reason is that there are no instances of McDonald's in their previous visits, which makes it less likely to recommend such a location as they have shown little interest or activity around it. The absence of similar patterns or repeated interactions with other fast-food chains suggests that recommendations like McDonald's may not resonate well with this user.
}
\end{tcolorbox}

\begin{tcolorbox}
\textbf{Consumer B (rating = 1)}

\textbf{Visit history:} Chipotle Mexican Grill; Dollar Tree; Unique Men's Haircut; Everest Cuisine; In-N-Out Burger

\textbf{Place of interest:} McDonald's

\textbf{Positive explanation:}
\emph{
Based on the user's visit history, which suggests they have an affinity for casual dining and quick service restaurants (e.g., Chipotle, In-N-Out), McDonald's is a natural fit as it offers similar characteristics. Additionally, McDonald's is one of the most popular fast-food chains globally, making it likely to appeal to someone who has already demonstrated preferences for other popular, accessible eateries.
}

\textbf{Negative explanation:}
\emph{
Based on the user's past visits, it appears that they tend to frequent casual dining establishments with unique or niche offerings (e.g., Chipotle, Unique Men's Haircut). McDonald's does not fit this pattern and may seem like an out-of-character suggestion, making it less likely to appeal to their preferences.
}
\end{tcolorbox}

The same place is evaluated through different preference dimensions. For Consumer A, McDonald's is framed as consistent with a preference for convenience and routine consumption. For Consumer B, it is framed as inconsistent with a preference for distinctive or niche experiences. Thus, the explanation adapts to consumer-specific tastes and does not rely on item-level features alone.

\noindent \emph{\textbf{Example C2: Walmart rated 5 by a fast-food regular, rated 1 by a unique-store lover, and rated 3 by an errand-runner/traveler}}

\begin{tcolorbox}
\textbf{Consumer A (rating = 5)}

\textbf{Visit history:} Starbucks; Monte Carlo Restaurant \& Bar; In-N-Out Burger; In-N-Out Burger; San Antonio Center

\textbf{Place of interest:} Walmart

\textbf{Positive explanation:}
\emph{
The frequent visits to In-N-Out Burger, a fast-food chain with multiple locations, suggests that the user is likely looking for quick and convenient dining options. Walmart, as a retail giant often located near highways or residential areas, could potentially offer food courts or convenience stores nearby, catering to users seeking affordable meal solutions during their daily routines.
}

\textbf{Negative explanation:}
\emph{
The reason ``Walmart'' wouldn't be a suitable recommendation is that it doesn't align with the user's preferences, which are focused on food and dining establishments rather than general merchandise stores like Walmart. The user's history suggests they enjoy visiting restaurants and cafes, making recommendations related to those categories more likely to resonate with them.
}
\end{tcolorbox}

\begin{tcolorbox}
\textbf{Consumer B (rating = 1)}

\textbf{Visit history:} America's Tire; Cost Plus World Market; New Krungthai; La Espuela Mexican Restaurant \& Cantina; Shiva's

\textbf{Place of interest:} Walmart

\textbf{Positive explanation:}
\emph{
The similarity in names between ``America's Tire'' and Walmart, as well as the presence of other retail stores like Cost Plus World Market on their list, suggest that they may have an interest in shopping or consumer goods. Therefore, recommending Walmart to this user could be a good fit based on their existing preferences.
}

\textbf{Negative explanation:}
\emph{
The reason ``Walmart'' wouldn't be a suitable recommendation is that it doesn't align with the user's interests and preferences, which seem to favor unique stores like America's Tire or cultural destinations like New Krungthai. The user has shown no indication of being drawn to big-box retailers like Walmart in their past visits, making it an unlikely match based on their profile.
}
\end{tcolorbox}

\begin{tcolorbox}
\textbf{Consumer C (rating = 3)}

\textbf{Visit history:} Teaspoon; Mantra India - Mountain View; In-N-Out Burger; San Antonio Center; Enterprise Rent-A-Car

\textbf{Place of interest:} Walmart

\textbf{Positive explanation:}
\emph{
The presence of ``In-N-Out Burger'' and ``Enterprise Rent-A-Car'', which are often associated with road trips or travel, suggests that the user may have visited Walmart as part of their travels. Additionally, Walmart is a common pit stop along highways in many parts of the country, making it a plausible destination based on the user's historical location data.
}

\textbf{Negative explanation:}
\emph{
The reason is that Walmart, as an entity with multiple locations across different cities and countries, does not have any direct connection to the specific locations mentioned in the user's history (Mountain View, San Francisco). Therefore, it may not be relevant or appealing to their interests and preferences.
}
\end{tcolorbox}

Even for the same place, explanations emphasize different dimensions: convenience (Consumer A), preference for unique or specialized stores (Consumer B), and travel-related context (Consumer C). These examples highlight that RecPIE does not rely on a single, fixed rationale for a place, but instead generates consumer-contingent reasoning paths. In other words, our prediction-informed explanations encode consumer-specific reasoning over personalized preference dimensions, allowing the same item to be justified differently across consumers. This flexibility enables RecPIE to capture heterogeneity in consumer preferences that is difficult to represent with static feature-based approaches.


\subsection{Embedding Analysis: Faster Representation Learning}
\label{appen:similarity}

To further illustrate how RecPIE accelerates learning, we analyze the evolution of embedding similarity between representative place pairs across training epochs. Figure~\ref{fig:similarity} compares RecPIE with leading sequential recommender models, BERT4Rec and SASRec, using cosine similarity as a measure of semantic alignment.

In Figure~\ref{fig:simiarlity:1}, we track the cosine similarity between In-N-Out Burger and Five Guys, two restaurants offering a similar menu. RecPIE reaches a similarity score of about 0.89 within two epochs, and within about five epochs when trained with just 25\% of the data. By contrast, both BERT4Rec and SASRec require roughly 8--10 epochs to reach comparable levels. This pattern confirms that prediction-informed explanations in RecPIE substantially accelerate its ability to identify conceptually similar items, thereby improving data efficiency.

Figure~\ref{fig:similarity:2} examines the pair of Kirin Chinese Restaurant and 99 Ranch Market, which are both related to Asian cuisine but dissimilar in category. Here, RecPIE quickly lowers the embedding similarity toward 0.55, correctly distinguishing between a sit-down restaurant and a grocery retailer. Competing models, however, converge more slowly and remain above 0.60 for several epochs, indicating that they struggle to disentangle fine-grained category boundaries early in training.

Finally, Figure~\ref{fig:similarity:3} presents results for Mountain View Farmers Market and Kirin Chinese Restaurant, two unrelated entities. RecPIE immediately separates their embeddings, reducing similarity below 0.25 within the first three epochs, while BERT4Rec and SASRec remain above 0.28 in the first three epochs and decline to about 0.25 only after about eight epochs. The rapid decline highlights the framework's ability to encode discriminative representations even with limited data.

Together, these results visually confirm our findings, as RecPIE learns meaningful relationships between places up to four times faster and achieves comparable accuracy using only a fraction of the training data. The explanation-augmented learning process allows the model to approach the true data-generating relationships earlier, accelerate convergence, and improve generalization to new consumers and places.

\begin{figure}[hbtp!]
    \begin{subfigure}[b]{0.45\textwidth}
        \centering
        \includegraphics[width=\textwidth]{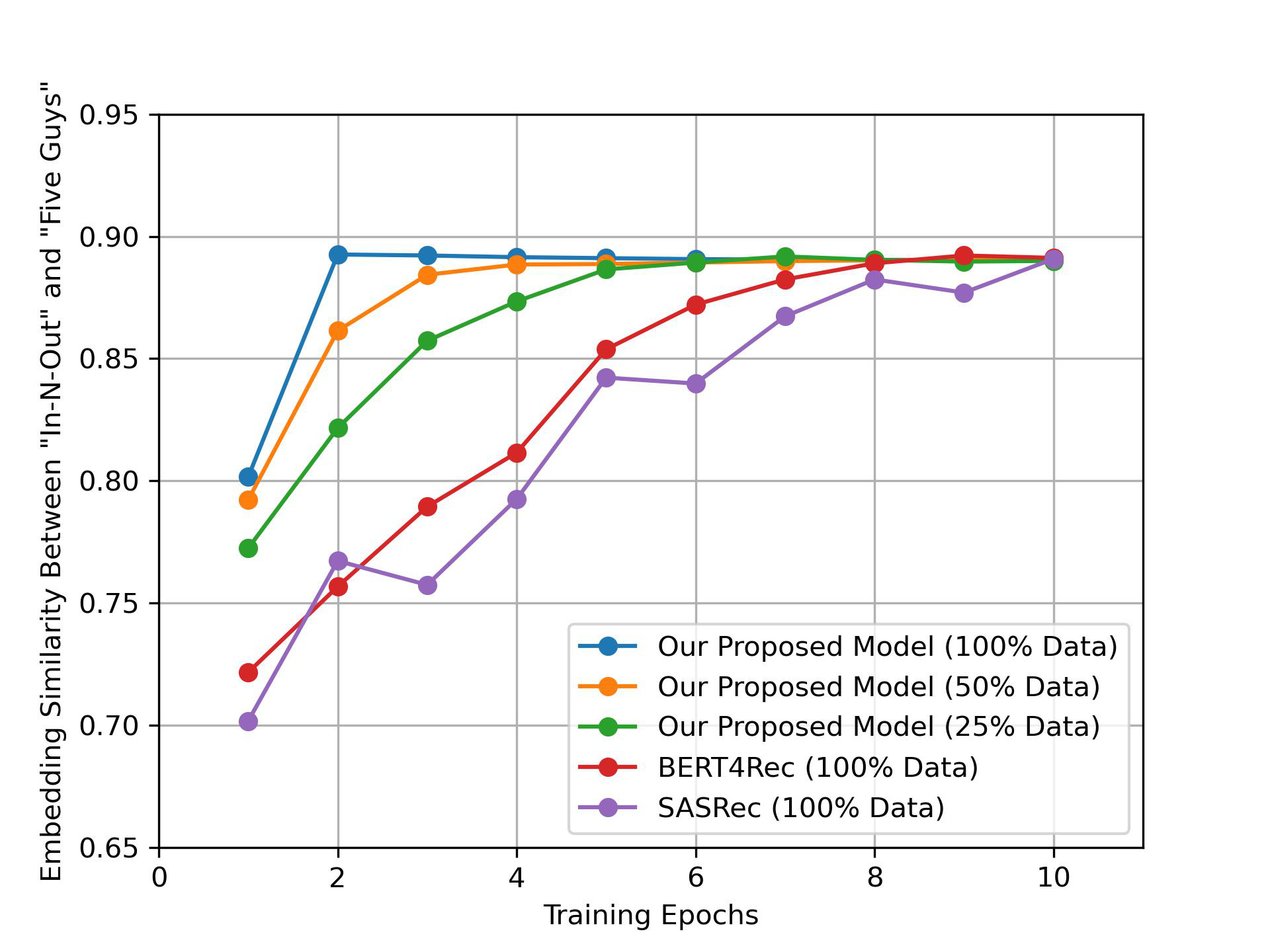}
        \caption{Similarity between similar entities \\ (In-N-Out Burger and Five Guys).}
        \label{fig:simiarlity:1}
    \end{subfigure}
    \hspace{0.5mm}
     \centering
    \begin{subfigure}[b]{0.45\textwidth}
        \centering
        \includegraphics[width=\textwidth]{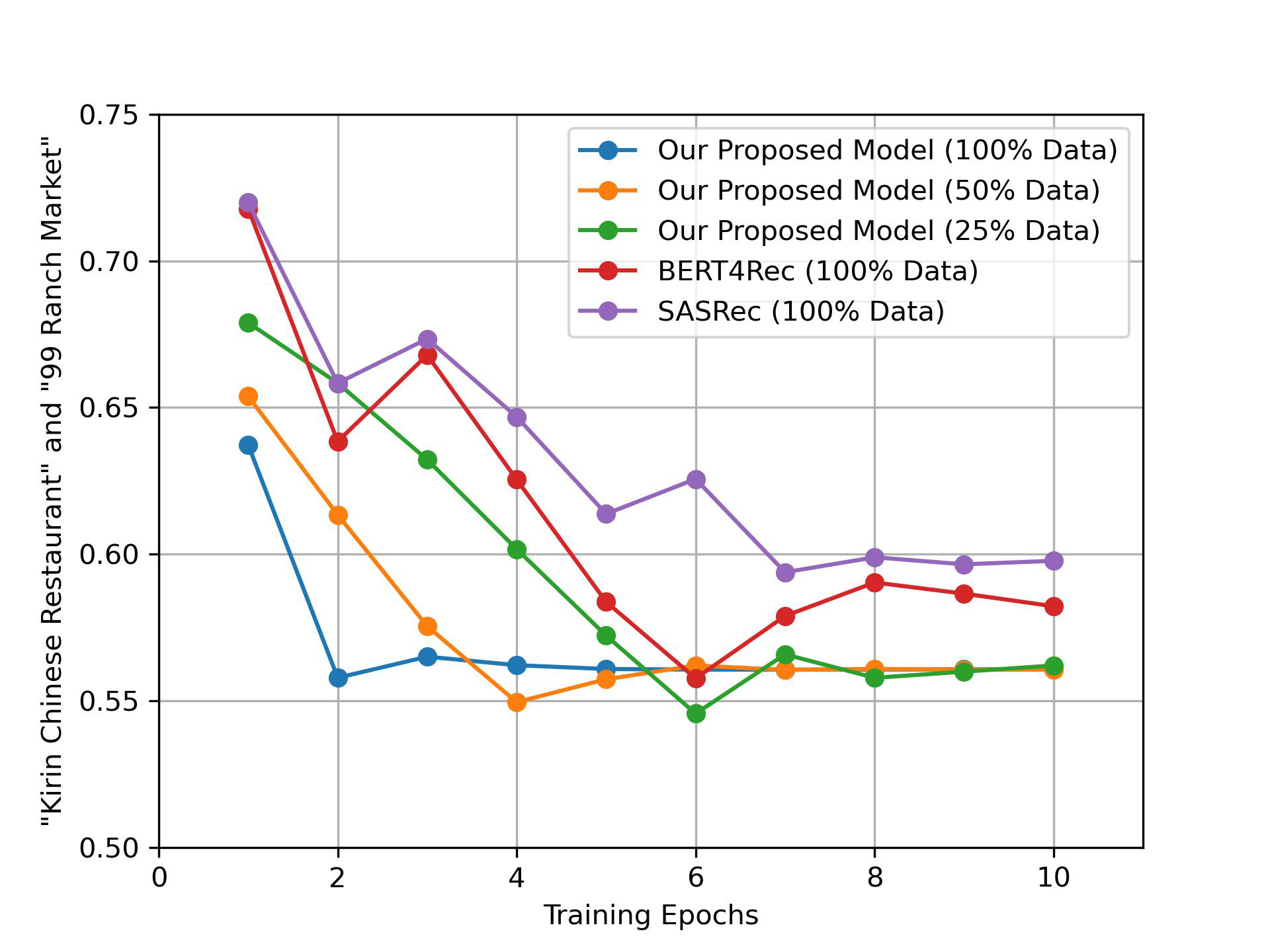}
        \caption{Similarity between related but distinct entities \\ (Kirin Chinese Restaurant and 99 Ranch Market).}
        \label{fig:similarity:2}
    \end{subfigure}
    \begin{subfigure}[b]{0.45\textwidth}
        \centering
        \includegraphics[width=\textwidth]{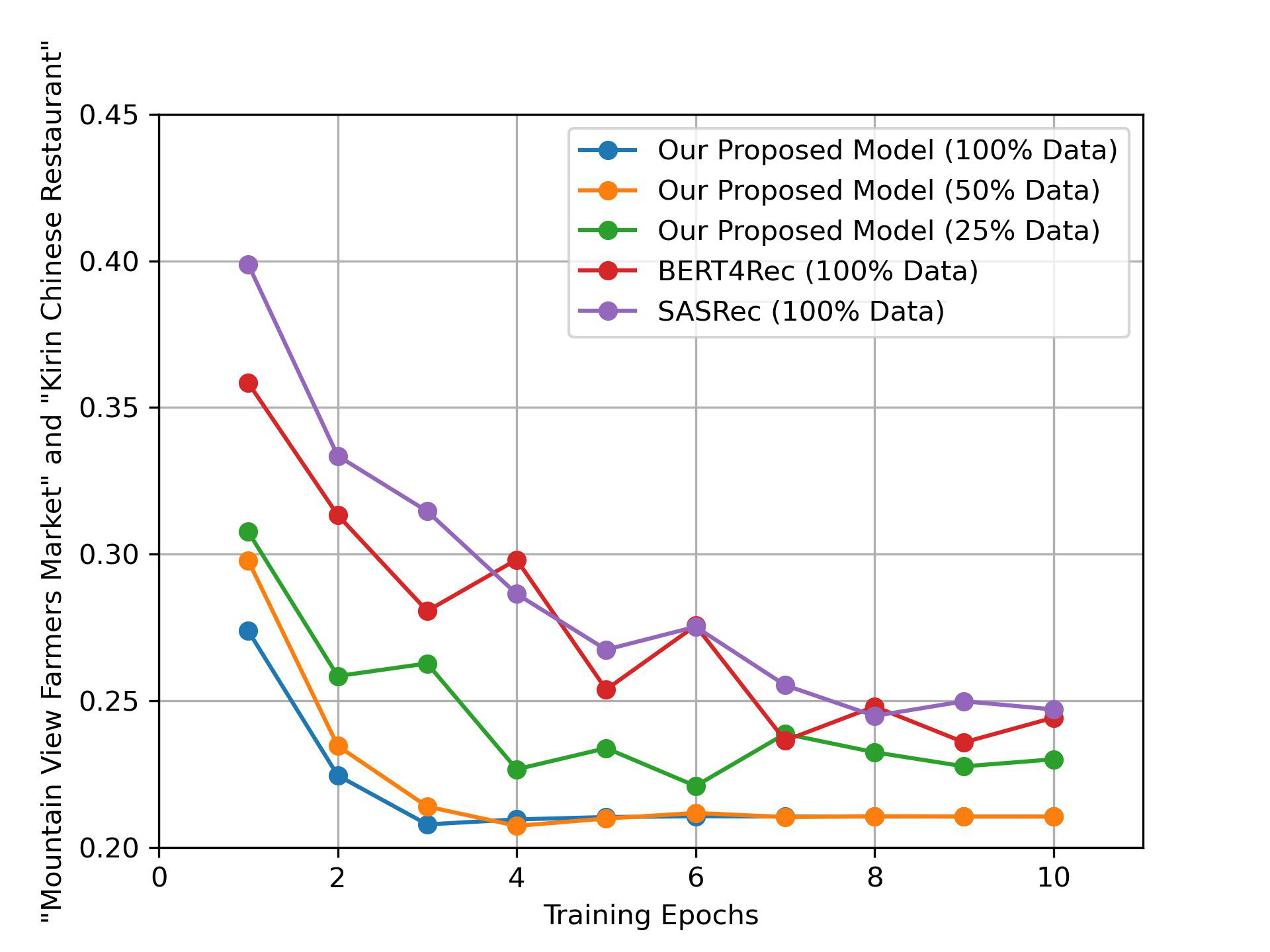}
        \caption{Similarity between unrelated entities \\ (Farmers Market and Kirin Chinese Restaurant).}
        \label{fig:similarity:3}
    \end{subfigure}
   \caption{(Color online) Evolution of embedding similarity during training.
    RecPIE converges faster to semantically meaningful relationships than sequential baselines, rapidly increasing similarity for related items and decreasing similarity for unrelated items.}
  \label{fig:similarity}
\end{figure}

\subsection{Alternative Prompts}
\label{appen:prompts}

In this section, we present the prompts that are used to construct different variants of our proposed RecPIE in Tables~\ref{tab:googlemaps_llm} and~\ref{tab:googlemaps_generation}.  

\noindent (1) Summarization Prompt (Table~\ref{tab:googlemaps_llm})
 \begin{quote}
\emph{
$\mathbf{X}^{(k)} \coloneqq$ ``$\ve_i$ $\oplus$ E(The consumer's consumption history includes \{history\}.
Given the history of the places this consumer has been to, can you provide a summary of the consumer's [positive/negative] preference?)''}
\end{quote}

\noindent (2) Aspect Terms Only (Table~\ref{tab:googlemaps_generation})
\begin{quote}
\emph{
$\mathbf{X}^{(k)} \coloneqq$ ``$\ve_i$ $\oplus$ E(Given the places this consumer has been to \{place\_name\_sequence\}, can you generate a series of aspect terms that represent the most important properties of the candidate place \{place\_name\} that the consumer might consider?)''}
\end{quote}

\noindent (3) Positive Explanation Prompt (Table~\ref{tab:googlemaps_generation})
\begin{quote}
\emph{
$\mathbf{X}^{(k)} \coloneqq$ ``$\ve_i$ $\oplus$ E(Given the places this consumer has been to \{place\_name\_sequence\}, can you provide a reason why this consumer may enjoy the following recommended place \{place\_name\}?)''}
\end{quote}

\noindent (4) General Explanation Prompt (Table~\ref{tab:googlemaps_generation})
\begin{quote}
\emph{
$\mathbf{X}^{(k)} \coloneqq$ ``$\ve_i$ $\oplus$ E(Given the places this consumer has been to \{place\_name\_sequence\}, can you provide a reason why this consumer may or may not enjoy the following recommended place \{place\_name\}?)''}
\end{quote}


\subsection{Role of Contrastive Explanations}
\label{appen:dual_explanation}

\subsubsection{Attention weights on positive and negative explanations.} 
\label{appen:attention_analysis}

RecPIE generates both \emph{positive} and \emph{negative} explanations, corresponding to reasons why a consumer may or may not visit a place. To quantify the relative contributions of these two types of explanations to the recommendation task, we conduct an \emph{attention weight analysis} based on the self-attention layer in the Recommendation Component.

Let $\alpha_{u,i}$ denote the attention weight from element $u$ to element $i$, following the notation in Section~\ref{sec:framework_dnn_loss}. We define the average attention assigned to the positive and negative explanation embeddings as
\begin{equation}
\label{eq:alpha_pos_neg}
\begin{aligned}
\bar{\alpha}_{\mathrm{pos}} &= \frac{1}{|I|} \sum_{i \in I} \alpha_{i,\mathrm{pos}}, \\
\bar{\alpha}_{\mathrm{neg}} &= \frac{1}{|I|} \sum_{i \in I} \alpha_{i,\mathrm{neg}},
\end{aligned}
\end{equation}
where $I = \{\mathrm{pos}, \mathrm{neg}, c, p, \mathrm{seq}\}$ indexes all components in the concatenated input $X_{\text{input}}$ for the Recommendation Component. Intuitively, $\bar{\alpha}_{\mathrm{pos}}$ and $\bar{\alpha}_{\mathrm{neg}}$ measure how much the Recommendation Component attends to the positive and negative explanation embeddings generated by the Explanation Component, respectively, when forming its final prediction, therefore capturing their relative importance in the recommendation decision.

Figure~\ref{fig:pie_googlemap} reports the distribution of attention weights across all input components. The two explanation embeddings together account for approximately 39\% of the total attention weight, highlighting the substantial role that prediction-informed explanations play in driving model predictions, relative to the consumer, place, and sequence embeddings.

To further understand how positive and negative explanations are used, we examine attention distributions separately for positive and negative examples. We define positive examples as observations with 4- or 5-star ratings, and negative examples as those with 1-, 2-, or 3-star ratings. Figure~\ref{fig:positive:google} shows that for positive examples, RecPIE assigns significantly more attention to positive explanations than to negative ones, with the attention distribution for positive explanations shifted to the right. In contrast, Figure~\ref{fig:negative:google} shows that for negative examples, attention shifts toward negative explanations, with positive explanation attention correspondingly reduced.

These patterns provide direct evidence that RecPIE makes effective use of \emph{contrastive explanations}. When a place is predicted to receive a high rating, the model selectively focuses on positive explanations that capture reasons the consumer may like the place. Conversely, when a place is predicted to receive a low rating, the model places greater emphasis on negative explanations that capture reasons the consumer may not like the place. By jointly generating and learning from both types of explanations, RecPIE can adaptively determine which signals are most informative for each prediction. This mechanism explains why contrastive explanations are critical for accurate recommendations and why RecPIE consistently outperforms alternative LLM-based variants that do not incorporate both positive and negative reasoning.


\subsubsection{The need for contrastive explanations.} 
\label{appen:need_both}

We conduct additional analyses to understand how explanations, and in particular the joint use of \emph{positive} and \emph{negative} explanations, contribute to the performance gains of RecPIE. To this end, we compare RecPIE with three alternative variants of the Explanation Component that remove elements of contrastive reasoning:
\begin{itemize}
    \item \textbf{Aspect Terms Only}: The LLM generates a small set of aspect terms describing salient properties of the candidate place, without producing full natural-language explanations or explicit reasoning.
    \item \textbf{Positive Explanations Only}: The LLM generates only positive explanations (i.e., reasons the consumer may like the place), removing negative explanations.
    \item \textbf{General Explanations Only}: The LLM first infers whether the consumer is likely to like the place and then generates a single, non-contrastive explanation, without distinguishing between positive and negative reasoning.
\end{itemize}

The exact prompts (together with the personalized soft prompts) used for these variants are reported in Appendix~\ref{appen:prompts}. Table~\ref{tab:googlemaps_generation} summarizes the results. None of the three variants matches the performance of the full RecPIE model with contrastive explanation generation. In particular, removing negative explanations or collapsing positive and negative reasoning into a single explanation leads to consistent degradation in RMSE, MAE, and AUC.

These results highlight the importance of \emph{contrastive explanations}. By explicitly modeling both reasons to like and reasons to dislike an item, RecPIE provides richer and more discriminative signals to the recommender system, enabling more accurate predictions than explanation variants that rely on only one-sided or non-contrastive reasoning.

\begin{table}[htbp]
\centering
\footnotesize
\setlength\extrarowheight{2pt}
\begin{tabular}{lccc}
\toprule
Method & RMSE $\downarrow$ & MAE $\downarrow$ & AUC $\uparrow$ \\
\midrule
\textbf{RecPIE (Ours)} 
& \textbf{0.2976***} (0.0011) 
& \textbf{0.1976***} (0.0010) 
& \textbf{0.6576***} (0.0021) \\

Aspect Terms Only 
& 0.3044 (0.0011) 
& 0.2014 (0.0010) 
& 0.6468 (0.0021) \\

Positive Explanations Only 
& 0.3011 (0.0011) 
& 0.1996 (0.0010) 
& 0.6520 (0.0021) \\

General Explanations Only 
& 0.3088 (0.0015) 
& 0.2054 (0.0015) 
& 0.6398 (0.0031) \\
\bottomrule
\end{tabular}
\caption{
\textbf{Effect of explanation variants on recommendation performance.}
RecPIE with contrastive positive and negative explanations outperforms all alternative explanation-generation variants. ***p$<$0.01
}
\label{tab:googlemaps_generation}
\end{table}

\subsection{Potential Economic Value}
\label{appen:economic_value}
Evidence from large-scale online experiments at major digital platforms indicates that offline accuracy gains can carry through to business outcomes \citep{zhang2023empowering, chen2024shopping, zhou2018deep, wang2024limaml, gunawardana2022evaluating}. In particular, \citet{li2024variety} report that a 3\% AUC improvement on a major video streaming platform at Alibaba led to a 2.29\% increase in click-through rate and video views, generating approximately \$30 million in additional annual revenue, and Netflix has noted that even a 0.1\% improvement in recommendation performance yields sizable economic value \citep{gomez2015netflix}. Applying that elasticity to a hypothetical platform with 100 million monthly active users and an average revenue per user of \$10 per month, an annual revenue base of roughly \$12 billion, RecPIE's 1.67\% AUC improvement would correspond to a lift on the order of \$150 million per year. Note that this calculation rests on three assumptions, namely that an offline improvement in rating prediction carries through proportionally to downstream engagement as observed in \citet{li2024variety}, that per-interaction values and base rates observed on other platforms transfer to the POI setting, and that the observed improvements persist in the long term as observed in \citet{wang2025recommending}. We also note that our data contain no clicks, recommendation-induced visits, engagement, retention, or advertising revenue, so the calculation is illustrative rather than an empirically validated estimate. Accurately quantifying RecPIE's economic impact would require a long-term field experiment.


\section{Additional Experiments on the Google Maps Dataset}
\label{appen:ablation_robustness}

\subsection{Robustness Checks}
\label{appen:robusness}

We conduct three additional robustness checks on the Google Maps dataset to examine whether RecPIE's performance is sensitive to specific implementation choices, including the explanation embedding model, prompt wording, and recommendation architecture. Table~\ref{tab:robustness_checks} summarizes the results.

\textbf{Alternative explanation embedding model.} We replace the separate pretrained text encoder (DistilBERT) with the intrinsic embeddings produced by the LLM during explanation generation (``LLM intrinsic explanation embeddings''), which are then fed directly into the Recommendation Component (Section~\ref{sec:framework_dnn_cons_embed}). RecPIE's original design achieves lower RMSE and MAE, while using the LLM's intrinsic embeddings yields slightly higher AUC. These results suggest that RecPIE's performance is relatively robust to the choice of explanation embedding model, although the separate encoder offers advantages in prediction error.

\textbf{Alternative prompt wording.} We replace the original explanation-generation prompts with an alternative formulation. For example, the prompt for positive explanations is replaced with:
\begin{quote}
\emph{
$\mathbf{X}^{(k)} \coloneqq$ ``$\ve_i$ $\oplus$ E(Can you explain the reason why this consumer enjoys this place \{place\_name\}, based on the list of previous places she enjoyed \{place\_history\}? Answer with one sentence with the following format: The consumer enjoys this place because ...)''}
\end{quote}
The resulting performance is close to the original RecPIE, with no statistically significant differences across the three metrics. This suggests that RecPIE's gains are not highly sensitive to the specific wording of the explanation-generation prompts.

\textbf{Alternative recommendation architecture.} We replace RecPIE's DNN Recommendation Component with a neural collaborative filtering (NCF) architecture \citep{he2017neural}, which uses learnable consumer and place embeddings and a three-layer multilayer perceptron (MLP) with hidden dimensions [32, 16, 8], without the original attention mechanism. We keep the number of trainable parameters approximately unchanged. The resulting model performs comparably to the original RecPIE and continues to outperform the strongest baselines in Table~\ref{tab:googlemap_results} across all three metrics, suggesting that RecPIE's gains do not depend on a particular recommendation architecture.

Together, these results demonstrate the robustness of RecPIE to alternative choices of explanation encoding, prompt wording, and recommender architecture.

\begin{table}[htbp]
\centering
\footnotesize
\setlength\extrarowheight{2pt}
\setlength\tabcolsep{5pt}
\begin{tabular}{lccc}
\toprule
Method & RMSE $\downarrow$ & MAE $\downarrow$ & AUC $\uparrow$ \\
\midrule
RecPIE (Original)
& \textbf{0.2976} (0.0011)
& \textbf{0.1976} (0.0010)
& 0.6576 (0.0021) \\
\midrule
LLM intrinsic explanation embeddings
& 0.3001 (0.0011)
& 0.2001 (0.0011)
& \textbf{0.6613} (0.0021) \\

Alternative prompt wording
& 0.2991 (0.0011)
& 0.1990 (0.0010)
& 0.6549 (0.0021) \\

Alternative DNN architecture (NCF)
& 0.2999 (0.0011)
& 0.1992 (0.0010)
& 0.6548 (0.0021) \\
\bottomrule
\end{tabular}
\caption{\textbf{Robustness checks on the Google Maps dataset.}}
\label{tab:robustness_checks}
\end{table}

\subsubsection{Ranking Metrics.}
\label{appen:ranking}

The main analysis evaluates rating prediction, which is the objective RecPIE optimizes and a commonly used objective for POI recommendation. In the broader recommender system literature, ranking evaluation metrics that are focused on a list, such as Hit Rate and NDCG, are also commonly adopted. To confirm that the gains carry over to a ranking-oriented evaluation, Table~\ref{tab:ranking_paper} reports standard ranking metrics on Google Maps, where for each held-out interaction we rank the true next place against 100 negative places sampled uniformly at random. RecPIE outperforms all 21 baselines on every ranking metric, with relative gains of 2.31\% to 5.41\% over the best baseline in each metric. In addition, we find that across the 21 baselines the ranking metrics largely agree with AUC in Table~\ref{tab:googlemap_results}, with Pearson correlations between 0.68 and 0.87 and Spearman rank correlations between 0.73 and 0.82, which is expected since both evaluate how well the predicted scores order the candidate places.

\begin{table}[htbp]
\centering
\footnotesize
\setlength\extrarowheight{2pt}
\begin{tabular}{lcccc}
\toprule
Model & HR@5 & NDCG@5 & HR@10 & NDCG@10 \\
\midrule
\textbf{RecPIE (Ours)} & \textbf{0.4423***} (0.0014) & \textbf{0.3137***} (0.0015) & \textbf{0.6012***} (0.0017) & \textbf{0.3820***} (0.0017) \\
\midrule
\multicolumn{5}{l}{\textit{Recommendation-focused baselines}} \\
RecSAVER    & 0.3968 (0.0026) & 0.2998 (0.0024) & 0.5791 (0.0024) & 0.3623 (0.0024) \\
LLMRG       & 0.3982 (0.0029) & 0.2987 (0.0024) & 0.5789 (0.0023) & 0.3611 (0.0024) \\
SASRec      & 0.3810 (0.0013) & 0.2958 (0.0012) & 0.5698 (0.0013) & 0.3577 (0.0013) \\
DIN         & 0.3844 (0.0012) & 0.2981 (0.0012) & 0.5703 (0.0014) & 0.3596 (0.0014) \\
TallRec     & 0.4185 (0.0026) & 0.3001 (0.0023) & 0.5811 (0.0022) & \underline{0.3665} (0.0024) \\
BERT4Rec    & 0.3837 (0.0012) & 0.2970 (0.0012) & 0.5714 (0.0014) & 0.3588 (0.0014) \\
UniSRec     & 0.3638 (0.0012) & 0.2822 (0.0012) & 0.5667 (0.0014) & 0.3523 (0.0014) \\
LLM4POI     & 0.3687 (0.0026) & 0.2879 (0.0024) & 0.5675 (0.0023) & 0.3537 (0.0023) \\
\midrule
\multicolumn{5}{l}{\textit{Explanation-focused baselines}} \\
ExpCTR      & \underline{0.4196} (0.0029) & \underline{0.3013} (0.0026) & \underline{0.5876} (0.0026) & 0.3661 (0.0024) \\
LLM2ER-EQR  & 0.4123 (0.0028) & 0.3009 (0.0026) & 0.5853 (0.0024) & 0.3655 (0.0024) \\
RXR         & 0.4017 (0.0031) & 0.3001 (0.0024) & 0.5830 (0.0024) & 0.3655 (0.0024) \\
A3NCF       & 0.3739 (0.0047) & 0.2734 (0.0038) & 0.5622 (0.0041) & 0.3476 (0.0038) \\
SULM        & 0.3748 (0.0045) & 0.2738 (0.0037) & 0.5625 (0.0039) & 0.3478 (0.0038) \\
AARM        & 0.3755 (0.0047) & 0.2751 (0.0037) & 0.5622 (0.0039) & 0.3473 (0.0039) \\
MMALFM      & 0.3732 (0.0047) & 0.2750 (0.0038) & 0.5638 (0.0039) & 0.3481 (0.0038) \\
ANR         & 0.3750 (0.0045) & 0.2744 (0.0039) & 0.5635 (0.0039) & 0.3481 (0.0038) \\
MTER        & 0.3788 (0.0045) & 0.2748 (0.0038) & 0.5622 (0.0039) & 0.3473 (0.0038) \\
AMCF        & 0.3791 (0.0045) & 0.2744 (0.0038) & 0.5632 (0.0041) & 0.3473 (0.0038) \\
PETER       & 0.3964 (0.0024) & 0.2893 (0.0029) & 0.5795 (0.0021) & 0.3568 (0.0021) \\
UCEPic      & 0.3928 (0.0023) & 0.2882 (0.0026) & 0.5783 (0.0019) & 0.3554 (0.0021) \\
PARSRec     & 0.3911 (0.0024) & 0.2895 (0.0024) & 0.5791 (0.0019) & 0.3551 (0.0019) \\
\bottomrule
\end{tabular}
\caption{Ranking metrics on Google Maps (true item vs.\ 100 uniformly sampled negatives). The best baseline in each metric is underlined. Standard deviations in parentheses; $^{***}$ denotes significance at $p<0.01$.}
\label{tab:ranking_paper}
\end{table}

\subsection{Value of Prediction-Informed Explanations}
\label{appen:explanation_value}

A natural question is whether RecPIE's gains arise from its \emph{prediction-informed explanations}, rather than from simply incorporating additional LLM-generated text or model parameters. Table~\ref{tab:explanation_value_paper} compares RecPIE against four alternatives. First, \emph{Llama~3.1 direct recommendation} replaces the DNN recommender with direct LLM rating prediction, testing whether the LLM alone can achieve comparable accuracy. Second, \emph{Base-LLM explanations} removes PPO fine-tuning, testing the value of our usefulness reward design (Section~\ref{sec:framework_ppo}), which optimizes explanations using downstream predictive feedback. Third, \emph{Free learnable parameters} replaces explanation embeddings with trainable parameters of the same dimension, testing whether additional model capacity alone accounts for the gains. Finally, \emph{consumption history summarization} (which is also presented in Section~\ref{res_understanding_reasoning}) replaces personalized explanations with LLM-generated summaries of consumption histories, testing whether generic semantic information provides similar benefits.

RecPIE outperforms all four alternatives across all three metrics. In particular, its advantage over base-LLM explanations and history summarization highlights the value of our usefulness reward, i.e., learning explanations specifically for downstream prediction, rather than relying on generic LLM-generated content. Together, these findings support the importance of \emph{prediction-informed explanations} in improving recommendation accuracy.

We also would like to note that these ablations explain the framework's performance rather than separately identifying the source of the 61.5\% human preference rate observed in Section~\ref{sec:human_evaluation}.

\begin{table}[htbp]
\centering
\footnotesize
\setlength\extrarowheight{2pt}
\begin{tabular}{l ccc}
\toprule
Variant & RMSE $\downarrow$ & MAE $\downarrow$ & AUC $\uparrow$ \\
\midrule
Llama 3.1 direct recommendation (no DNN)            & 0.3477 (0.0041) & 0.2295 (0.0047) & 0.6103 (0.0069) \\
Base-LLM explanations (no PPO fine-tuning)          & 0.3098 (0.0013) & 0.2055 (0.0011) & 0.6410 (0.0023) \\
Free learnable parameters (same capacity, no LLM text)  & 0.3075 (0.0011) & 0.2038 (0.0010) & 0.6396 (0.0021) \\
Consumption history summarization (semantic text, no expl.) & 0.3008 (0.0011) & 0.1994 (0.0011) & 0.6542 (0.0023) \\
\midrule
\textbf{RecPIE (prediction-informed explanation)}   & \textbf{0.2976***} (0.0011) & \textbf{0.1976***} (0.0010) & \textbf{0.6576***} (0.0021) \\
\bottomrule
\end{tabular}
\caption{Value of prediction-informed explanations. ***p$<$0.01}
\label{tab:explanation_value_paper}
\end{table}

\subsubsection{Value of the usefulness reward.}
\label{appen:usefulness_reward}
The experiments above validate RecPIE's alternating training architecture to generate prediction-informed explanations. 

We next investigate whether its \emph{usefulness} reward, which guides explanation generation using downstream recommendation accuracy, provides advantages over alternative explanation-quality objectives while holding the alternating training architecture fixed. Specifically, following \citet{yu2025expctr}, we consider an \emph{LC-alignment} reward, which uses an LLM-generated binary judgment of whether an explanation is consistent with the consumer's observed behavior. We evaluate two variants: \emph{``Usefulness + LC-alignment''} adds the LC-alignment reward \emph{on top of} RecPIE's usefulness reward, while \emph{``LC-alignment only''} replaces RecPIE's usefulness reward completely with LC-alignment alone.

\begin{table}[htbp]
\centering
\footnotesize
\setlength\extrarowheight{2pt}
\setlength\tabcolsep{6pt}
\begin{tabular}{lccc}
\toprule
Metric
& \shortstack{RecPIE\\(Usefulness)}
& \shortstack{Usefulness +\\LC-alignment}
& \shortstack{LC-alignment\\Only} \\
\midrule
\multicolumn{4}{l}{\textit{Recommendation Accuracy}} \\
\midrule
RMSE $\downarrow$
& \textbf{0.2976} (0.0011)
& 0.3008 (0.0013)
& 0.3039 (0.0014) \\

MAE $\downarrow$
& \textbf{0.1976} (0.0010)
& 0.2005 (0.0013)
& 0.2025 (0.0013) \\

AUC $\uparrow$
& \textbf{0.6576} (0.0021)
& 0.6512 (0.0024)
& 0.6423 (0.0024) \\

\midrule
\multicolumn{4}{l}{\textit{Explanation Quality}} \\
\midrule
BLEURT $\uparrow$
& 0.3998 (0.0003)
& \textbf{0.4001} (0.0003)
& 0.3968 (0.0003) \\

Coverage $\uparrow$
& 0.3138 (0.0011)
& \textbf{0.3150} (0.0011)
& 0.3092 (0.0011) \\

Informativeness $\uparrow$
& \textbf{0.4997} (0.0011)
& 0.4969 (0.0012)
& 0.4937 (0.0012) \\

Fluency $\uparrow$
& \textbf{0.3836} (0.0009)
& 0.3810 (0.0010)
& 0.3795 (0.0010) \\
\bottomrule
\end{tabular}
\caption{\textbf{Value of the usefulness reward.}}
\label{tab:usefulness_reward}
\end{table}
Table~\ref{tab:usefulness_reward} reports the results. In terms of recommendation accuracy, adding the LC-alignment reward to RecPIE's usefulness reward slightly reduces recommendation accuracy across all three metrics, while replacing the usefulness reward with LC-alignment alone leads to further degradation in all three metrics. In terms of explanation quality, adding LC-alignment yields no consistent improvement (not statistically significant), while replacing the usefulness reward with LC-alignment reduces all four quality metrics. Together, these findings support our proposed \emph{usefulness reward}, which optimizes explanations for their \emph{usefulness} in improving prediction, rather than for an independently measured explanation-quality objective as in the existing XAI literature.

To further examine the value of the usefulness reward beyond RecPIE, we add RecPIE's usefulness reward to ExpCTR's original rewards. As shown in Table~\ref{tab:expctr_usefulness}, we find that the usefulness reward would be able to improve ExpCTR on all three recommendation accuracy metrics and all four explanation quality metrics, although ExpCTR trained with the usefulness reward still underperforms RecPIE.

\begin{table}[htbp]
\centering
\footnotesize
\setlength\extrarowheight{2pt}
\begin{tabular}{l cccc}
\toprule
\multicolumn{5}{l}{\textit{(a) Recommendation accuracy}} \\
Model & RMSE $\downarrow$ & MAE $\downarrow$ & AUC $\uparrow$ & \\
\midrule
\textbf{RecPIE (Ours)} & \textbf{0.2976} (0.0011) & \textbf{0.1976} (0.0010) & \textbf{0.6576} (0.0021) & \\
ExpCTR & 0.3077 (0.0015) & 0.2049 (0.0011) & 0.6408 (0.0026) & \\
ExpCTR + usefulness reward & 0.3021 (0.0011) & 0.2014 (0.0010) & 0.6463 (0.0021) & \\
\midrule
\multicolumn{5}{l}{\textit{(b) Explanation quality}} \\
Model & BLEURT $\uparrow$ & Coverage $\uparrow$ & Informativeness $\uparrow$ & Fluency $\uparrow$ \\
\midrule
\textbf{RecPIE (Ours)} & \textbf{0.3998} (0.0003) & \textbf{0.3138} (0.0011) & \textbf{0.4997} (0.0011) & \textbf{0.3836} (0.0009) \\
ExpCTR & 0.3826 (0.0039) & 0.3086 (0.0013) & 0.4701 (0.0017) & 0.3511 (0.0023) \\
ExpCTR + usefulness reward & 0.3914 (0.0037) & 0.3101 (0.0013) & 0.4802 (0.0017) & 0.3669 (0.0021) \\
\bottomrule
\end{tabular}
\caption{\textbf{Effect of training ExpCTR's model with RecPIE's usefulness reward.}}
\label{tab:expctr_usefulness}
\end{table}

\subsubsection{Value of the personalized learnable soft prompt.}
\label{appen:value_soft_prompt}


To examine the value of the personalized learnable soft prompt, we remove the soft prompt from the input of the Explanation Component while keeping the rest of RecPIE unchanged. As shown in Table~\ref{tab:soft_prompt}, removing the soft prompt increases RMSE from 0.2976 to 0.3041 and MAE from 0.1976 to 0.2033, and reduces AUC from 0.6576 to 0.6448, and all three differences are statistically significant. We find that the shared soft prompt would be able to improve recommendation accuracy significantly, since it allows the explanation generator to use the \emph{collaborative filtering information} (i.e., information about the preferences and behaviors of other consumers in addition to the focal consumer) that the Recommendation Component learns across consumers and to align the generated explanations with its predictions.

\begin{table}[htbp]
\centering
\footnotesize
\setlength\extrarowheight{2pt}
\begin{tabular}{l ccc}
\toprule
Variant & RMSE $\downarrow$ & MAE $\downarrow$ & AUC $\uparrow$ \\
\midrule
RecPIE without soft prompt & 0.3041 (0.0009) & 0.2033 (0.0009) & 0.6448 (0.0020) \\
\midrule
\textbf{RecPIE (shared soft prompt)} & \textbf{0.2976***} (0.0011) & \textbf{0.1976***} (0.0010) & \textbf{0.6576***} (0.0021) \\
\bottomrule
\end{tabular}
\caption{Value of the personalized learnable soft prompt. ***p$<$0.01}
\label{tab:soft_prompt}
\end{table}

\subsection{Robustness to Geographic Region}
\label{appen:wyoming}

Our main results focus on Mountain View, California, a technology-oriented region whose places may be well represented in LLM pretraining data. To examine whether RecPIE's gains depend on such geographic familiarity, we replicate the experiments in Wyoming, a geographically different and less familiar region. Table~\ref{tab:region_paper} reports RecPIE's relative improvements over the best-performing baseline in each region. RecPIE achieves consistent gains in both regions, with even larger improvements in Wyoming across all three recommendation accuracy metrics and all four explanation quality metrics. These findings demonstrate the geographic robustness of RecPIE, suggesting that its benefits do not depend on LLMs' familiarity with the Mountain View region.

To understand why the gains are larger in Wyoming, we computed some further summary statistics. We found that the Mountain View subset has 133,599 unique places and the Wyoming subset has 12,016 unique places, and consumers average roughly 160 ratings in Mountain View versus 24 ratings in Wyoming. So we think it is because (1) Wyoming has fewer unique places so it could be slightly easier to reason through, (2) explanations should help most where per-consumer data are scarce (related to the uncertainty results in Section~\ref{sec:harder_examples}), so RecPIE's gains can be larger for Wyoming.

\begin{table}[htbp]
\centering
\footnotesize
\setlength\extrarowheight{2pt}
\setlength\tabcolsep{8pt}
\begin{tabular}{lcc}
\toprule
Metric & Mountain View & Wyoming \\
\midrule
\multicolumn{3}{l}{\textit{Recommendation Accuracy}} \\
\midrule
RMSE & 3.28\% & \textbf{4.95\%} \\
MAE & 3.56\% & \textbf{4.43\%} \\
AUC & 1.67\% & \textbf{3.38\%} \\
\midrule
\multicolumn{3}{l}{\textit{Explanation Quality}} \\
\midrule
BLEURT & 0.45\% & \textbf{0.55\%} \\
Coverage & 1.26\% & \textbf{1.40\%} \\
Informativeness & 1.36\% & \textbf{1.50\%} \\
Fluency & 0.47\% & \textbf{0.55\%} \\
\bottomrule
\end{tabular}
\caption{\textbf{Robustness of RecPIE's improvements across geographic regions.}}
\label{tab:region_paper}
\end{table}


\section{Generalizability to Unseen Users, Places, and Regions}
\label{appen:coldstart}

Building on the heterogeneity and data-efficiency results in Sections~\ref{sec:harder_examples} and~\ref{sec:learning_effiency}, we further examine RecPIE's ability to generalize beyond the consumers, places, and regions observed during training, representing the long-standing \textbf{cold-start} challenges in recommender systems \citep{lika2014facing}. Specifically, we compare RecPIE against five strong baselines (ExpCTR, LLM2ER-EQR, TallRec, PETER, and BERT4Rec) across four settings: the \emph{warm-start} setting, \emph{unseen consumers}, \emph{unseen places}, and \emph{cross-region transfer} (training on Mountain View and evaluating on Wyoming without retraining). The results are reported in Table~\ref{tab:coldstart_paper}.

We clarify some implementation details. To construct the unseen consumers and places, we randomly select 5\% of the consumers and 5\% of the places in the Google Maps dataset, and we remove from the training set all ratings by these consumers and all ratings of these places. Whenever one of these places appears in the visitation history of a remaining rating, we replace it with a reserved out-of-vocabulary (OOV) token, so that no embedding of a held-out consumer or place is learned during training, while the names of the visited places remain available to the LLMs. The remaining ratings are split into training and test sets at the user-temporal level as in Section~\ref{sec:context}, and all methods are trained on the same training set. We then evaluate each method separately on the warm-start test set, on the ratings of the unseen consumers, and on the ratings of the unseen places. During training, we map consumers and places with fewer than five ratings in the training set to the same reserved OOV embedding, so that this embedding is learned from consumers and places with sparse histories. For these unseen cases, the consumer- or place-specific embeddings fall back to the reserved OOV embedding, while RecPIE can still generate explanations conditioned on the consumer's observed visitation history and the candidate place's name.

As shown in Table~\ref{tab:coldstart_paper}, although performance degrades for all methods in these more challenging prediction scenarios, RecPIE consistently outperforms every baseline across all three metrics. Moreover, its relative advantage is larger than in the warm-start setting: compared with the best baseline for each metric, RecPIE's RMSE reduction increases from \textbf{3.07\%} in the warm-start setting to \textbf{6.34\%} for unseen consumers, \textbf{6.33\%} for unseen places, and \textbf{5.02\%} for cross-region transfer. Its relative AUC improvements similarly increase from \textbf{1.66\%} to \textbf{3.1--4.1\%}. These findings suggest that RecPIE is particularly valuable when conventional recommender systems must generalize beyond observed consumer--place interactions, consistent with the potential benefits of prediction-informed explanations in data-sparse and out-of-distribution (OOD) settings.

\begin{table}[htbp]
\centering
\footnotesize
\setlength\extrarowheight{2pt}
\begin{tabular}{l ccc}
\toprule
Method & RMSE $\downarrow$ & MAE $\downarrow$ & AUC $\uparrow$ \\
\midrule
\multicolumn{4}{l}{\textit{Warm start (seen consumers and places)}} \\
\textbf{RecPIE (Ours)} & \textbf{0.2998} (0.0011) & \textbf{0.1989} (0.0010) & \textbf{0.6556} (0.0022) \\
ExpCTR   & 0.3093 (0.0016) & 0.2062 (0.0011) & 0.6399 (0.0028) \\
LLM2ER-EQR & 0.3127 (0.0013) & 0.2073 (0.0011) & 0.6404 (0.0025) \\
TallRec  & 0.3130 (0.0015) & 0.2091 (0.0012) & 0.6449 (0.0028) \\
PETER    & 0.3149 (0.0021) & 0.2079 (0.0013) & 0.6377 (0.0029) \\
BERT4Rec & 0.3136 (0.0011) & 0.2068 (0.0010) & 0.6391 (0.0019) \\
\midrule
\multicolumn{4}{l}{\textit{Unseen consumers}} \\
\textbf{RecPIE (Ours)} & \textbf{0.3148} (0.0023) & \textbf{0.2103} (0.0023) & \textbf{0.6412} (0.0030) \\
ExpCTR   & 0.3402 (0.0027) & 0.2269 (0.0021) & 0.6170 (0.0031) \\
LLM2ER-EQR & 0.3368 (0.0023) & 0.2301 (0.0017) & 0.6141 (0.0029) \\
TallRec  & 0.3361 (0.0025) & 0.2243 (0.0023) & 0.6220 (0.0029) \\
PETER    & 0.3442 (0.0029) & 0.2340 (0.0025) & 0.6055 (0.0037) \\
BERT4Rec & 0.3464 (0.0013) & 0.2303 (0.0011) & 0.6138 (0.0014) \\
\midrule
\multicolumn{4}{l}{\textit{Unseen places}} \\
\textbf{RecPIE (Ours)} & \textbf{0.3301} (0.0027) & \textbf{0.2262} (0.0025) & \textbf{0.6236} (0.0029) \\
ExpCTR   & 0.3544 (0.0031) & 0.2423 (0.0023) & 0.5942 (0.0031) \\
LLM2ER-EQR & 0.3556 (0.0031) & 0.2431 (0.0023) & 0.5966 (0.0029) \\
TallRec  & 0.3524 (0.0029) & 0.2411 (0.0023) & 0.5990 (0.0031) \\
PETER    & 0.3558 (0.0035) & 0.2452 (0.0027) & 0.5877 (0.0033) \\
BERT4Rec & 0.3588 (0.0016) & 0.2489 (0.0013) & 0.5928 (0.0018) \\
\midrule
\multicolumn{4}{l}{\textit{Cross-region (train on Mountain View, evaluate on Wyoming)}} \\
\textbf{RecPIE (Ours)} & \textbf{0.3594} (0.0038) & \textbf{0.2433} (0.0034) & \textbf{0.5968} (0.0044) \\
ExpCTR   & 0.3809 (0.0035) & 0.2565 (0.0037) & 0.5739 (0.0044) \\
LLM2ER-EQR & 0.3825 (0.0035) & 0.2576 (0.0036) & 0.5711 (0.0044) \\
TallRec  & 0.3784 (0.0039) & 0.2543 (0.0038) & 0.5773 (0.0044) \\
PETER    & 0.3844 (0.0043) & 0.2610 (0.0041) & 0.5683 (0.0044) \\
BERT4Rec & 0.3860 (0.0026) & 0.2589 (0.0023) & 0.5711 (0.0031) \\
\bottomrule
\end{tabular}
\caption{Generalization to unseen consumers, places, and regions.}
\label{tab:coldstart_paper}
\end{table}


\section{Generalizability to Additional Benchmark Datasets}
\label{appen:generalizability_three_datasets}
To further demonstrate the generalizability of RecPIE, apart from the Google Maps dataset that we use in the main paper, we also implement our RecPIE model, as well as other baselines, on the following three additional benchmark recommendation datasets:

\begin{itemize}
\item \textbf{Amazon Movie}: This dataset captures consumer purchasing behavior in the Movies \& TV category on Amazon \citep{ni2019justifying}. Collected in 2023, it contains 17,328,314 records from 6,503,429 consumers and 747,910 unique movies. Each record includes the consumer ID, movie ID, movie title, consumer rating (on a 1--5 scale), purchasing timestamp, consumer's past purchasing history (as a sequence of movie IDs), and product review provided by the consumer.
\item \textbf{Yelp Restaurant}: This dataset documents consumers' restaurant check-ins on the Yelp platform. It spans 11 metropolitan areas in the United States and comprises 6,990,280 check-in records from 1,987,897 consumers across 150,346 restaurants. Each record includes the consumer ID, restaurant ID, check-in timestamp, consumer rating (on a 1--5 scale), consumer's historical visits (as a sequence of restaurant IDs), and consumer review for the restaurant.
\item \textbf{TripAdvisor Hotel}: This dataset captures consumers' hotel stays on the TripAdvisor platform \citep{li2023personalized}. Collected in 2019, it contains 343,277 hotel stay records from 9,765 consumers and 6,280 unique hotels. Each record includes the consumer ID, hotel ID, check-in timestamp, consumer rating, the list of hotels previously visited by the consumer, and the review provided by the consumer for the hotel. 
\end{itemize}

We repeat the analyses of Section~\ref{sec:results} on each dataset, and our RecPIE model again obtains consistent and significant performance improvements over state-of-the-art baseline models across all three. We detail our experiment results in the following subsections.

\subsection{Recommendation Accuracy on the Three Additional Datasets}
Table~\ref{main_results} presents the main results for recommendation accuracy, where we can observe that our RecPIE model consistently outperforms the baseline models across all three evaluation metrics and three datasets. The table reports the same set of baselines as Table~\ref{tab:googlemap_results} of the main paper, drawn from the categories described in Appendix~\ref{appen:baselines}. Note that these three additional datasets no longer deal with the POI recommendation task, so the POI recommendation baseline LLM4POI is at a disadvantage in this setting. Specifically, RecPIE achieves improvements of 4--10\% in RMSE, 5--11\% in MAE, and 2--3\% in AUC compared to the best-performing baseline models. These results, along with the results that we have already presented in Table~\ref{tab:googlemap_results} of the main paper, further demonstrate the efficacy of our RecPIE model and the value of prediction-informed explanations in improving the recommendation performance.

It is worth noting that these gains are larger than those on Google Maps in Mountain View. We conjecture that this difference arises from the \emph{cross-category} nature of the Google Maps data, which contains places such as garages, restaurants, museums, and grocery stores and therefore requires the recommender system to perform \emph{cross-category generalization}. In contrast, each of the three additional datasets covers a \emph{single} category, namely hotels in TripAdvisor, restaurants in Yelp, and movies in Amazon Movie, where consumption histories can repeatedly reveal the same type of preference, such as preferences for quiet rooms, service, or amenities in hotels, for cuisines in restaurants, and for genres in movies. The more heterogeneous places on Google Maps would make useful explanations \emph{harder} to learn, which may explain the smaller gains there.


\begin{table}
\centering
\footnotesize
\setlength{\tabcolsep}{4pt}
\renewcommand{\arraystretch}{1.10}

\resizebox{0.9\textwidth}{!}{%
\begin{tabular}{lccc ccc ccc}
\toprule
& \multicolumn{3}{c}{\textbf{TripAdvisor}} & \multicolumn{3}{c}{\textbf{Yelp}} & \multicolumn{3}{c}{\textbf{Amazon Movie}} \\
\cmidrule(lr){2-4}\cmidrule(lr){5-7}\cmidrule(lr){8-10}
\textbf{Model}
& RMSE$\downarrow$ & MAE$\downarrow$ & AUC$\uparrow$
& RMSE$\downarrow$ & MAE$\downarrow$ & AUC$\uparrow$
& RMSE$\downarrow$ & MAE$\downarrow$ & AUC$\uparrow$ \\
\midrule

\textbf{RecPIE (Ours)}
& \textbf{0.1733} & \textbf{0.1333} & \textbf{0.7376}
& \textbf{0.2020} & \textbf{0.1620} & \textbf{0.7350}
& \textbf{0.1653} & \textbf{0.1165} & \textbf{0.7582} \\
& (0.0010) & (0.0008) & (0.0018)
& (0.0010) & (0.0009) & (0.0017)
& (0.0010) & (0.0009) & (0.0018) \\
\textit{\% Improved}
& \textit{+9.93\%} & \textit{+11.01\%} & \textit{+2.79\%}
& \textit{+9.74\%} & \textit{+6.47\%} & \textit{+3.01\%}
& \textit{+3.90\%} & \textit{+4.66\%} & \textit{+2.45\%} \\
\addlinespace[3pt]
\midrule

RecSAVER
& 0.1949 & 0.1512 & 0.7038
& 0.2279 & 0.1764 & 0.7088
& 0.1755 & 0.1242 & 0.7359 \\
& (0.0017) & (0.0013) & (0.0027)
& (0.0017) & (0.0014) & (0.0027)
& (0.0016) & (0.0013) & (0.0024) \\
\addlinespace[1pt]
LLMRG
& 0.1956 & 0.1514 & 0.7026
& 0.2268 & 0.1760 & 0.7091
& 0.1758 & 0.1246 & 0.7353 \\
& (0.0017) & (0.0013) & (0.0027)
& (0.0017) & (0.0014) & (0.0027)
& (0.0016) & (0.0013) & (0.0024) \\
\addlinespace[1pt]

SASRec
& 0.2089 & 0.1731 & 0.7005
& 0.2491 & 0.2135 & 0.6897
& 0.2176 & 0.1869 & 0.7025 \\
& (0.0007) & (0.0006) & (0.0015)
& (0.0011) & (0.0009) & (0.0016)
& (0.0013) & (0.0008) & (0.0013) \\
\addlinespace[1pt]
DIN
& 0.2022 & 0.1709 & 0.7076
& 0.2479 & 0.2116 & 0.6917
& 0.2155 & 0.1853 & 0.7046 \\
& (0.0009) & (0.0007) & (0.0017)
& (0.0009) & (0.0008) & (0.0015)
& (0.0009) & (0.0007) & (0.0013) \\
\addlinespace[1pt]
TallRec
& \underline{0.1924} & \underline{0.1498} & \underline{0.7176}
& \underline{0.2238} & \underline{0.1732} & \underline{0.7135}
& \underline{0.1720} & \underline{0.1222} & \underline{0.7401} \\
& (0.0017) & (0.0013) & (0.0027)
& (0.0017) & (0.0014) & (0.0027)
& (0.0016) & (0.0013) & (0.0024) \\
\addlinespace[3pt]
\midrule
BERT4Rec
& 0.2003 & 0.1701 & 0.7085
& 0.2460 & 0.2101 & 0.6928
& 0.2126 & 0.1832 & 0.7088 \\
& (0.0009) & (0.0006) & (0.0017)
& (0.0009) & (0.0008) & (0.0015)
& (0.0009) & (0.0008) & (0.0015) \\
\addlinespace[1pt]
UniSRec
& 0.2026 & 0.1720 & 0.7066
& 0.2448 & 0.2093 & 0.6956
& 0.2103 & 0.1810 & 0.7133 \\
& (0.0015) & (0.0010) & (0.0023)
& (0.0013) & (0.0011) & (0.0020)
& (0.0011) & (0.0009) & (0.0017) \\
\addlinespace[3pt]
LLM4POI
& 0.2011 & 0.1703 & 0.7091 
& 0.2411 & 0.2077 & 0.6988 
& 0.2073 & 0.1791 & 0.7165 \\
& (0.0015) & (0.0011) & (0.0023) 
& (0.0013) & (0.0011) & (0.0021) 
& (0.0011) & (0.0009) & (0.0017) \\
\addlinespace[1pt]
\midrule
ExpCTR
& 0.1933 & 0.1501 & 0.7170 
& 0.2244 & 0.1735 & 0.7126 
& 0.1725 & 0.1228 & 0.7395 \\
& (0.0017) & (0.0013) & (0.0027)
& (0.0017) & (0.0014) & (0.0027)
& (0.0016) & (0.0013) & (0.0024) \\
\addlinespace[1pt]
LLM2ER-EQR
& 0.1977 & 0.1526 & 0.7139 
& 0.2269 & 0.1746 & 0.7101 
& 0.1739 & 0.1241 & 0.7356 \\
& (0.0019) & (0.0017) & (0.0029) 
& (0.0019) & (0.0016) & (0.0029) 
& (0.0016) & (0.0014) & (0.0027) \\
\addlinespace[1pt]
RXR
& 0.2038 & 0.1778 & 0.6936 
& 0.2556 & 0.2134 & 0.6826 
& 0.2212 & 0.1877 & 0.7035 \\
& (0.0019) & (0.0013) & (0.0027)
& (0.0023) & (0.0016) & (0.0029)
& (0.0029) & (0.0018) & (0.0032) \\
\addlinespace[1pt]
A3NCF
& 0.2103 & 0.1811 & 0.6879
& 0.2607 & 0.2181 & 0.6785
& 0.2241 & 0.1903 & 0.6971 \\
& (0.0019) & (0.0013) & (0.0027)
& (0.0023) & (0.0016) & (0.0029)
& (0.0029) & (0.0018) & (0.0032) \\
\addlinespace[1pt]
SULM
& 0.2191 & 0.1872 & 0.6736
& 0.2823 & 0.2258 & 0.6614
& 0.2477 & 0.1980 & 0.6855 \\
& (0.0021) & (0.0013) & (0.0027)
& (0.0019) & (0.0015) & (0.0029)
& (0.0027) & (0.0019) & (0.0027) \\
\addlinespace[1pt]
AARM
& 0.2083 & 0.1803 & 0.6901
& 0.2582 & 0.2162 & 0.6801
& 0.2162 & 0.1845 & 0.7032 \\
& (0.0019) & (0.0014) & (0.0030)
& (0.0021) & (0.0015) & (0.0029)
& (0.0027) & (0.0018) & (0.0029) \\
\addlinespace[1pt]
MMALFM
& 0.2117 & 0.1820 & 0.6894
& 0.2591 & 0.2167 & 0.6801
& 0.2301 & 0.1931 & 0.6931 \\
& (0.0019) & (0.0014) & (0.0029)
& (0.0020) & (0.0016) & (0.0030)
& (0.0028) & (0.0020) & (0.0036) \\
\addlinespace[1pt]
ANR
& 0.2083 & 0.1804 & 0.6905
& 0.2575 & 0.2145 & 0.6817
& 0.2275 & 0.1915 & 0.6960 \\
& (0.0017) & (0.0014) & (0.0027)
& (0.0021) & (0.0017) & (0.0031)
& (0.0026) & (0.0018) & (0.0026) \\
\addlinespace[1pt]
MTER
& 0.2099 & 0.1825 & 0.6889
& 0.2614 & 0.2169 & 0.6809
& 0.2283 & 0.1906 & 0.6967 \\
& (0.0019) & (0.0014) & (0.0029)
& (0.0021) & (0.0016) & (0.0031)
& (0.0026) & (0.0017) & (0.0026) \\
\addlinespace[1pt]
AMCF
& 0.2088 & 0.1755 & 0.6989
& 0.2501 & 0.2123 & 0.6928
& 0.2376 & 0.1863 & 0.7035 \\
& (0.0019) & (0.0013) & (0.0027)
& (0.0016) & (0.0013) & (0.0023)
& (0.0013) & (0.0010) & (0.0019) \\
\addlinespace[1pt]
PETER
& 0.1996 & 0.1715 & 0.7078
& 0.2423 & 0.2071 & 0.7003
& 0.2099 & 0.1770 & 0.7226 \\
& (0.0019) & (0.0013) & (0.0027)
& (0.0015) & (0.0013) & (0.0022)
& (0.0013) & (0.0010) & (0.0019) \\
\addlinespace[1pt]
UCEPic
& 0.2035 & 0.1723 & 0.7066
& 0.2477 & 0.2099 & 0.7018
& 0.2228 & 0.1801 & 0.7080 \\
& (0.0015) & (0.0011) & (0.0023)
& (0.0015) & (0.0012) & (0.0023)
& (0.0011) & (0.0009) & (0.0017) \\
\addlinespace[1pt]
PARSRec
& 0.2008 & 0.1703 & 0.7080
& 0.2471 & 0.2106 & 0.6923
& 0.2133 & 0.1837 & 0.7069 \\
& (0.0009) & (0.0007) & (0.0017)
& (0.0009) & (0.0008) & (0.0015)
& (0.0009) & (0.0007) & (0.0013) \\
\addlinespace[1pt]
\bottomrule
\end{tabular}%
}

\caption{Recommendation performance on three datasets. ``\% Improved'' reports the gains of RecPIE (Ours) over the best-performing baseline (underlined). Parentheses report standard deviations.}
\label{main_results}
\end{table}


\subsection{Explanation Quality on the Three Additional Datasets}
For the quality of the generated explanations, the results shown in Table~\ref{main_results_reason} demonstrate that RecPIE is also capable of producing explanations with significantly better quality compared to the baselines for these three datasets, at a level similar to the gains reported in Table~\ref{tab:googlemap_reason} of the main paper.

\begin{table}
\centering
\footnotesize
\setlength{\tabcolsep}{4pt}
\renewcommand{\arraystretch}{1.15}

\resizebox{0.99\textwidth}{!}{%
\begin{tabular}{lcccc cccc cccc}
\toprule
& \multicolumn{4}{c}{\textbf{TripAdvisor}} & \multicolumn{4}{c}{\textbf{Yelp}} & \multicolumn{4}{c}{\textbf{Amazon Movie}} \\
\cmidrule(lr){2-5}\cmidrule(lr){6-9}\cmidrule(lr){10-13}
\textbf{Model}
& BLEURT & Coverage & Informativeness & Fluency
& BLEURT & Coverage & Informativeness & Fluency
& BLEURT & Coverage & Informativeness & Fluency \\
\midrule

\textbf{RecPIE (Ours)}
& \textbf{0.4196} & \textbf{0.3784} & \textbf{0.5744} & \textbf{0.4193}
& \textbf{0.4386} & \textbf{0.3984} & \textbf{0.5936} & \textbf{0.4380}
& \textbf{0.4487} & \textbf{0.3580} & \textbf{0.5276} & \textbf{0.4003} \\
& (0.0003) & (0.0011) & (0.0011) & (0.0009)
& (0.0003) & (0.0011) & (0.0011) & (0.0009)
& (0.0004) & (0.0012) & (0.0011) & (0.0010) \\
\addlinespace[2pt]

TallRec
& \underline{0.4078} & 0.3668 & \underline{0.5583} & \underline{0.4158}
& 0.4330 & 0.3958 & 0.5901 & 0.4358
& 0.4410 & 0.3479 & 0.5155 & 0.3530 \\
& (0.0004) & (0.0011) & (0.0011) & (0.0009)
& (0.0003) & (0.0011) & (0.0011) & (0.0009)
& (0.0004) & (0.0012) & (0.0011) & (0.0010) \\
\addlinespace[2pt]

ExpCTR
& 0.4071 & 0.3650 & 0.5571 & 0.4150 
& \underline{0.4332} & \underline{0.3961} & \underline{0.5910} & \underline{0.4361}
& \underline{0.4417} & \underline{0.3488} & \underline{0.5170} & \underline{0.3536} \\
& (0.0004) & (0.0011) & (0.0011) & (0.0009)
& (0.0003) & (0.0011) & (0.0011) & (0.0009)
& (0.0004) & (0.0011) & (0.0011) & (0.0009) \\
\addlinespace[2pt]
LLM2ER-EQR
& 0.4002 & 0.3617 & 0.5523 & 0.4112 
& 0.4302 & 0.3933 & 0.5882 & 0.4337 
& 0.4386 & 0.3468 & 0.5144 & 0.3505 \\
& (0.0004) & (0.0011) & (0.0011) & (0.0010)
& (0.0004) & (0.0011) & (0.0011) & (0.0010)
& (0.0004) & (0.0011) & (0.0011) & (0.0010) \\
\addlinespace[2pt]
RXR
& 0.3832 & 0.3425 & 0.5338 & 0.3986 
& 0.4111 & 0.3729 & 0.5629 & 0.4146 
& 0.4127 & 0.3230 & 0.5003 & 0.3303 \\
& (0.0006) & (0.0013) & (0.0013) & (0.0013) 
& (0.0006) & (0.0013) & (0.0013) & (0.0013) 
& (0.0006) & (0.0011) & (0.0013) & (0.0011) \\
\addlinespace[2pt]

A3NCF
& 0.3529 & 0.3539 & 0.5116 & 0.3638 
& 0.3836 & 0.3503 & 0.5406 & 0.3729 
& 0.3716 & 0.2928 & 0.4542 & 0.2965 \\
& (0.0270) & (0.0330) & (0.0390) & (0.0310) 
& (0.0290) & (0.0370) & (0.0370) & (0.0310) 
& (0.0290) & (0.0330) & (0.0390) & (0.0290) \\
\addlinespace[2pt]
SULM
& 0.3516 & 0.3544 & 0.5108 & 0.3646 
& 0.3844 & 0.3510 & 0.5417 & 0.3738 
& 0.3708 & 0.2933 & 0.4530 & 0.2969 \\
& (0.0290) & (0.0370) & (0.0370) & (0.0290) 
& (0.0290) & (0.0370) & (0.0370) & (0.0290) 
& (0.0290) & (0.0330) & (0.0370) & (0.0290) \\
\addlinespace[2pt]
AARM
& 0.3521 & 0.3540 & 0.5115 & 0.3641 
& 0.3846 & 0.3505 & 0.5420 & 0.3732 
& 0.3717 & 0.2931 & 0.4534 & 0.2967 \\
& (0.0290) & (0.0370) & (0.0370) & (0.0288) 
& (0.0291) & (0.0370) & (0.0373) & (0.0290) 
& (0.0290) & (0.0332) & (0.0370) & (0.0290) \\
\addlinespace[2pt]
MMALFM
& 0.3509 & 0.3541 & 0.5101 & 0.3639 
& 0.3828 & 0.3495 & 0.5402 & 0.3720 
& 0.3691 & 0.2812 & 0.4518 & 0.2950 \\
& (0.0290) & (0.0370) & (0.0370) & (0.0290) 
& (0.0290) & (0.0370) & (0.0370) & (0.0290) 
& (0.0290) & (0.0330) & (0.0370) & (0.0290) \\
\addlinespace[2pt]
ANR
& 0.3477 & 0.3510 & 0.5079 & 0.3598 
& 0.3781 & 0.3466 & 0.5370 & 0.3688 
& 0.3660 & 0.2780 & 0.4477 & 0.2917 \\
& (0.0270) & (0.0330) & (0.0390) & (0.0310) 
& (0.0290) & (0.0370) & (0.0370) & (0.0310) 
& (0.0270) & (0.0330) & (0.0390) & (0.0290) \\
\addlinespace[2pt]
MTER
& 0.3489 & 0.3533 & 0.5091 & 0.3605 
& 0.3789 & 0.3475 & 0.5388 & 0.3701 
& 0.3679 & 0.2791 & 0.4492 & 0.2926 \\
& (0.0310) & (0.0370) & (0.0390) & (0.0330) 
& (0.0310) & (0.0370) & (0.0390) & (0.0330) 
& (0.0310) & (0.0370) & (0.0390) & (0.0310) \\
\addlinespace[2pt]
AMCF
& 0.3502 & 0.3537 & 0.5102 & 0.3616 
& 0.3797 & 0.3488 & 0.5398 & 0.3710 
& 0.3685 & 0.2802 & 0.4499 & 0.2930 \\
& (0.0290) & (0.0370) & (0.0390) & (0.0330) 
& (0.0310) & (0.0370) & (0.0390) & (0.0330) 
& (0.0290) & (0.0370) & (0.0390) & (0.0310) \\
\addlinespace[2pt]
PETER
& 0.3679 & \underline{0.3730} & 0.4762 & 0.3772
& 0.4147 & 0.3770 & 0.5327 & 0.4012
& 0.4025 & 0.3026 & 0.4748 & 0.3264 \\
& (0.0004) & (0.0011) & (0.0012) & (0.0010)
& (0.0003) & (0.0012) & (0.0011) & (0.0009)
& (0.0004) & (0.0012) & (0.0011) & (0.0010) \\
\addlinespace[2pt]
UCEPic
& 0.3633 & 0.3703 & 0.4745 & 0.3751 
& 0.4120 & 0.3750 & 0.5301 & 0.3988 
& 0.4003 & 0.3003 & 0.4721 & 0.3239 \\
& (0.0006) & (0.0013) & (0.0013) & (0.0011)
& (0.0004) & (0.0013) & (0.0013) & (0.0011)
& (0.0006) & (0.0013) & (0.0012) & (0.0010) \\
\addlinespace[2pt]
PARSRec
& 0.3601 & 0.3677 & 0.4710 & 0.3711 
& 0.4088 & 0.3712 & 0.5269 & 0.3952 
& 0.3970 & 0.2968 & 0.4688 & 0.3201 \\
& (0.0060) & (0.0130) & (0.0130) & (0.0110)
& (0.0040) & (0.0130) & (0.0130) & (0.0110)
& (0.0060) & (0.0130) & (0.0120) & (0.0100) \\
\addlinespace[2pt]
\bottomrule
\end{tabular}%
}

\caption{Reason generation performance on three datasets (treating user reviews as ground truth). The best baseline in each metric is underlined, and parentheses report standard deviations.}
\label{main_results_reason}
\end{table}


\subsection{Understanding the Improvements on the Three Additional Datasets}
In this subsection, we present additional analyses to decompose and better understand the significant performance gains observed with RecPIE in our generalizability analysis.


\subsubsection{Improved learning efficiency.} 
Following the approach in Section~\ref{sec:learning_effiency}, we randomly sample subsets of the three datasets, keeping 15\%, 25\%, and 50\% of the original training data, and train RecPIE on these subsets while keeping the same test set for evaluation. The results are presented in Table~\ref{generalizability_efficiency}, where we find that with 25\% of the training data RecPIE already surpasses PETER \citep{li2021personalized} trained on the full data across all three metrics and all three datasets, and that with 15\% of the training data it still performs on par with it. These findings mirror those in Table~\ref{tab:googlemaps_efficiency} of the main paper and validate the improved learning efficiency of RecPIE, in line with the conceptual motivation in Section~\ref{sec:conceptual_motivation}.

\begin{table}
\centering
\footnotesize
\setlength{\tabcolsep}{4pt}
\renewcommand{\arraystretch}{1.15}

\resizebox{0.85\textwidth}{!}{%
\begin{tabular}{llccc ccc ccc}
\toprule
& & \multicolumn{3}{c}{\textbf{TripAdvisor}} & \multicolumn{3}{c}{\textbf{Yelp}} & \multicolumn{3}{c}{\textbf{Amazon Movie}} \\
\cmidrule(lr){3-5}\cmidrule(lr){6-8}\cmidrule(lr){9-11}
\textbf{Model} & \textbf{Training data}
& RMSE$\downarrow$ & MAE$\downarrow$ & AUC$\uparrow$
& RMSE$\downarrow$ & MAE$\downarrow$ & AUC$\uparrow$
& RMSE$\downarrow$ & MAE$\downarrow$ & AUC$\uparrow$ \\
\midrule

RecPIE & 100\% 
& \textbf{0.1733} & \textbf{0.1333} & \textbf{0.7376}
& \textbf{0.2020} & \textbf{0.1620} & \textbf{0.7350}
& \textbf{0.1653} & \textbf{0.1165} & \textbf{0.7582} \\
& 
& (0.0010) & (0.0008) & (0.0018)
& (0.0010) & (0.0009) & (0.0017)
& (0.0010) & (0.0009) & (0.0018) \\
\addlinespace[2pt]

RecPIE & 50\%
& 0.1849 & 0.1426 & 0.7189
& 0.2116 & 0.1695 & 0.7193
& 0.1721 & 0.1250 & 0.7349 \\
& 
& (0.0017) & (0.0013) & (0.0029)
& (0.0018) & (0.0017) & (0.0031)
& (0.0021) & (0.0017) & (0.0033) \\
\addlinespace[2pt]

RecPIE & 25\%
& 0.1927 & 0.1559 & 0.7121
& 0.2246 & 0.1798 & 0.7104
& 0.1847 & 0.1403 & 0.7263 \\
&
& (0.0026) & (0.0021) & (0.0041)
& (0.0027) & (0.0027) & (0.0047)
& (0.0039) & (0.0028) & (0.0049) \\
\addlinespace[2pt]

RecPIE & 15\%
& 0.1968 & 0.1606 & 0.7052
& 0.2297 & 0.1865 & 0.7029
& 0.1935 & 0.1516 & 0.7185 \\
&
& (0.0036) & (0.0030) & (0.0054)
& (0.0039) & (0.0038) & (0.0068)
& (0.0066) & (0.0044) & (0.0063) \\
\addlinespace[4pt]
\midrule

ExpCTR & 100\%
& 0.1933 & 0.1501 & 0.7170
& 0.2244 & 0.1735 & 0.7126
& 0.1725 & 0.1228 & 0.7395 \\
&
& (0.0017) & (0.0013) & (0.0027)
& (0.0017) & (0.0014) & (0.0027)
& (0.0016) & (0.0013) & (0.0024) \\
\addlinespace[2pt]

TallRec & 100\%
& 0.1924 & 0.1498 & 0.7176
& 0.2238 & 0.1732 & 0.7135
& 0.1720 & 0.1222 & 0.7401 \\
&
& (0.0017) & (0.0013) & (0.0027)
& (0.0017) & (0.0014) & (0.0027)
& (0.0016) & (0.0013) & (0.0024) \\
\addlinespace[2pt]

PETER & 100\%
& 0.1996 & 0.1715 & 0.7078
& 0.2423 & 0.2071 & 0.7003
& 0.2099 & 0.1770 & 0.7226 \\
&
& (0.0019) & (0.0013) & (0.0027)
& (0.0015) & (0.0013) & (0.0022)
& (0.0013) & (0.0010) & (0.0019) \\
\bottomrule
\end{tabular}%
}

\caption{Recommendation performance across three datasets, where RecPIE is trained on varying percentages of the training data and the baselines are trained on the full training data. Parentheses report standard deviations.}
\label{generalizability_efficiency}
\end{table}

\subsubsection{The gain is from the LLM's reasoning capability.}  
Following the settings in Section~\ref{res_understanding_reasoning}, we conduct additional experiments using different LLMs with varying reasoning capabilities on these three datasets. As shown in Table~\ref{llm_models}, the performance of RecPIE with Llama~3.1 is significantly better than that of RecPIE with Llama~3, Mixtral-8$\times$7B, Vicuna-7B-v1.5, Qwen2-7B, or GPT-2. This aligns with our observations in Table~\ref{tab:googlemaps_llm} of the main paper and the reasoning capability leaderboard at \url{https://huggingface.co/spaces/allenai/ZebraLogic}, where Llama~3.1 demonstrates the highest reasoning capabilities among the tested models. These results suggest that better \emph{reasoning} capabilities in LLMs translate into improved performance within the RecPIE framework. 

Moreover, RecPIE significantly outperforms an alternative approach that uses LLMs directly for recommendations without generating explicit explanations (``Llama 3.1 direct recommendation'' row in Table~\ref{llm_models}). Profile augmentation, meanwhile, does not have a tangible impact on the recommendation performance in our experiment (``RecPIE w/ profile augmentation'' row in Table~\ref{llm_models}). Furthermore, as shown in the last row of Table~\ref{llm_models} (``Consumption history summarization''), the performance of using an LLM for summarization is significantly worse than that of RecPIE using an LLM for explanations. All of these findings are consistent with those reported in Section~\ref{res_understanding_reasoning} and with the interpretation that the performance gains observed with RecPIE are primarily driven by the LLMs' reasoning capabilities, \emph{not} their external dataset knowledge or summarization skills.

\begin{table}
\centering
\footnotesize
\setlength{\tabcolsep}{4pt}
\renewcommand{\arraystretch}{1.10}

\resizebox{1.00\textwidth}{!}{%
\begin{tabular}{lccc ccc ccc}
\toprule
& \multicolumn{3}{c}{\textbf{TripAdvisor}} & \multicolumn{3}{c}{\textbf{Yelp}} & \multicolumn{3}{c}{\textbf{Amazon Movie}} \\
\cmidrule(lr){2-4}\cmidrule(lr){5-7}\cmidrule(lr){8-10}
\textbf{Setting / Model}
& RMSE$\downarrow$ & MAE$\downarrow$ & AUC$\uparrow$
& RMSE$\downarrow$ & MAE$\downarrow$ & AUC$\uparrow$
& RMSE$\downarrow$ & MAE$\downarrow$ & AUC$\uparrow$ \\
\midrule

\multicolumn{10}{l}{\textit{RecPIE with different LLM backbones}} \\
\addlinespace[2pt]

RecPIE w/ Llama 3.1 (Ours)
& 0.1733 & 0.1333 & 0.7376
& 0.2020 & 0.1620 & 0.7350
& 0.1653 & 0.1165 & 0.7582 \\
& (0.0010) & (0.0008) & (0.0018)
& (0.0010) & (0.0009) & (0.0017)
& (0.0010) & (0.0009) & (0.0018) \\
\addlinespace[2pt]

RecPIE w/ Llama 3
& 0.1834 & 0.1391 & 0.7260
& 0.2166 & 0.1697 & 0.7210
& 0.1695 & 0.1203 & 0.7472 \\
& (0.0010) & (0.0008) & (0.0018)
& (0.0010) & (0.0009) & (0.0017)
& (0.0010) & (0.0009) & (0.0018) \\
\addlinespace[2pt]

RecPIE w/ Mixtral-8$\times$7B
& 0.1810 & 0.1362 & 0.7271
& 0.2163 & 0.1693 & 0.7218
& 0.1691 & 0.1199 & 0.7480 \\
& (0.0010) & (0.0008) & (0.0018)
& (0.0010) & (0.0009) & (0.0017)
& (0.0010) & (0.0009) & (0.0018) \\
\addlinespace[2pt]

RecPIE w/ Vicuna-7B-v1.5
& 0.1849 & 0.1402 & 0.7243
& 0.2175 & 0.1703 & 0.7196
& 0.1703 & 0.1210 & 0.7455 \\
& (0.0010) & (0.0008) & (0.0018)
& (0.0010) & (0.0009) & (0.0017)
& (0.0010) & (0.0009) & (0.0018) \\
\addlinespace[2pt]

RecPIE w/ Qwen2-7B
& 0.1866 & 0.1420 & 0.7202
& 0.2193 & 0.1724 & 0.7170
& 0.1727 & 0.1235 & 0.7419 \\
& (0.0010) & (0.0008) & (0.0018)
& (0.0010) & (0.0009) & (0.0017)
& (0.0010) & (0.0009) & (0.0018) \\
\addlinespace[2pt]

RecPIE w/ GPT-2
& 0.1940 & 0.1582 & 0.7169
& 0.2211 & 0.1801 & 0.7144
& 0.1799 & 0.1304 & 0.7288 \\
& (0.0014) & (0.0010) & (0.0023)
& (0.0015) & (0.0013) & (0.0023)
& (0.0015) & (0.0013) & (0.0024) \\
\addlinespace[3pt]
\midrule

\multicolumn{10}{l}{\textit{Baselines / ablations}} \\
\addlinespace[2pt]

Llama 3.1 direct recommendation
& 0.2233 & 0.1838 & 0.6735
& 0.2976 & 0.2402 & 0.6528
& 0.2680 & 0.2055 & 0.6736 \\
& (0.0033) & (0.0020) & (0.0036)
& (0.0044) & (0.0029) & (0.0046)
& (0.0046) & (0.0031) & (0.0055) \\
\addlinespace[2pt]

RecPIE w/ profile augmentation
& 0.1723 & 0.1320 & 0.7381
& 0.2013 & 0.1619 & 0.7363
& 0.1644 & 0.1159 & 0.7590 \\
& (0.0013) & (0.0011) & (0.0021)
& (0.0012) & (0.0012) & (0.0021)
& (0.0013) & (0.0012) & (0.0021) \\
\addlinespace[2pt]

Consumption history summarization
& 0.1971 & 0.1700 & 0.7055
& 0.2409 & 0.2051 & 0.6990
& 0.2077 & 0.1751 & 0.7236 \\
& (0.0011) & (0.0008) & (0.0018)
& (0.0011) & (0.0009) & (0.0017)
& (0.0011) & (0.0010) & (0.0018) \\
\bottomrule
\end{tabular}%
}

\caption{Recommendation performance across three datasets using LLMs with varying reasoning capability. Parentheses report standard deviations.}
\label{llm_models}
\end{table}


\subsubsection{``Harder'' examples benefit more from RecPIE.} 
\label{appen:hte_three_datasets}
Following the settings in Section~\ref{sec:harder_examples}, we created plots for the three datasets in Figure~\ref{fig:uncertainty}, where the x-axis represents the normalized uncertainty level (scaled between 0 and 1 using min-max normalization \citep{patro2015normalization}), and the y-axis represents the performance improvement of our RecPIE over the best-performing baseline (measured by RMSE). As shown by the regression lines in Figures~\ref{fig:uncertainty:amazon}, \ref{fig:uncertainty:tripadvisor}, and \ref{fig:uncertainty:yelp}, there is a statistically significant \emph{positive correlation} between uncertainty and performance improvements, consistent with our observation in Figure~\ref{fig:uncertainty:googlemap} and the conceptual motivation in Section~\ref{sec:conceptual_motivation}, validating the insight that our RecPIE is more beneficial for examples that are ``harder'' or more uncertain.

\begin{figure}[hbtp!]
    \begin{subfigure}[b]{0.32\textwidth}
        \centering
        \includegraphics[width=\textwidth]{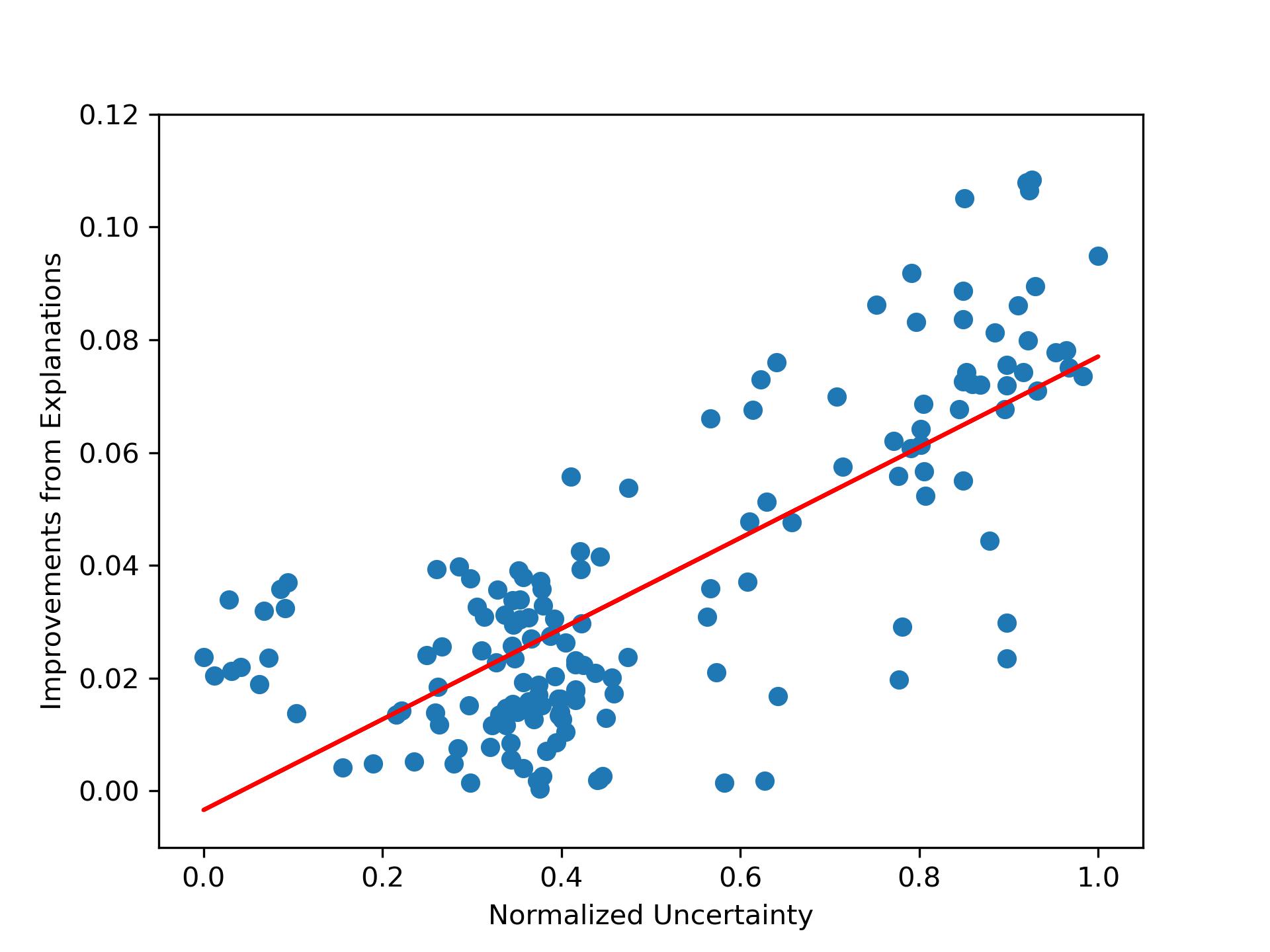}
        \caption{Amazon Movie.}
        \label{fig:uncertainty:amazon}
    \end{subfigure}
     \centering
    \begin{subfigure}[b]{0.32\textwidth}
        \centering
        \includegraphics[width=\textwidth]{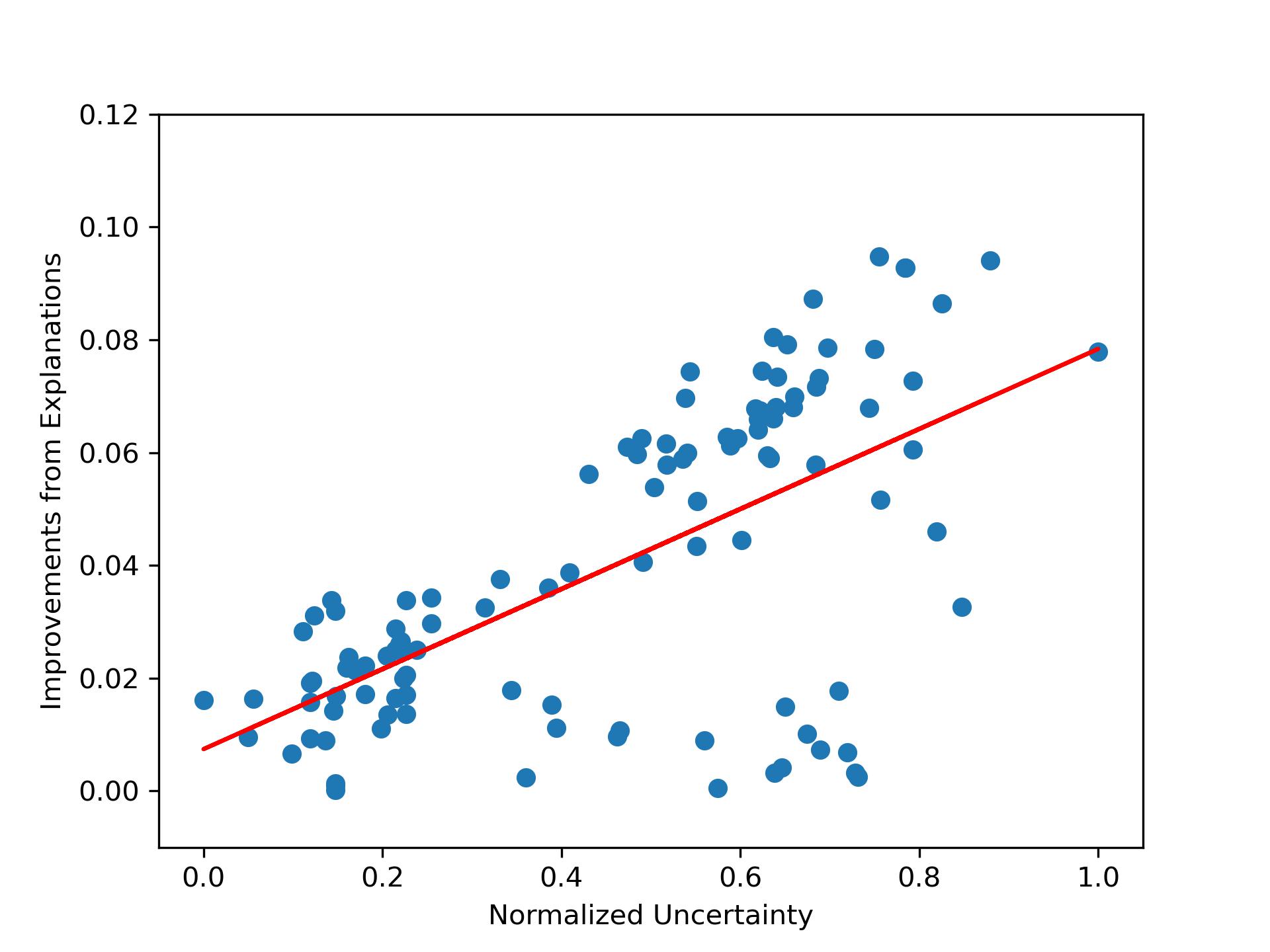}
        \caption{TripAdvisor Dataset.}
        \label{fig:uncertainty:tripadvisor}
    \end{subfigure}
    \begin{subfigure}[b]{0.32\textwidth}
        \centering
        \includegraphics[width=\textwidth]{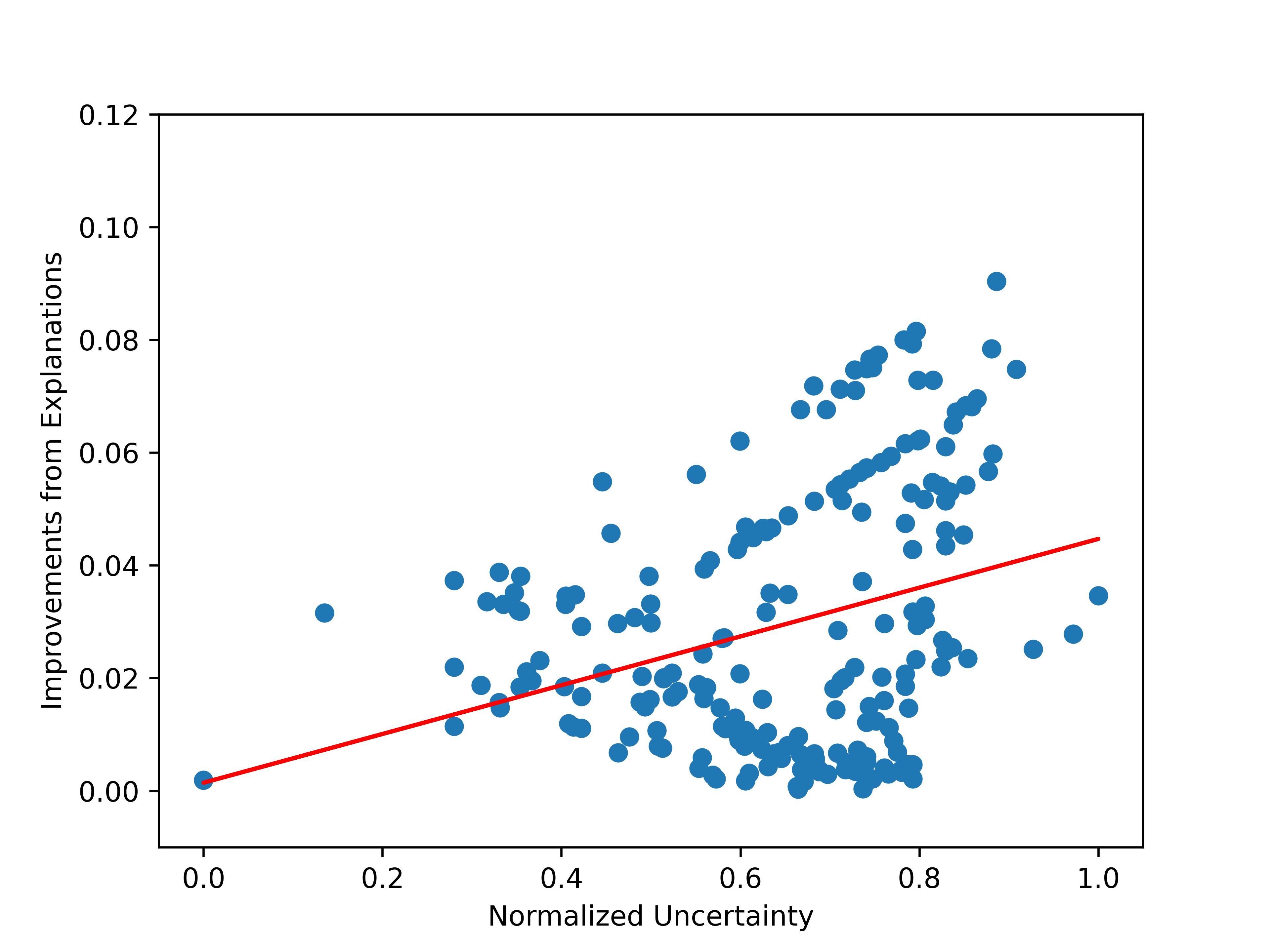}
        \caption{Yelp Dataset.}
        \label{fig:uncertainty:yelp}
    \end{subfigure}
   \caption{(Color online) Performance improvement of RecPIE against (normalized) prediction uncertainty.}
  \label{fig:uncertainty}
\end{figure}

\subsection{Role of Contrastive Explanations on the Three Additional Datasets}
\label{appen:role_pos_neg}

\subsubsection{Attention weights across the three datasets.}

Following the settings in Appendix~\ref{appen:attention_analysis}, we conduct an \emph{attention weight analysis} to quantify the contributions of positive and negative explanations to the recommendation task for these three datasets. We visualize the distribution of attention values on positive explanations ($\bar{\alpha}_{\mathrm{pos}}$) and negative explanations ($\bar{\alpha}_{\mathrm{neg}}$) for all three datasets in Figure~\ref{fig:attention:generalizability}. Similar to Appendix~\ref{appen:attention_analysis}, the results across all three datasets consistently show that when a product receives a high rating, RecPIE assigns \emph{more} attention to positive explanations than negative ones, while for products receiving a low rating, RecPIE assigns more attention to negative explanations. We also plot the distribution of attention values on other input components (i.e., consumer, product, and sequence embeddings) and find that the two explanation embeddings together account for a substantial share (30--40\%) of total attention values.

\begin{figure}[hbtp!]
    \begin{subfigure}[b]{0.32\textwidth}
        \centering
        \includegraphics[width=\textwidth]{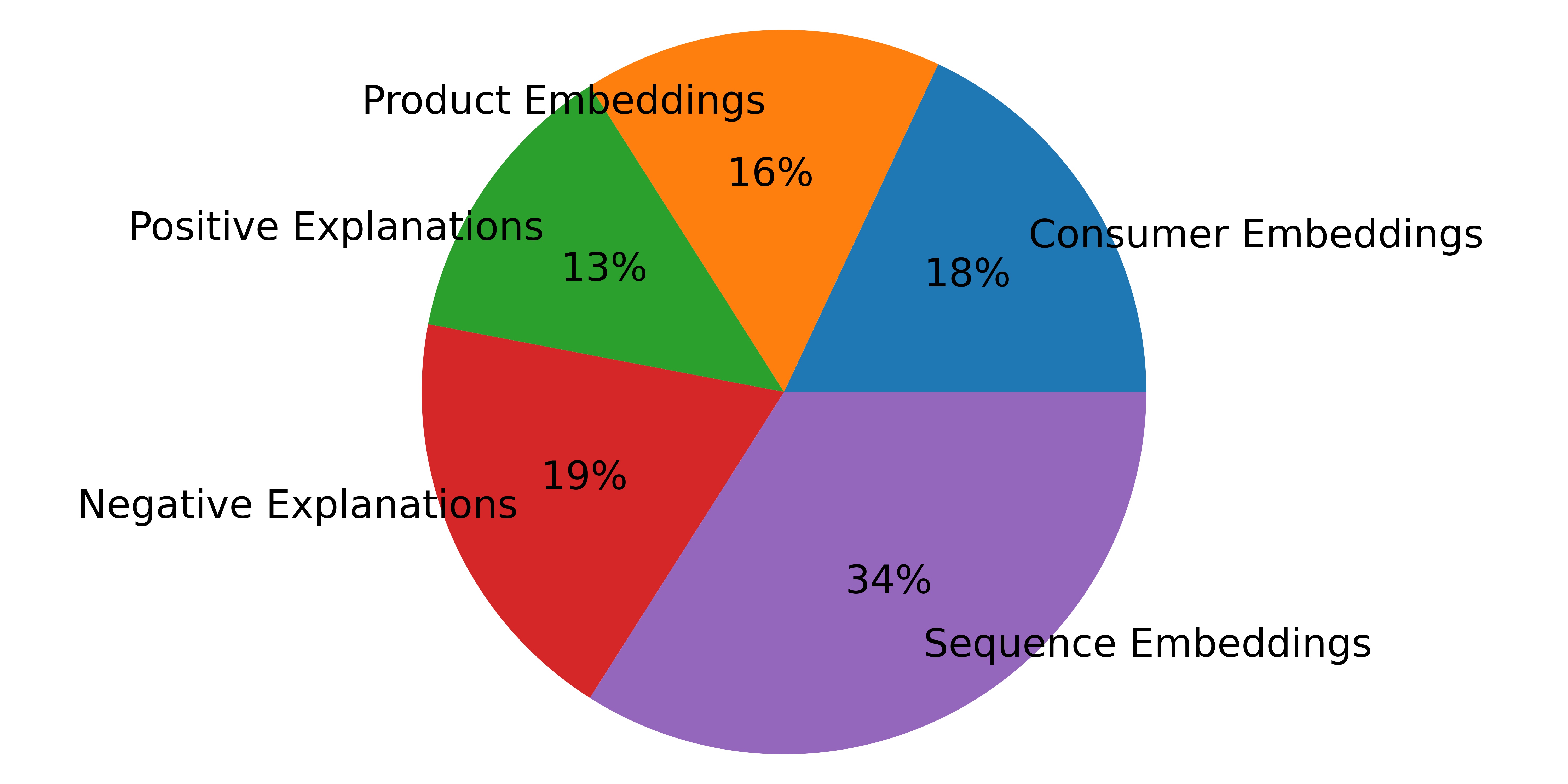}
        \caption{Attention value distribution. \\ (TripAdvisor Dataset)}
        \label{fig:pie_tripadvisor}
    \end{subfigure}
    \begin{subfigure}[b]{0.33\textwidth}
        \centering
        \includegraphics[width=\textwidth]{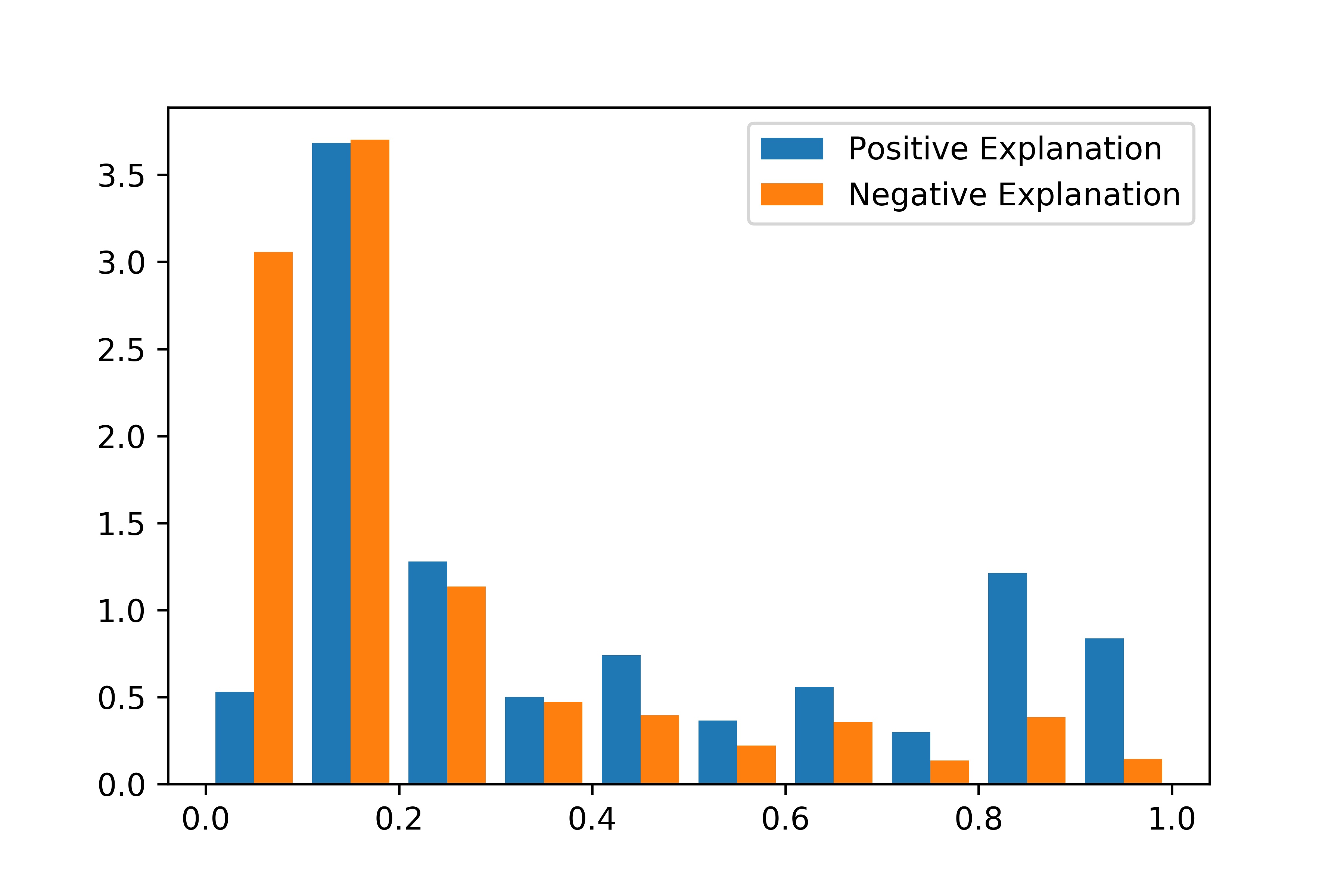}
        \caption{Positive examples. \\ (TripAdvisor Dataset)}
        \label{fig:positive:tripadvisor}
    \end{subfigure}
     \centering
    \begin{subfigure}[b]{0.33\textwidth}
        \centering
        \includegraphics[width=\textwidth]{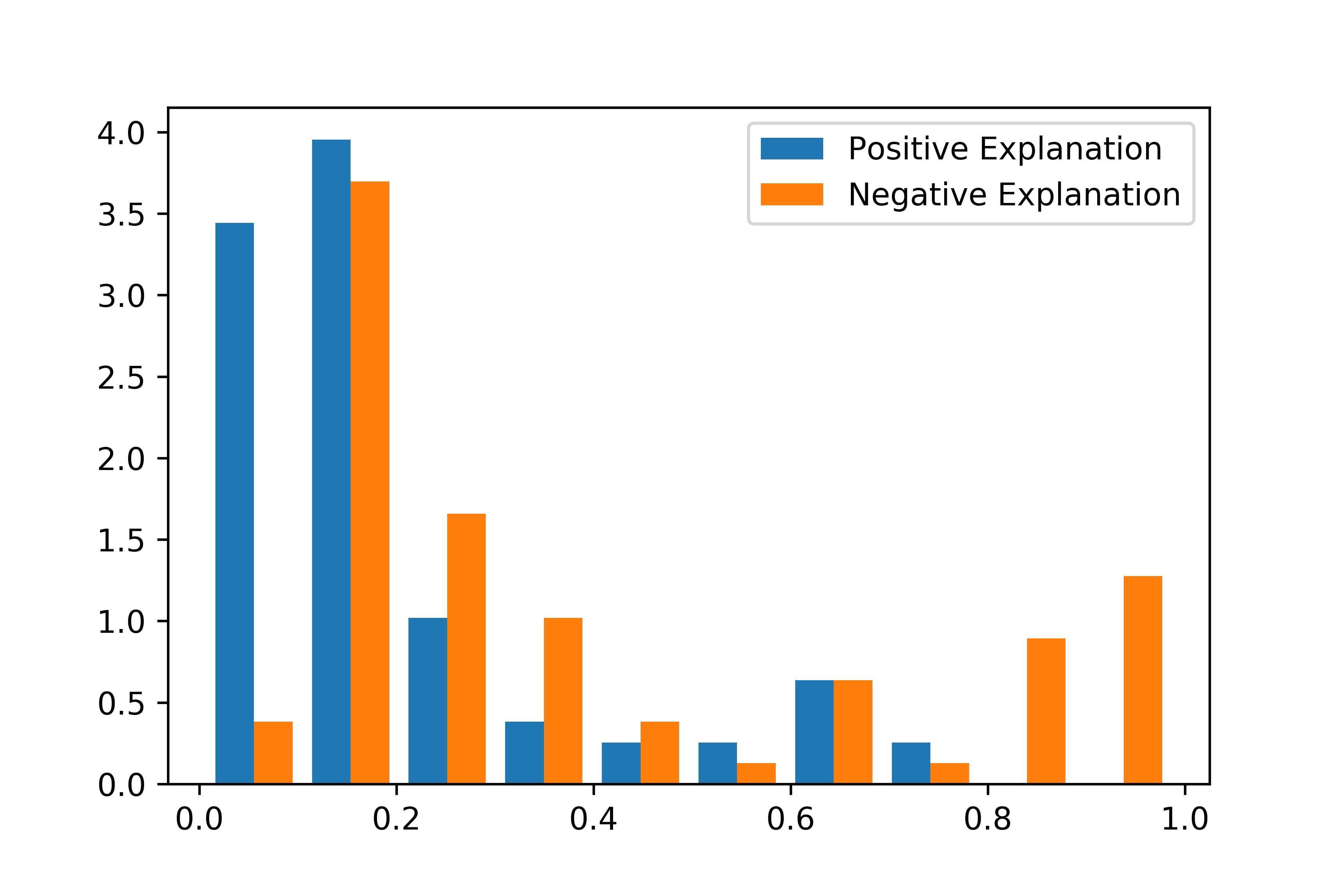}
        \caption{Negative examples. \\ (TripAdvisor Dataset)}
        \label{fig:negative:tripadvisor}
    \end{subfigure}
    \hspace{0.5mm}
    \begin{subfigure}[b]{0.32\textwidth}
        \centering
        \includegraphics[width=\textwidth]{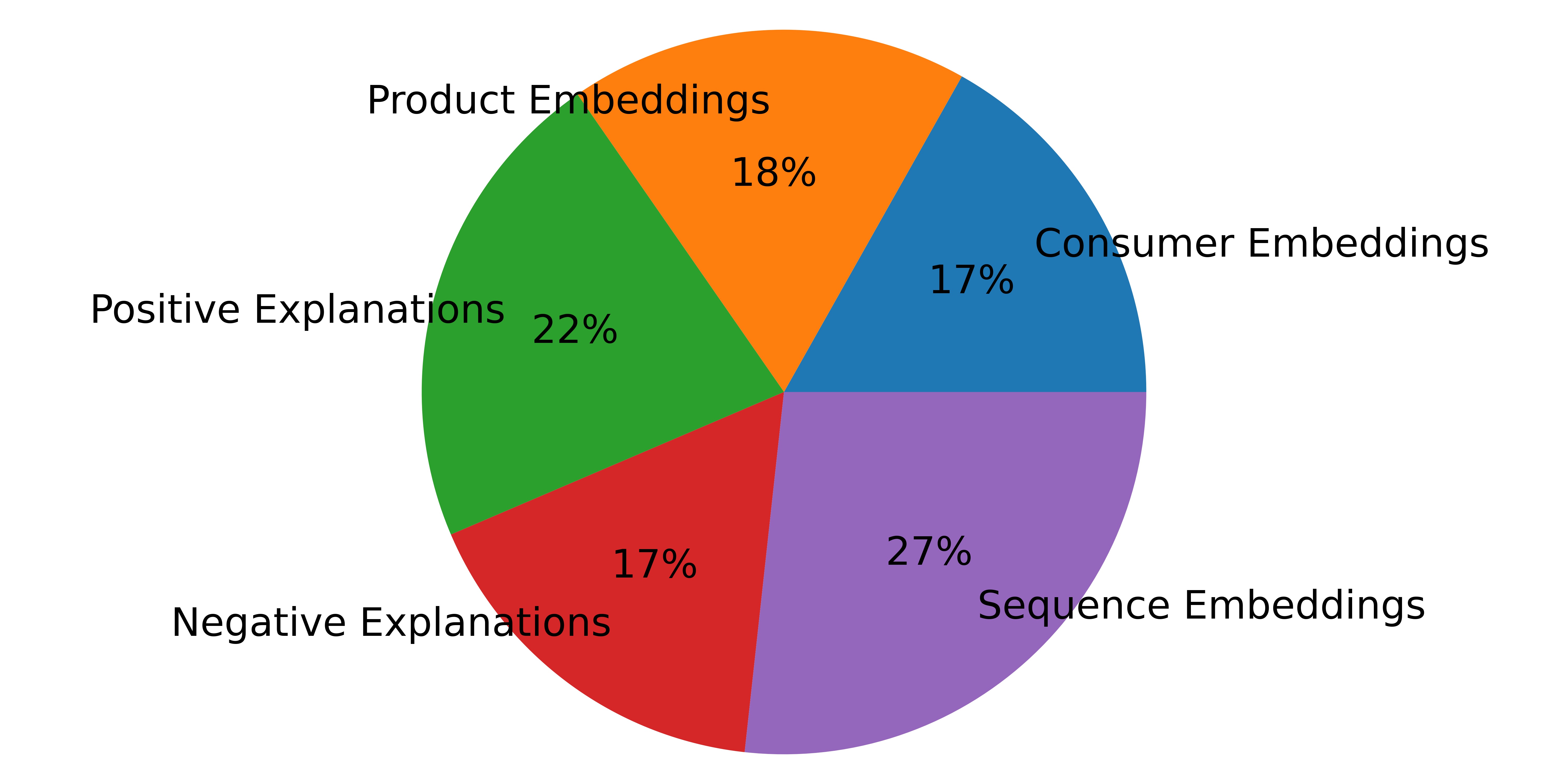}
        \caption{Attention value distribution. \\ (Yelp Dataset)}
        \label{fig:pie_yelp}
    \end{subfigure}
    \begin{subfigure}[b]{0.33\textwidth}
        \centering
        \includegraphics[width=\textwidth]{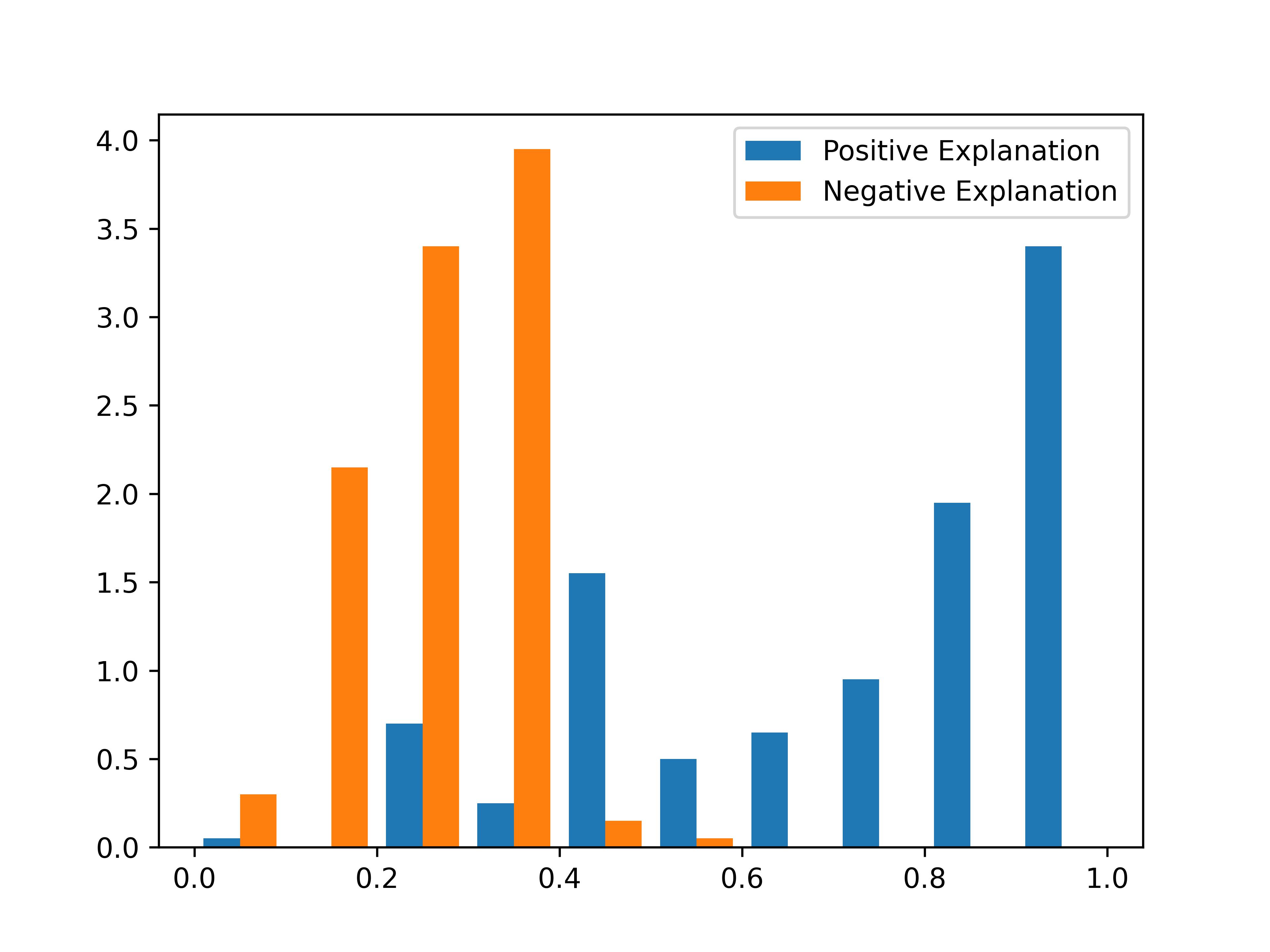}
        \caption{Positive examples. \\ (Yelp Dataset)}
        \label{fig:positive:yelp}
    \end{subfigure}
     \centering
    \begin{subfigure}[b]{0.33\textwidth}
        \centering
        \includegraphics[width=\textwidth]{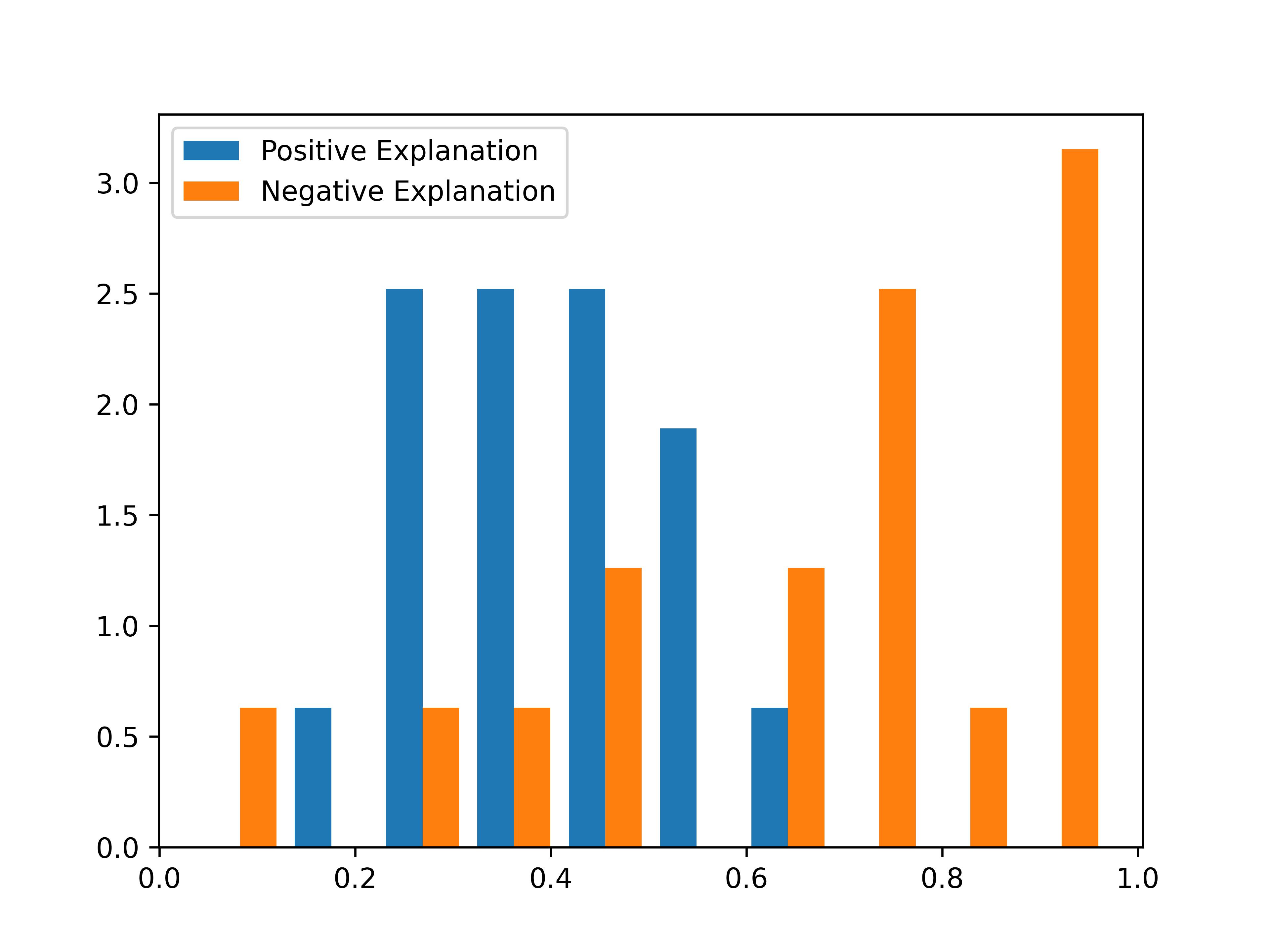}
        \caption{Negative examples. \\ (Yelp Dataset)}
        \label{fig:negative:yelp}
    \end{subfigure}
    \hspace{0.5mm}
    \begin{subfigure}[b]{0.32\textwidth}
        \centering
        \includegraphics[width=\textwidth]{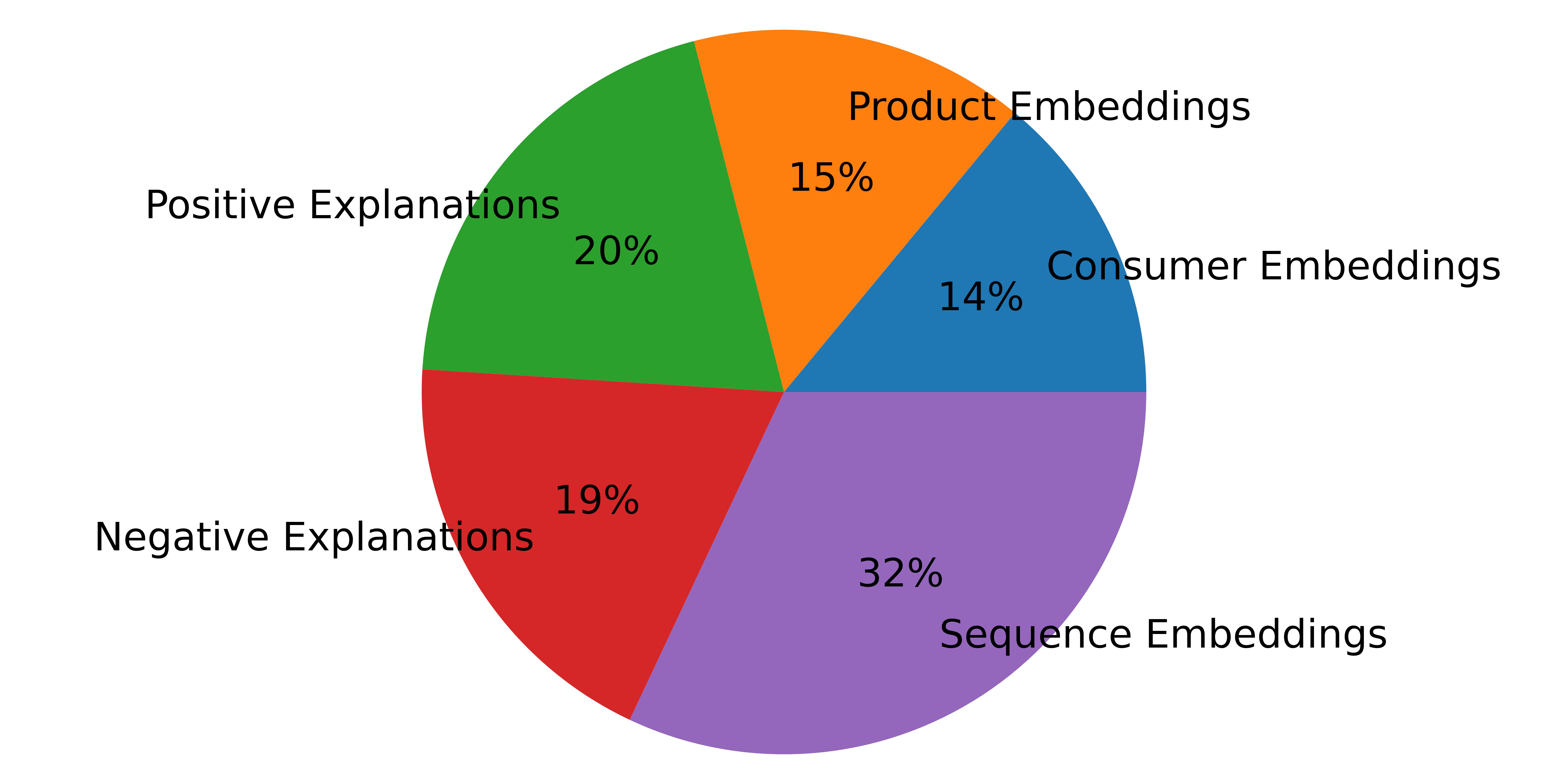}
        \caption{Attention value distribution. \\ (Amazon Movie)}
        \label{fig:pie_amazon}
    \end{subfigure}
    \begin{subfigure}[b]{0.33\textwidth}
        \centering
        \includegraphics[width=\textwidth]{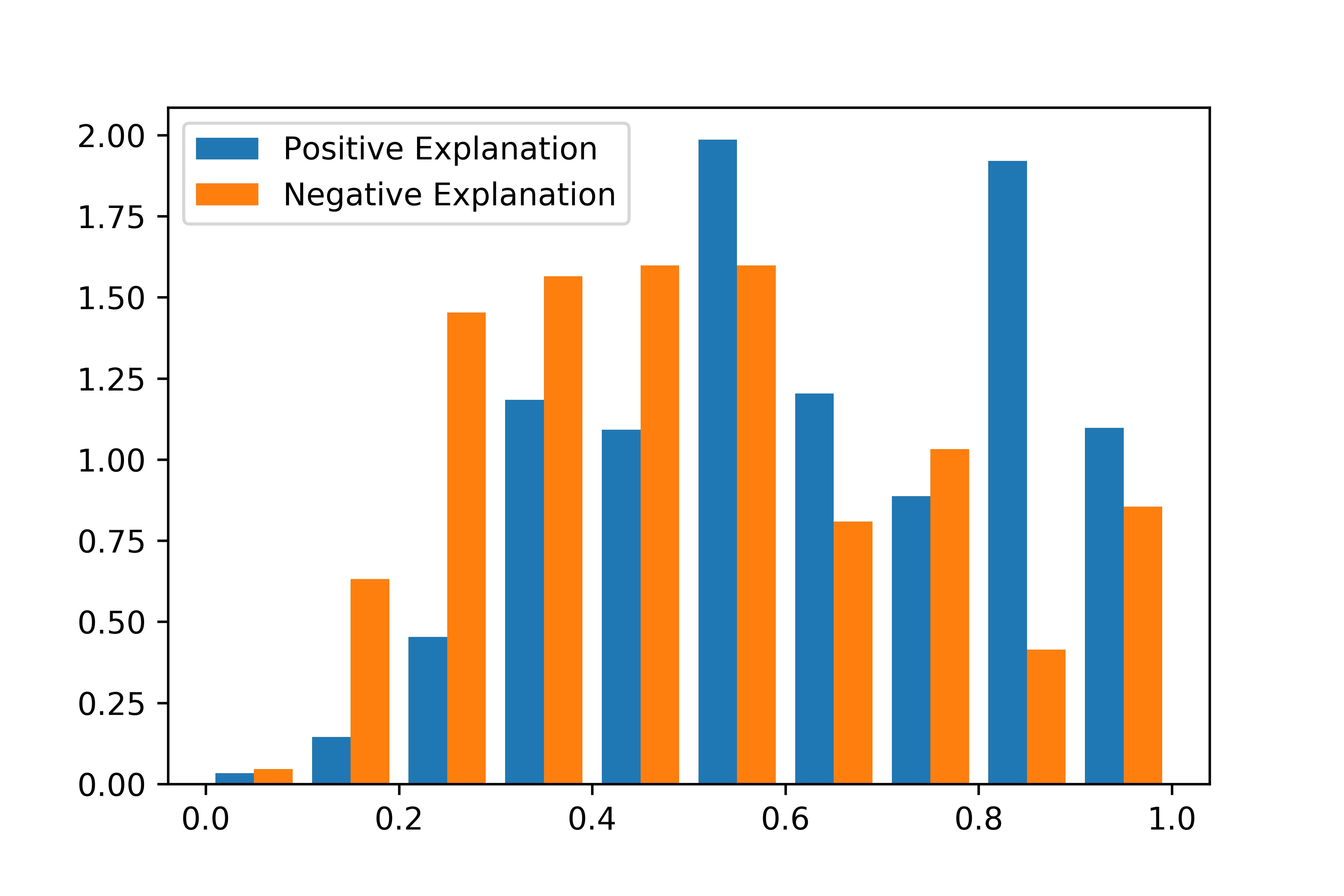}
        \caption{Positive examples. \\ (Amazon Movie)}
        \label{fig:positive:amazon}
    \end{subfigure}
     \centering
    \begin{subfigure}[b]{0.33\textwidth}
        \centering
        \includegraphics[width=\textwidth]{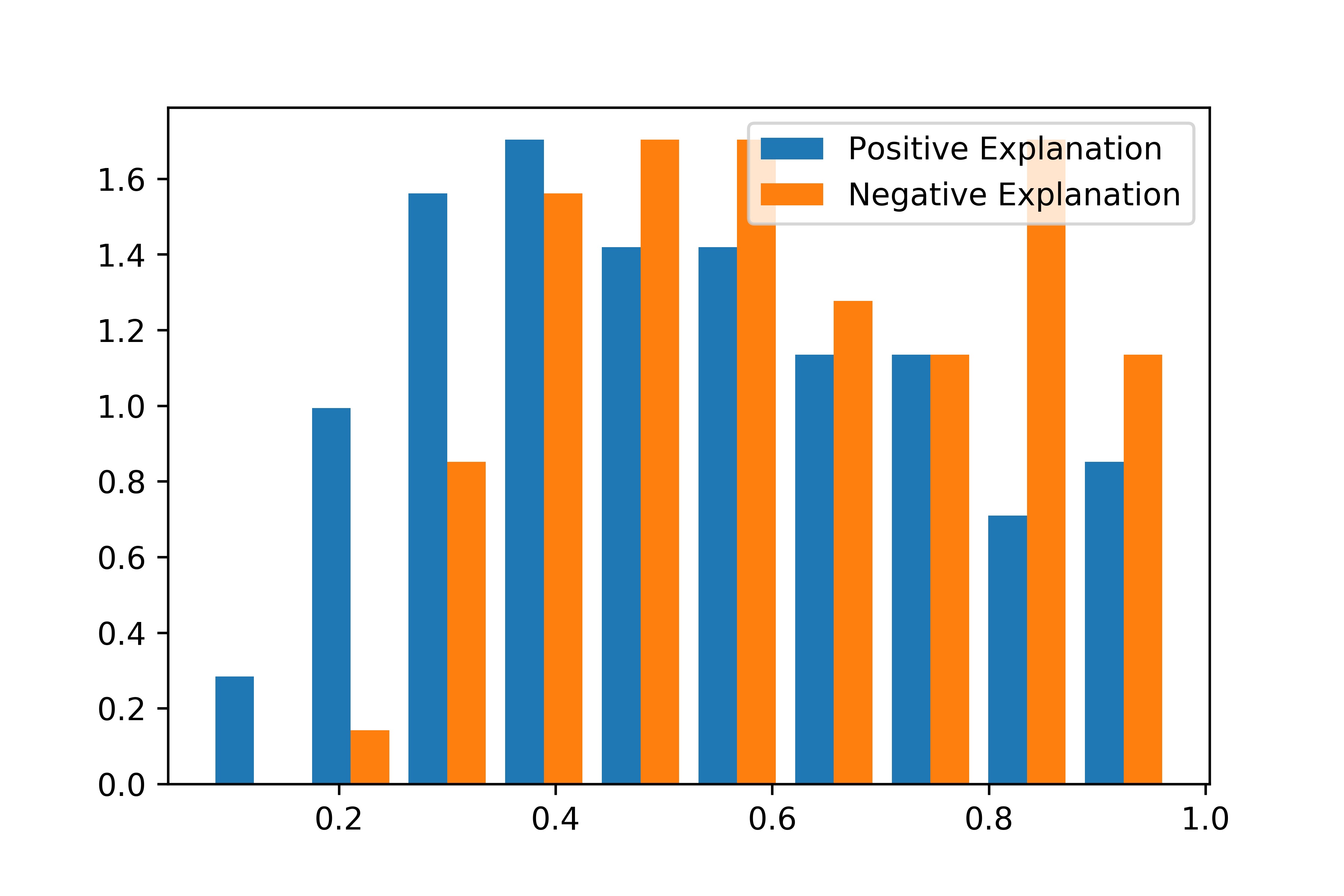}
        \caption{Negative examples. \\ (Amazon Movie)}
        \label{fig:negative:amazon}
    \end{subfigure}
   \caption{(Color online) Attention value distribution, and the distribution of attention values on positive and negative explanations for positive and negative examples in three datasets.}
  \label{fig:attention:generalizability}
\end{figure}

\subsubsection{The need for both explanation types across the three datasets.}
Following the setting in Appendix~\ref{appen:dual_explanation}, we compared three variants of the contrastive-explanation generator in RecPIE for these three datasets. The results, summarized in Table~\ref{generation} and similar to those in Table~\ref{tab:googlemaps_generation}, show that none of these variants matches the performance of our proposed RecPIE with the contrastive-explanation generator. This confirms the value of incorporating both positive and negative explanations to improve predictive performance.

\begin{table}
\centering
\footnotesize
\setlength{\tabcolsep}{4pt}
\renewcommand{\arraystretch}{1.10}

\resizebox{0.95\textwidth}{!}{%
\begin{tabular}{lccc ccc ccc}
\toprule
& \multicolumn{3}{c}{\textbf{TripAdvisor}}
& \multicolumn{3}{c}{\textbf{Yelp}}
& \multicolumn{3}{c}{\textbf{Amazon Movie}} \\
\cmidrule(lr){2-4}\cmidrule(lr){5-7}\cmidrule(lr){8-10}
\textbf{Setting}
& RMSE$\downarrow$ & MAE$\downarrow$ & AUC$\uparrow$
& RMSE$\downarrow$ & MAE$\downarrow$ & AUC$\uparrow$
& RMSE$\downarrow$ & MAE$\downarrow$ & AUC$\uparrow$ \\
\midrule

\textbf{RecPIE (Ours)}
& \textbf{0.1733} & \textbf{0.1333} & \textbf{0.7376}
& \textbf{0.2020} & \textbf{0.1620} & \textbf{0.7350}
& \textbf{0.1653} & \textbf{0.1165} & \textbf{0.7582} \\
& (0.0010) & (0.0008) & (0.0018)
& (0.0010) & (0.0009) & (0.0017)
& (0.0010) & (0.0009) & (0.0018) \\
\addlinespace[3pt]

Aspect terms only
& 0.1975 & 0.1709 & 0.7071
& 0.2413 & 0.2053 & 0.6991
& 0.2083 & 0.1757 & 0.7243 \\
& (0.0010) & (0.0008) & (0.0018)
& (0.0010) & (0.0009) & (0.0017)
& (0.0011) & (0.0010) & (0.0018) \\
\addlinespace[2pt]

Positive explanations only
& 0.1928 & 0.1480 & 0.7258
& 0.2179 & 0.1708 & 0.7168
& 0.1703 & 0.1209 & 0.7456 \\
& (0.0010) & (0.0008) & (0.0018)
& (0.0010) & (0.0009) & (0.0018)
& (0.0010) & (0.0009) & (0.0018) \\
\addlinespace[2pt]

General explanations only
& 0.1961 & 0.1633 & 0.7006
& 0.2499 & 0.2279 & 0.6710
& 0.2136 & 0.1797 & 0.7076 \\
& (0.0014) & (0.0010) & (0.0023)
& (0.0015) & (0.0016) & (0.0023)
& (0.0015) & (0.0013) & (0.0027) \\
\bottomrule
\end{tabular}%
}

\caption{Recommendation performance across three datasets using LLMs for alternative generation tasks. Parentheses report standard deviations.}
\label{generation}
\end{table}

\section{Evidence on Faithfulness of RecPIE's Explanations}
\label{appen:faithfulness_evidence}

The goal of RecPIE is to optimize explanations for their \emph{usefulness} in improving recommendation accuracy. As discussed in Section~\ref{sec:usefulness_faithfulness}, usefulness does not necessarily imply faithfulness. Nevertheless, we conduct additional evaluations in this section to examine the faithfulness of generated explanations in two respects. The first is \emph{model-side faithfulness} \citep{lyu2024faithful}, which measures whether an explanation reflects how the Recommendation Component arrives at its prediction. The second is \emph{user-side faithfulness}, which measures whether an explanation reflects the reasons behind the consumer's own choice. This is sometimes also referred to as \emph{plausibility} in the XAI literature \citep{lyu2024faithful}. We obtain these metrics for RecPIE and compare them to the top-performing baselines that generate explanations, ExpCTR, LLM2ER-EQR, TallRec, and PETER.

\subsection{Model-Side Faithfulness}
\label{sec:faithfulness_model}

Following the notation of Section~\ref{sec:framework_ppo}, $k$ denotes an observation in the test set of Section~\ref{sec:context_exp_setup}. We additionally write $H_k$ for its consumption history and $\hat{y}_k(H, \mZ)$ for the rating predicted when the history is $H$ and the explanation is $\mZ$, where $\mZ_k = (\mZ^{\mathrm{pos}}_k, \mZ^{\mathrm{neg}}_k)$ for RecPIE and $\mZ_k$ is the single explanation that each baseline generates with the same prompt $\mathbf{X}^{(k)}$. We select the following faithfulness metrics according to \citet{lyu2024faithful}, each of which serves as a \emph{necessary condition} that a faithful explanation should pass:
 
\subsubsection{Input ablation consistency.}
If an explanation faithfully identifies decision-relevant inputs, removing a place mentioned in the explanation should affect the prediction more than removing an unmentioned place. For each observation $k$ whose explanation mentions at least one place in $H_k$ and leaves at least one other unmentioned, let $M_k$ denote the set of mentioned places. For each $h\in M_k$, we randomly select an unmentioned place $u_h$ from $H_k$. Holding the explanation fixed, we compute
\begin{align*}
D_{\mathrm{mention}}
&= \frac{1}{|K|}\sum_{k\in K}\frac{1}{|M_k|}
\sum_{h\in M_k}
\left|
\hat{y}_k(H_k\setminus\{h\},\mZ_k)
-\hat{y}_k(H_k,\mZ_k)
\right|,\\
D_{\mathrm{other}}
&= \frac{1}{|K|}\sum_{k\in K}\frac{1}{|M_k|}
\sum_{h\in M_k}
\left|
\hat{y}_k(H_k\setminus\{u_h\},\mZ_k)
-\hat{y}_k(H_k,\mZ_k)
\right|,
\end{align*}
where $K$ denotes the set of eligible observations. A place counts as mentioned when its full name appears in the explanation, ignoring capitalization; for RecPIE, we consider either the positive or negative explanation. We report $D_{\mathrm{mention}}/D_{\mathrm{other}}$, where values above one indicate that mentioned places have greater influence on predictions than unmentioned places.

\subsubsection{Input sensitivity.}
If explanations reflect the model's predictive behavior, changes in the input that substantially affect predictions should also produce larger changes in explanations. For each test observation, we randomly replace one place in the consumer's history with a place visited by another consumer, obtaining a modified history $H'_k$ and a regenerated explanation $\mZ'_k$. We measure changes in prediction and explanation as
\begin{equation*}
X_k =
\left|
\hat{y}_k(H'_k,\mZ'_k)
-\hat{y}_k(H_k,\mZ_k)
\right|,
\qquad
Y_k =
1-\cos\left(g(\mZ_k),g(\mZ'_k)\right),
\end{equation*}
where $g(\cdot)$ denotes the DistilBERT text encoder as used in RecPIE, which is applied to the concatenated positive and negative explanations for RecPIE or the single explanation for each baseline. We report Pearson and Spearman correlations between $X_k$ and $Y_k$, averaged over ten independent runs with different random seeds. Higher correlations indicate greater consistency between changes in explanations and predictions.

\subsubsection{OOD predictive power.}
We additionally examine whether RecPIE maintains predictive accuracy when consumer- or place-specific parameters cannot be learned from training data. Specifically, we evaluate previously unseen consumers and places, whose learned embeddings fall back to reserved out-of-vocabulary representations, while explanations can still be generated from available visitation histories and candidate place names. As we show in Appendix~\ref{appen:coldstart}, RecPIE achieves the best predictive performance among the evaluated methods, with RMSE of 0.3148 and AUC of 0.6412 for unseen consumers, and RMSE of 0.3301 and AUC of 0.6236 for unseen places. These findings are consistent with explanations providing informative signals beyond learned consumer- and place-specific parameters, although with the caveat that predictive performance alone does not establish faithfulness.

Table~\ref{tab:faithfulness_paper} summarizes the input ablation and sensitivity results. RecPIE achieves the highest input ablation consistency ratio (1.71), indicating that removing places mentioned in its explanations changes predictions substantially more than removing unmentioned places. It also achieves among the highest input sensitivity correlations, with a Pearson correlation of 0.798 and a Spearman correlation of 0.783, suggesting that changes in explanations are highly correlated with changes in predictions. Together with the OOD results, these findings provide complementary evidence consistent with the faithfulness of RecPIE's explanations to the recommender's predictive behavior.

\begin{table}[htbp]
\centering
\footnotesize
\setlength\extrarowheight{2pt}
\setlength\tabcolsep{5pt}
\begin{tabular}{lccc}
\toprule
Method &
\shortstack{Input ablation consistency\\
$D_{\mathrm{mention}}/D_{\mathrm{other}}$ $\uparrow$} &
\shortstack{Input sensitivity\\Pearson $\uparrow$} &
\shortstack{Input sensitivity\\Spearman $\uparrow$} \\
\midrule
RecPIE (Ours) & \textbf{1.71} (0.06) & 0.798 (0.013) & 0.783 (0.012) \\
ExpCTR & 1.35 (0.05) & 0.734 (0.019) & 0.723 (0.017) \\
LLM2ER-EQR & 1.16 (0.04) & 0.748 (0.017) & 0.745 (0.017) \\
TallRec & 1.49 (0.05) & \textbf{0.801} (0.013) & \textbf{0.791} (0.012) \\
PETER & 1.04 (0.04) & 0.644 (0.027) & 0.623 (0.025) \\
\bottomrule
\end{tabular}
\caption{\textbf{Model-side faithfulness.}
Input ablation consistency and input sensitivity of generated explanations. Parenthetical values are standard deviations. Best values for each metric are highlighted in bold.}
\label{tab:faithfulness_paper}
\end{table}

\subsection{User-Side Faithfulness}
\label{sec:faithfulness_user}
We next examine \emph{user-side faithfulness}: whether generated explanations align with the reasons underlying consumers' preferences. Because consumers' true motivations are typically unobserved, we consider three complementary evaluations: (i) comparison with user-elicited reasons where such ground truths are available, (ii) recovery of common-sense preference signals (``white-box recovery''), and (iii) human-perceived informativeness.

\subsubsection{Ground-Truth Reasons: REASONER Benchmark.}
\label{appen:reasoner}
We evaluate RecPIE on REASONER \citep{chen2023reasoner}, a benchmark designed for explainable recommendation that provides human-annotated reasons elicited directly from consumers for individual interactions. Unlike conventional recommendation datasets, REASONER enables us to compare generated explanations with the reasons consumers themselves reported. It also helps alleviate a limitation in the Google Maps data, where user reviews are treated as ground-truth explanations, but they may reflect user feedback \emph{after} visiting the place and not necessarily the reasoning behind their original decisions to visit.

Table~\ref{tab:reasoner_rec_paper} reports recommendation accuracy against the full set of baselines, and Table~\ref{tab:reasoner_faith_paper} reports explanation quality using the same linguistic quality metrics as in Section~\ref{sec:exp_linguistic}. Note that REASONER is not a sequential recommendation task, so sequential models such as DIN, SASRec, and BERT4Rec may be disadvantaged in this benchmark.

RecPIE outperforms all baselines in recommendation accuracy, achieving relative improvements of 6.89\% in RMSE, 5.94\% in MAE, and 3.24\% in AUC over the best baseline for each metric. Its explanations also achieve the highest scores across all four explanation-quality metrics, indicating closer alignment with consumers' elicited reasons. Notably, RecPIE outperforms ExpCTR and LLM2ER-EQR, two recent reinforcement-learning-based explainable recommenders, on both prediction and explanation quality. These findings suggest that optimizing explanations for the \emph{usefulness} reward can also improve their alignment with consumers' reported motivations.

\begin{table}[htbp]
\centering
\footnotesize
\setlength\extrarowheight{2pt}
\begin{tabular}{l ccc}
\toprule
\textbf{Method} & RMSE $\downarrow$ & MAE $\downarrow$ & AUC $\uparrow$ \\
\midrule
\textbf{RecPIE (Ours)}
& \textbf{0.2202}$^{***}$ (0.0037) & \textbf{0.1758}$^{***}$ (0.0029) & \textbf{0.7586}$^{***}$ (0.0067) \\

\midrule
\multicolumn{4}{l}{\textit{Recommendation-focused baselines}} \\

RecSAVER & 0.2390 (0.0029) & 0.1961 (0.0027) & \underline{0.7348} (0.0063) \\
LLMRG    & 0.2391 (0.0041) & 0.1964 (0.0033) & 0.7338 (0.0074) \\
SASRec   & 0.2722 (0.0018) & 0.2172 (0.0016) & 0.6363 (0.0054) \\
DIN      & 0.2868 (0.0019) & 0.2254 (0.0017) & 0.6261 (0.0054) \\
TallRec  & \underline{0.2365} (0.0041) & 0.1962 (0.0033) & 0.7211 (0.0073) \\
BERT4Rec & 0.2660 (0.0017) & 0.2130 (0.0015) & 0.6457 (0.0053) \\
UniSRec  & 0.2454 (0.0015) & 0.1968 (0.0014) & 0.7063 (0.0058) \\
LLM4POI  & 0.2390 (0.0041) & 0.1959 (0.0035) & 0.7343 (0.0077) \\

\midrule
\multicolumn{4}{l}{\textit{Explanation-focused baselines}} \\

ExpCTR     & 0.2368 (0.0043) & 0.1898 (0.0035) & 0.7237 (0.0077) \\
LLM2ER-EQR & 0.2389 (0.0037) & 0.1922 (0.0031) & 0.7204 (0.0065) \\
RXR        & 0.2689 (0.0033) & 0.2246 (0.0029) & 0.5827 (0.0059) \\
A3NCF    & 0.2627 (0.0014) & 0.2222 (0.0013) & 0.5428 (0.0052) \\
SULM     & 0.2375 (0.0014) & 0.1922 (0.0013) & 0.7320 (0.0045) \\
AARM     & 0.2378 (0.0014) & 0.1932 (0.0012) & 0.7309 (0.0046) \\
MMALFM   & 0.2368 (0.0014) & 0.1914 (0.0012) & 0.7342 (0.0045) \\
ANR      & 0.2644 (0.0017) & 0.2073 (0.0015) & 0.6763 (0.0050) \\
MTER     & 0.2375 (0.0014) & 0.1923 (0.0013) & 0.7319 (0.0045) \\
AMCF     & 0.2616 (0.0018) & 0.2022 (0.0015) & 0.7031 (0.0047) \\
PETER    & 0.2474 (0.0017) & \underline{0.1869} (0.0015) & 0.7191 (0.0046) \\
UCEPic   & 0.2866 (0.0022) & 0.2120 (0.0018) & 0.6765 (0.0050) \\
PARSRec  & 0.2796 (0.0021) & 0.2072 (0.0017) & 0.6774 (0.0049) \\
\bottomrule
\end{tabular}
\caption{Recommendation accuracy on the REASONER dataset. RecPIE (Ours) is shown at the top; the best baseline in each metric is underlined; $^{***}$ denotes significance at $p<0.01$ relative to the best baseline.}
\label{tab:reasoner_rec_paper}
\end{table}

\begin{table}[htbp]
\centering
\footnotesize
\setlength\extrarowheight{2pt}
\begin{tabular}{l cccc}
\toprule
Model & BLEURT $\uparrow$ & Coverage $\uparrow$ & Informativeness $\uparrow$ & Fluency $\uparrow$ \\
\midrule
\textbf{RecPIE (Ours)} & \textbf{0.4178***} (0.0049) & \textbf{0.2979***} (0.0029) & \textbf{0.4873***} (0.0032) & \textbf{0.3771***} (0.0026) \\
\midrule
\multicolumn{5}{l}{\textit{Baselines that are able to generate explanations}} \\
TallRec & \underline{0.3928} (0.0041) & 0.2803 (0.0021) & \underline{0.4625} (0.0025) & \underline{0.3555} (0.0022) \\
ExpCTR & 0.3911 (0.0043) & 0.2811 (0.0021) & 0.4613 (0.0027) & 0.3530 (0.0022) \\
LLM2ER-EQR & 0.3604 (0.0039) & 0.2775 (0.0019) & 0.4404 (0.0021) & 0.3411 (0.0022) \\
RXR & 0.2486 (0.0037) & 0.1965 (0.0017) & 0.3208 (0.0017) & 0.2930 (0.0017) \\
A3NCF & 0.2437 (0.0021) & 0.1937 (0.0013) & 0.3174 (0.0011) & 0.2878 (0.0009) \\
SULM & 0.2624 (0.0023) & 0.2262 (0.0013) & 0.3226 (0.0013) & 0.2976 (0.0009) \\
AARM & 0.2621 (0.0023) & 0.2253 (0.0013) & 0.3221 (0.0013) & 0.2973 (0.0009) \\
MMALFM & 0.2698 (0.0023) & 0.2315 (0.0013) & 0.3277 (0.0013) & 0.3019 (0.0009) \\
ANR & 0.2601 (0.0023) & 0.2478 (0.0013) & 0.3258 (0.0011) & 0.3001 (0.0009) \\
MTER & 0.2557 (0.0021) & 0.2470 (0.0015) & 0.3301 (0.0011) & 0.3001 (0.0009) \\
AMCF & 0.2646 (0.0023) & 0.2429 (0.0015) & 0.3239 (0.0011) & 0.2983 (0.0009) \\
PETER & 0.3873 (0.0039) & \underline{0.2822} (0.0021) & 0.4379 (0.0023) & 0.3499 (0.0022) \\
UCEPic & 0.3625 (0.0041) & 0.2771 (0.0021) & 0.4174 (0.0023) & 0.3427 (0.0022) \\
PARSRec & 0.2329 (0.0036) & 0.1886 (0.0019) & 0.3101 (0.0020) & 0.2727 (0.0017) \\
\bottomrule
\end{tabular}
\caption{Linguistic quality of generated explanations on the REASONER dataset. The best baseline in each metric is underlined; $^{***}$ denotes significance at $p<0.01$ relative to the best baseline.}
\label{tab:reasoner_faith_paper}
\end{table}

\subsubsection{White-Box Recovery.}
\label{appen:common_sense}
When the places that drive a recommendation are known in advance, a faithful explanation should identify them. We identify manually, using common sense, a series of transparent cases in the Google Maps dataset in which the important places in the consumption history are evident, for example, a consumer whose last five visits include Joy Sushi and Yugen Ramen and who is recommended Gochi Japanese Fusion Tapas, where the two Japanese restaurants are the important places. We examine whether RecPIE's explanations recover such obvious preference signals in these ``white-box'' recommendation cases. 

Let $C$ denote the set of selected cases and $P_c$ the set of historical places supporting the common-sense inference for case $c$. We measure the fraction of cases in which the explanation identifies at least one such place:
\begin{equation*}
\mathrm{WB} =
\frac{1}{|C|}\sum_{c\in C}
\mathbbm{1}\left[
\text{the explanation for case }c
\text{ mentions at least one place in }P_c
\right].
\end{equation*}
For RecPIE, we evaluate the positive explanation.

Table~\ref{tab:common_sense_paper} presents the four cases and RecPIE's generated explanations. RecPIE and all four baselines successfully identify the relevant historical places in every case, achieving 100\% recovery. This provides a basic sanity check that RecPIE preserves readily identifiable preference signals, although it does not distinguish RecPIE from the baselines.

\begin{table}[htbp]
\centering
\scriptsize
\setlength\extrarowheight{0pt}
\setlength\tabcolsep{4pt}
\begin{tabular}{@{}p{2.2cm} p{3.6cm} p{2.1cm} p{7.3cm}@{}}
\toprule
Recommended candidate & Consumer's last 5 visits & Common-sense explanation & RecPIE's generated explanation \\
\midrule
Gochi Japanese Fusion Tapas & 99 Ranch Market; \textbf{Joy Sushi}; In-N-Out Burger; Chez TJ; \textbf{Yugen Ramen} & affinity for Japanese cuisine & Their visits to Joy Sushi and Yugen Ramen show a clear appreciation for fresh, \textbf{authentic Japanese flavors} and thoughtful presentation. Gochi Japanese Fusion Tapas offers a blend of traditional techniques with inventive twists that will satisfy their taste for quality and creativity. \\
McDonald's & Grant Park Plaza; Cuesta Park; Century Cinema 16; \textbf{Burger King}; \textbf{Panda Express} & affinity for fast food & The consumer's history of visiting Burger King and Panda Express suggests a routine centered on \textbf{fast, no-frills meals}. McDonald's fits this pattern with its consistent speed, familiar menu, and easy access, making it a natural next step in their convenience-driven eating habits. \\
Sports Page & Samovar; \textbf{Steins Beer Garden \& Restaurant}; \textbf{St. Stephen's Green}; McDonald's; San Antonio Center & affinity for bars \& pubs & The consumer's visits to Steins Beer Garden \& Restaurant and St. Stephen's Green suggest a preference for \textbf{relaxed, community-oriented spaces where conversation flows as freely as the drinks}. Sports Page offers a similar laid-back energy with its friendly crowd and casual setup, a natural fit for someone who enjoys sharing a pint in a lively, neighborly environment. \\
Bajis Cafe & \textbf{Starbucks}; \textbf{Peet's Coffee}; Giant's Liquor \& Services; Doppio Zero Mountain View; Total Wine \& More & affinity for coffeehouses & The consumer's routine at Starbucks and Peet's suggests a consistent craving for \textbf{familiar coffee experiences}, and Bajis Cafe offers a local twist with its bold flavors and welcoming atmosphere. It is a chance to stay loyal to the coffeehouse habit while discovering something new and nearby. \\
\bottomrule
\end{tabular}
\caption{Common-sense POI cases (Google Maps) where the reason to recommend is inferable. Boldface marks the visits that drive the common-sense inference.}
\label{tab:common_sense_paper}
\end{table}

\subsubsection{Human-Perceived Informativeness.}
An explanation that is faithful to the user should match the intuition of a person making the same decision.
Finally, we examine whether RecPIE's explanations align with human judgments of what constitutes an informative explanation. In the human study described in Section~\ref{sec:human_evaluation}, participants evaluate explanations generated by five methods for three recommendation scenarios, selecting the most informative explanation in each scenario.

For method $m$, we compute the average selection rate:
\begin{equation*}
\mathrm{HP}_m =
\frac{1}{3}\sum_{s=1}^{3}
\frac{n_{s,m}}{n_s},
\end{equation*}
where $n_s$ denotes the number of valid responses in scenario $s$, and $n_{s,m}$ denotes the number selecting method $m$.

As reported in Section~\ref{sec:human_evaluation}, RecPIE's explanations are selected as the most informative in 61.5\% of cases, substantially exceeding the 20\% random-selection benchmark and the strongest baseline, PETER (16.6\%). Although human-perceived informativeness does not necessarily establish faithfulness to consumers' true motivations, it provides complementary evidence that RecPIE's explanations are meaningful and convincing to human evaluators.

Taken together, these results suggest that RecPIE's prediction-informed explanations can align with both the recommender's predictive behavior and consumers' stated reasons, even though faithfulness is not explicitly optimized. Thus, optimizing explanations for predictive usefulness need not come at the expense of these other desirable explanation qualities.

\end{APPENDIX}

\AddToHook{enddocument/afteraux}{%
  \immediate\write18{cp output.aux a_main.aux}%
}	
\end{document}